\documentclass[apj,revtex4]{emulateapj}
\usepackage{ulem}
\usepackage{color}
\usepackage{CJKutf8}
\usepackage[dvipsnames]{xcolor}
\usepackage[colorlinks=true,citecolor=blue,linkcolor=blue]{hyperref}
\usepackage{pdflscape}
\usepackage{rotating}
\usepackage{dashrule}
\usepackage{soul}

\begin{document}

\newcommand {\dem}{DEM\,L\,71\ }
\newcommand {\ha}{H$\alpha$\ }
\newcommand {\hb}{H$\beta$\ }
\newcommand {\nii}{[\ion{N}{2}]\ }
\newcommand {\hii}{\ion{H}{2}\ }
\newcommand {\kms}{km~s$^{-1}$\ }
\newcommand {\oiii}{[\ion{O}{3}]\ }
\newcommand {\sii}{[\ion{S}{2}]\ }
\newcommand {\hst}{{\it{HST}}\ }
\newcommand {\chandra}{{\it{Chandra}}\ }
\newcommand {\xray}{X-ray\ }
\newcommand {\msun}{$M_\odot$}
\newcommand{\specialcell}[2][c]{%
  \begin{tabular}[#1]{@{}c@{}}#2\end{tabular}}

\newcommand{\tabledashline}{ 
\hdashrule[0.5ex][x]{2cm}{0.1pt}{0.5mm} & \hdashrule[0.5ex][x]{3.5cm}{0.1pt}{0.5mm} & \hdashrule[0.5ex][x]{5.5cm}{0.1pt}{0.5mm} & \hdashrule[0.5ex][x]{1cm}{0.1pt}{0.5mm} & \hdashrule[0.5ex][x]{4.5cm}{0.1pt}{0.5mm} & \hdashrule[0.5ex][x]{0.5cm}{0.1pt}{0.5mm} & \hdashrule[0.5ex][x]{4.5cm}{0.1pt}{0.5mm}\\}
\newcommand{\tabledashlinegap}{ 
\hdashrule[0.5ex][x]{2cm}{0.5pt}{1.5mm} & \hdashrule[0.5ex][x]{3.5cm}{0.5pt}{1.5mm} & \hdashrule[0.5ex][x]{5.5cm}{0.5pt}{1.5mm} & \hdashrule[0.5ex][x]{1cm}{0.5pt}{1.5mm} & \hdashrule[0.5ex][x]{4.5cm}{0.5pt}{1.5mm} & \hdashrule[0.5ex][x]{0.5cm}{0.5pt}{1.5mm} & \hdashrule[0.5ex][x]{4.5cm}{0.5pt}{1.5mm}\\}

\defcitealias{edwards2012}{EPS2012}
\defcitealias{pagnotta2015}{PS2015}

\title{Search for Surviving Companions of Progenitors of Young LMC Type Ia Supernova Remnants 
}
\author{Chuan-Jui Li \begin{CJK}{UTF8}{bsmi}(李傳睿)\end{CJK}\altaffilmark{1,2}, 
Wolfgang E. Kerzendorf\altaffilmark{3}, You-Hua Chu \begin{CJK}{UTF8}{bsmi}(朱有花)\end{CJK}\altaffilmark{2,4},\\
Ting-Wan Chen \begin{CJK}{UTF8}{bsmi}(陳婷琬)\end{CJK}\altaffilmark{5}, 
Tuan Do\altaffilmark{6}, Robert A.\ Gruendl\altaffilmark{4}, Abigail Holmes\altaffilmark{4},\\
R. Ishioka\altaffilmark{2}, 
Bruno Leibundgut\altaffilmark{3}, Kuo-Chuan Pan \begin{CJK}{UTF8}{bsmi}(潘國全)\end{CJK}\altaffilmark{7}, Paul M.\ Ricker\altaffilmark{4}, Daniel Weisz\altaffilmark{8} \\}
\affil{\\ $^1$ Graduate Institute of Astrophysics, National Taiwan University, Taipei 10617, Taiwan, R.O.C.\ \\
$^2$ Institute of Astronomy and Astrophysics, Academia Sinica, P.O. Box 23-141, Taipei 10617, Taiwan, R.O.C.\ 
\\ cjli@asiaa.sinica.edu.tw, yhchu@asiaa.sinica.edu.tw\\
$^3$ European Southern Observatory, Karl-Schwarzschild-Straße 2, 85748 Garching bei München, Germany\\
$^4$ Department of Astronomy, University of Illinois at Urbana-Champaign, 1002 West Green Street, \\
Urbana,IL 61801, U.S.A. \\
$^5$ Max-Planck-Institut f{\"u}r Extraterrestrische Physik, Giessenbachstra\ss e 1, 85748, Garching, Germany
\\
$^6$ Physics and Astronomy Department, University of California, Los Angeles, CA 90095-1547, USA\\
$^7$ Department of Physics and Institute of Astronomy, National Tsing Hua University, Hsinchu 30013, Taiwan, R.O.C.
\\
$^8$ Department of Astronomy, University of California, 501 Campbell Hall \#3411, Berkeley, CA 94720-3411, U.S.A.}
%======================================================================
% abstract
%======================================================================
%

\begin{abstract}
We have used two methods to search for surviving companions of Type Ia supernova progenitors in three Balmer-dominated supernova remnants (SNRs)
in the Large Magellanic Cloud: 0519--69.0, 0505--67.9 (DEM\,L71), and 0548--70.4.
In the first method, we use the {\it{Hubble Space Telescope}} photometric 
measurements of stars to construct color-magnitude diagrams (CMDs), and compare 
positions of stars in the CMDs with those expected from theoretical post-impact 
evolution of surviving main sequence or helium star companions. 
No obvious candidates of surviving companion are identified in this photometric 
search.  Future models for surviving red giant companions or with different 
explosion mechanisms are needed for thorough comparisons with these observations
in order to make more definitive conclusions.
In the second method, we use Multi-Unit  
Spectroscopic Explorer (MUSE) observations of 0519--69.0 and DEM\,L71
to carry out spectroscopic analyses of stars in order to use
large peculiar radial velocities as diagnostics of surviving companions.   
We find a star in 0519--69.0 and a star in DEM\,L71 moving at radial 
velocities of 182 $\pm$ 0 \kms and 213 $\pm$ 0 km~s$^{-1}$, more than 2.5$\sigma$ from 
the mean radial velocity of the underlying stellar population,
264 \kms and 270 km~s$^{-1}$, respectively.
These stars need higher-quality spectra to investigate their abundances and 
rotation velocities to determine whether they are indeed surviving companions
of the SN progenitors.

\end{abstract}

\subjectheadings{ISM: supernova remnants --- Magellanic Clouds --- ISM: individual (SNR 0519--69.0, SNR DEM\,L71, SNR 0548--70.4)}

%======================================================================
\section{Introduction}  \label{sec:intro}
%======================================================================

Type Ia Supernovae (SNe Ia) are important standardizable candles for determining 
cosmological distances.  While they are known to be thermonuclear explosions 
of carbon-oxygen white dwarfs (WDs) that have reached $\sim$ the Chandrasekhar 
mass limit, the exact origins of their progenitor systems 
are uncertain  (see 
\citealt{wang2012,maoz2014,ruiz-lapuente2014,wang2018,ruiz-lapuente2018a} for reviews). 
Two contrasting origins have been suggested: a double 
degenerate (DD) origin that results from the merger of two WDs 
\citep{iben1984, webbink1984}, and a single degenerate (SD) origin in 
which a WD accretes material from a non-degenerate companion 
\citep{whelan1973, nomoto1982}. 

In the DD case, both WDs are destroyed and no stellar remnant is expected.
In the SD case, the non-degenerate companion may be a main-sequence (MS) star \citep{ivanova2004, wang2010},
 a red giant (RG) \citep{hachisu1999, hachisu2008}, or a helium star \citep{wang2009b, bildsten2007}; 
the companion's surface material 
may be stripped by the SN blast but its dense core can survive.
Surviving companions may be spectroscopically identified by 
distinguishing characteristics, such as high translational velocities, high 
rotational velocities, and elevated metallicities \citep{ruiz-lapuente2004, 
gonzalez2009, gonzalez2012, kerzendorf2014}. %\citep{russell1992}.
Surviving companions may also be identified photometrically through their 
positions in the color-magnitude diagrams (CMDs) in comparison with models
of the post-impact evolution of surviving companions \citep[e.g.,][]{marietta2000,
pan2014}.  
If a surviving companion is identified near the explosion center 
of a young Type Ia supernova remnant (SNR), a SD origin can
be affirmed \citep{ruiz-lapuente1997, canal2001}.  

%--------------------------------------------
%             Figure 1

\begin{figure}[h]  
\epsscale{1.1}
%\centering
\hspace{-0.5cm}
\plotone{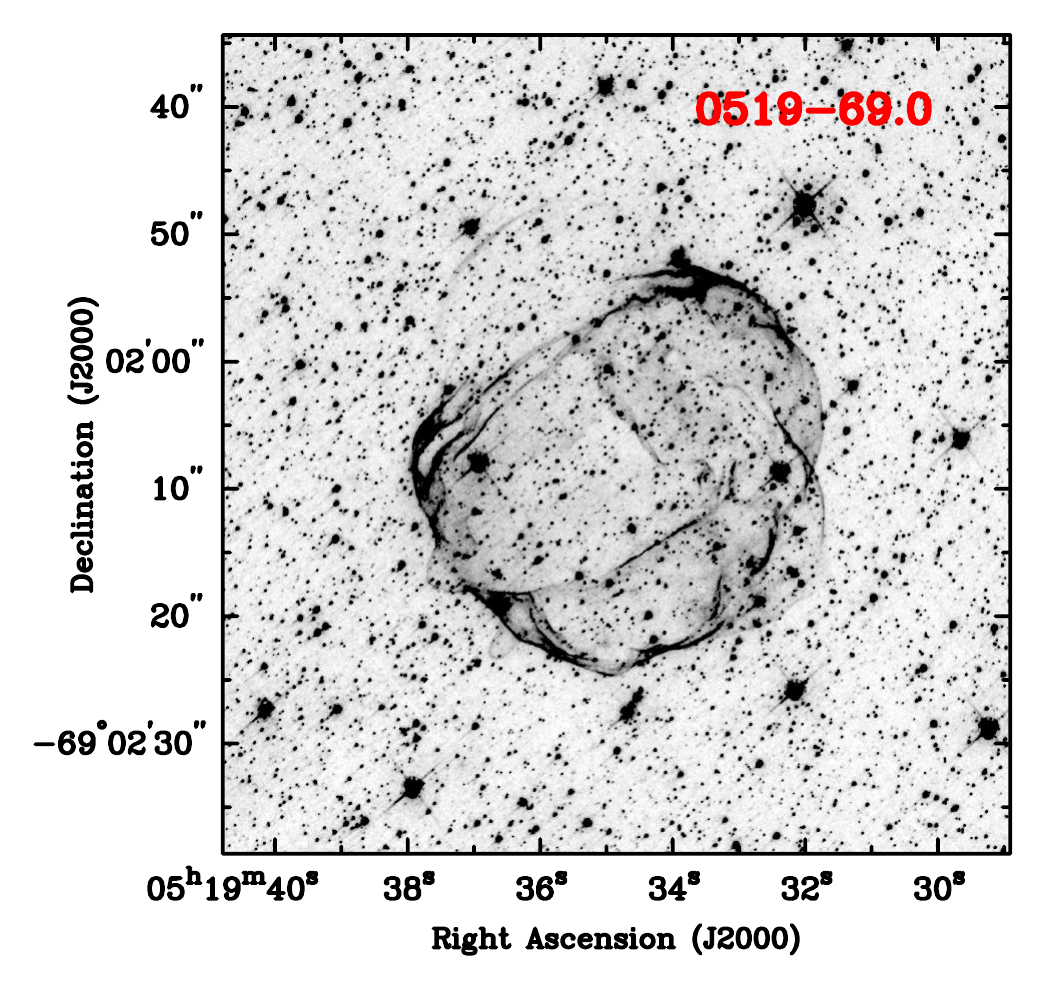}
\caption{High-resolution \ha (F658N) image of SNR  0519--69.0 obtained with the \hst ACS/WFC.}
\label{figure:0519}
\end{figure}

%--------------------------------------------
%             Figure 2

\begin{figure}[h]     
\epsscale{1.1}
%\centering
\hspace{-0.5cm}
\plotone{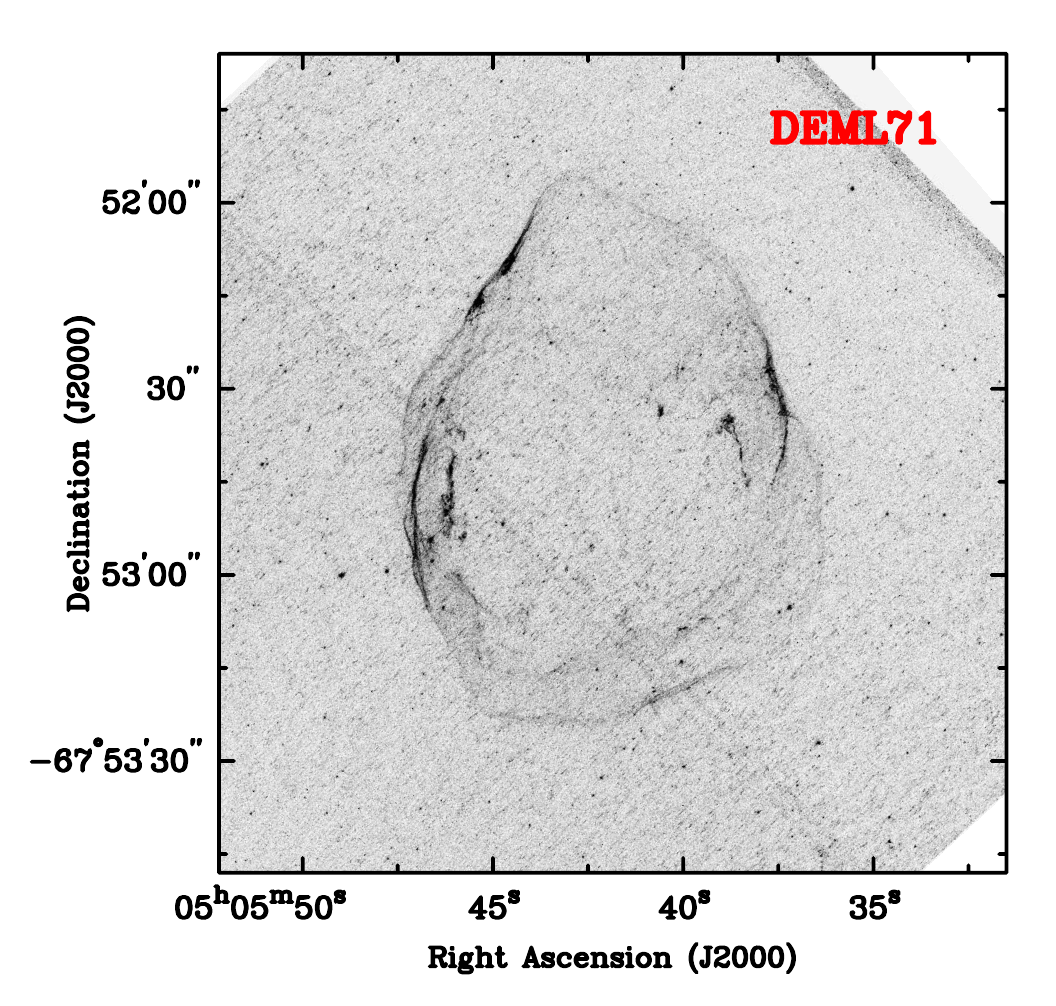}
\caption{High-resolution \ha (F656N) image of DEM L71 obtained with the \hst WFC3/UVIS.}
\label{figure:deml71}
\end{figure}

%--------------------------------------------
%             Figure 3

\begin{figure}[]    
\epsscale{1.1}
%\centering
\hspace{-0.5cm}
\plotone{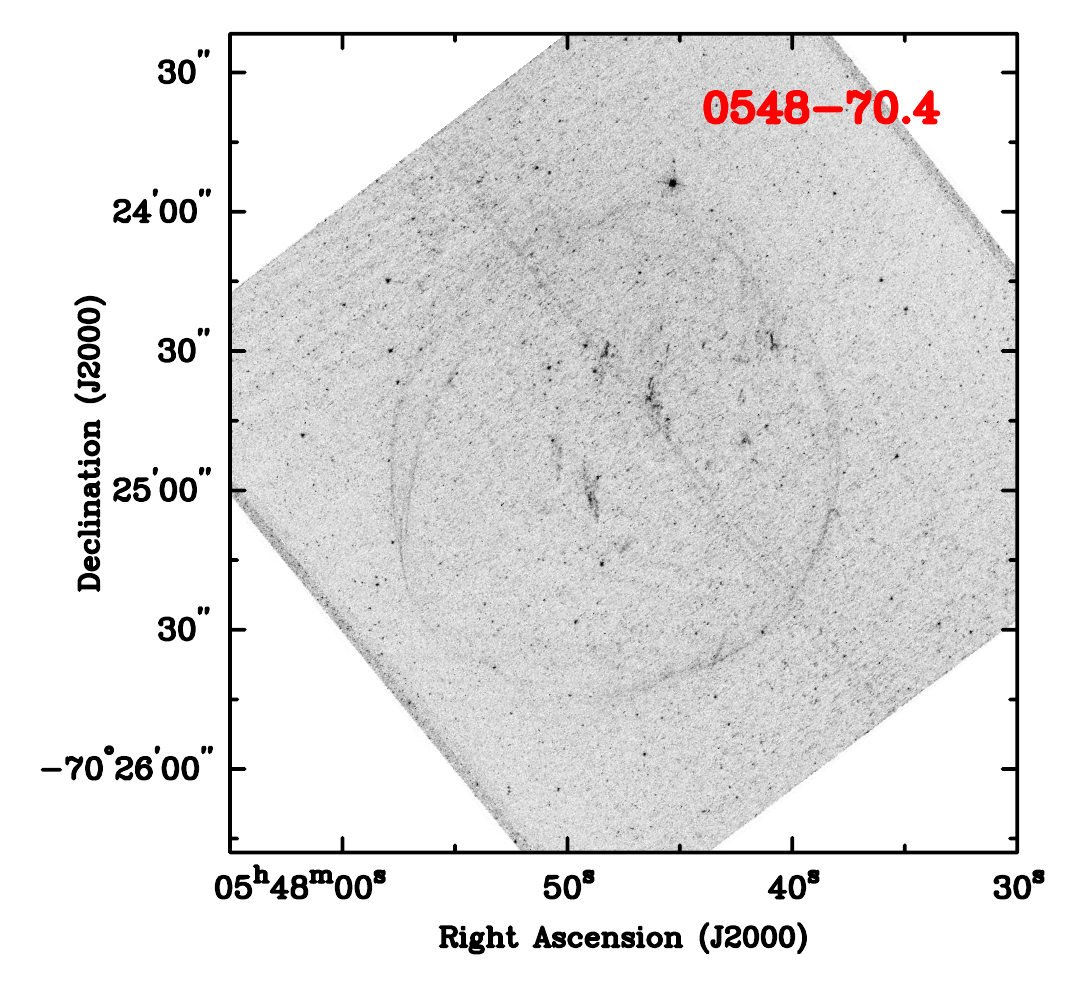}
\caption{High-resolution \ha (F656N) image of SNR 0548--70.4 obtained with the \hst WFC3/UVIS.}
\label{figure:0548}
\end{figure}

%--------------------------------------------
%             Table 1  
  
\begin{deluxetable*}{ccccccccc}
\tabletypesize{\scriptsize}
%\rotate
\tablewidth{0pc}
\tablecaption{\hst Observations}
\tablehead{
SNR & R.A. & Decl. & Filter  & Instrument & Date  & t$_{exp}$ (s) & PI & Proposal ID
}
\startdata 
0519--69.0 & 5:19:34.80 & -69:02:09.54 & F550M & ACS/WFC & 2011 Apr 21 & 750 & Hughes & 12017\\
                  &         &                  & F658N & ACS/WFC & 2011 Apr 21 & 4757 & Hughes & 12017\\
                  &         &                & F475W & WFC3/UVIS & 2014 Feb 21 & 1070 & Chu & 13282 \\
                  &         &                & F814W & WFC3/UVIS & 2014 Feb 21 & 1174  & Chu & 13282\\
DEM L71  & 5:05:41.70 & -67:52:39.90 & F475W & WFC3/UVIS & 2014 Mar 04 & 1050 & Chu & 13282\\
                  &         &                & F555W & WFC3/UVIS & 2014 Mar 04 & 1117  & Chu & 13282\\
                  &         &                & F656N  & WFC3/UVIS & 2014 Mar 05 & 1350  & Chu & 13282\\
                  &         &               & F814W & WFC3/UVIS & 2014 Mar 05 & 1050  & Chu & 13282 \\
0548--70.4 & 5:47:48.50 & -70:24:53.32 & F475W & WFC3/UVIS & 2013 Sep 20 & 1050 & Chu & 13282\\
                  &         &                & F555W & WFC3/UVIS & 2013 Sep 20 & 1170 & Chu & 13282 \\
                  &         &               & F656N  & WFC3/UVIS & 2013 Sep 20 &1350  & Chu & 13282 \\
                  &         &               & F814W & WFC3/UVIS & 2013 Sep 20 & 1111   & Chu & 13282
\enddata
%\tablecomments{The \hst observations were obtained in Program 13282 (PI: Chu)\\}
\label{table:hst}
\end{deluxetable*}

Type Ia SN progenitors' surviving companions have been searched for
in Galactic SNRs
Tycho \citep{ruiz-lapuente2004, fuhrmann2005, ihara2007, kerzendorf2009, 
gonzalez2009, kerzendorf2013, bedin2014, kerzendorf2018b, ruiz-lapuente2019}, 
SN 1006 \citep{gonzalez2012, kerzendorf2012, kerzendorf2018a}, and Kepler 
\citep{kerzendorf2014, ruiz-lapuente2018b}; however, none have been unambiguously identified 
near explosion centers of these SNRs.  Instead, a dense circumstellar 
medium (CSM) is detected in the SNR Kepler, and this dense CSM has been
used to argue for a SD origin, as it represents mass loss from the binary system
of the SN progenitor \citep{vandenbergh1973, vandenbergh1977, dennefeld1982, 
blair1991, williams2012}.
From X-ray studies, around 8--9 Type Ia SNRs are known in our Galaxy\citep{yamaguchi2014, martinez-rodriquez2018},
though only a smaller sample of those are suitable for searches of SN progenitors’ surviving companions.
Both the identification of Type Ia SNRs and the search for their SN progenitors' surviving companions in the Galaxy are hampered by the confusion and extinction in the Galactic plane, compounded by the uncertain distances to the SNRs.

%%%%%%%%  Centers  %%%%%%%%

%--------------------------------------------
%             Figure 4

\begin{figure}[t]   
\epsscale{1.2}
\vspace*{-0.5cm}\plotone{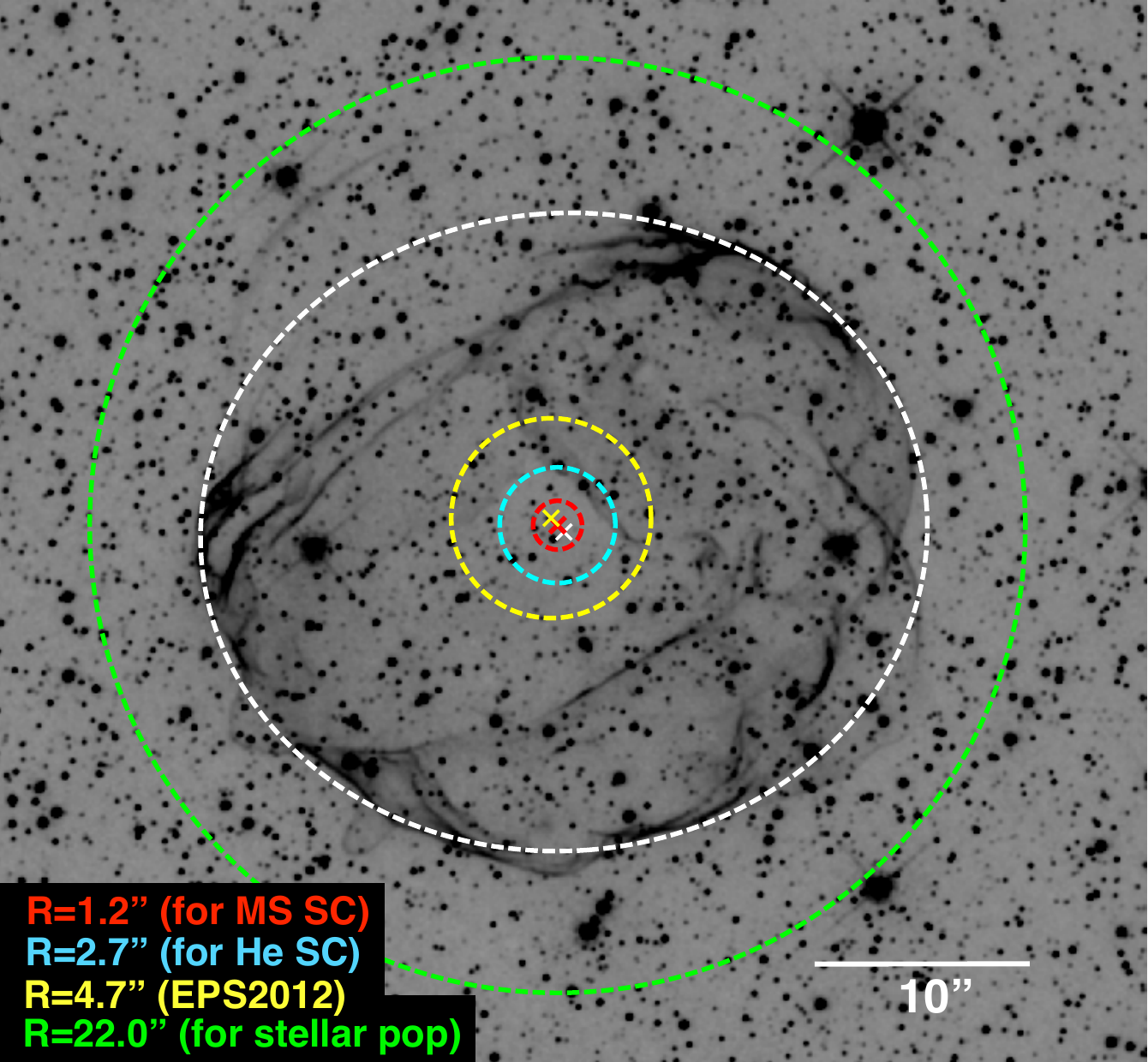}
\caption{
\ha image of SNR 0519--69.0.
%The Balmer-dominated shell is visually fitted with a $\sim$ 17\farcs1 and 15\farcs0 ellipse (dashed white) centered at 05$^{\mathrm{h}}$19$^{\mathrm{m}}$34$^{\mathrm{s}}$.72, $-$69$^\circ$02$'$07\farcs57 (J2000). 
The white cross marks the center of the fitted Balmer-dominated ellipse (dashed white),
the yellow cross marks the site of SN explosion of \citetalias{edwards2012},
and the red cross mark our adopted site of SN explosion, at
05$^{\mathrm{h}}$19$^{\mathrm{m}}$34$^{\mathrm{s}}$.77, 
$-$69$^\circ$02$'$07\farcs25 (J2000).  
The dashed red circle 
illustrates the maximum projected runaway distance for an MS surviving companion, 
and the dashed cyan circle represent the maximum projected runaway distance for a
helium star surviving companion; the dashed yellow circle marks the search radius 
of \citetalias{edwards2012}. The stars within 22\arcsec\ from the center 
are used to study the background stellar population (the dashed 
green circle).}
\label{figure:center0519}
\end{figure}

%%%%%%%%  Stars in central region of SNR 0519, Ha + RGB  %%%%%%%%

%--------------------------------------------
%             Figure 

\begin{figure*}   
\epsscale{1.2}
%\vspace*{-1.5cm}
\hspace*{0.4cm}
\includegraphics[height=8.5cm]{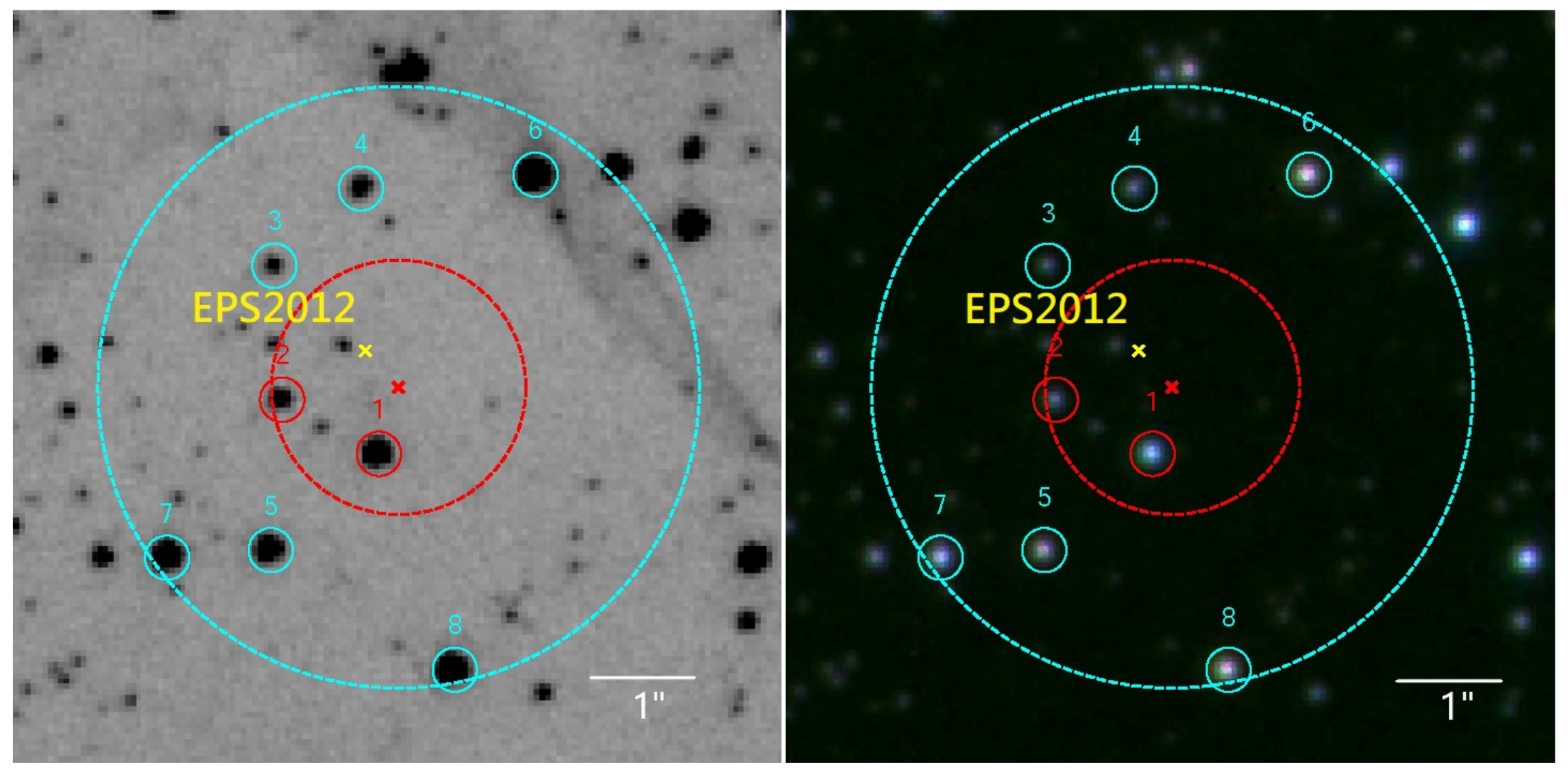}
\caption{
Left: \ha image of SNR 0519--69.0. 
Right: color-composite image of SNR 0519-69, with the F475W image in blue, 
the F550M image in green, and F814W image in red. The red cross marks the adopted explosion center at 
05$^{\mathrm{h}}$19$^{\mathrm{m}}$34$^{\mathrm{s}}$.77, 
$-$69$^\circ$02$'$07\farcs25 (J2000)
in this work.
The yellow cross marks the explosion center in \citetalias{edwards2012}.
The stars with V $<$ 23 within 2\farcs7 from the center 
are marked and numbered in both the \ha image and the color-composite image. 
The dashed red and cyan circles over the images illustrate our 1\farcs2 (0.3 pc) and 2\farcs7 (0.7 pc)
runaway distances for the MS and helium star surviving companions.}
\label{figure:stars0519}
\end{figure*}

%%%%%%%%  CMDs  of SNR 0519 %%%%%%%%

%--------------------------------------------
%             Figure 

\begin{figure*}
\epsscale{1.2}
\hspace*{-0.5cm}\plotone{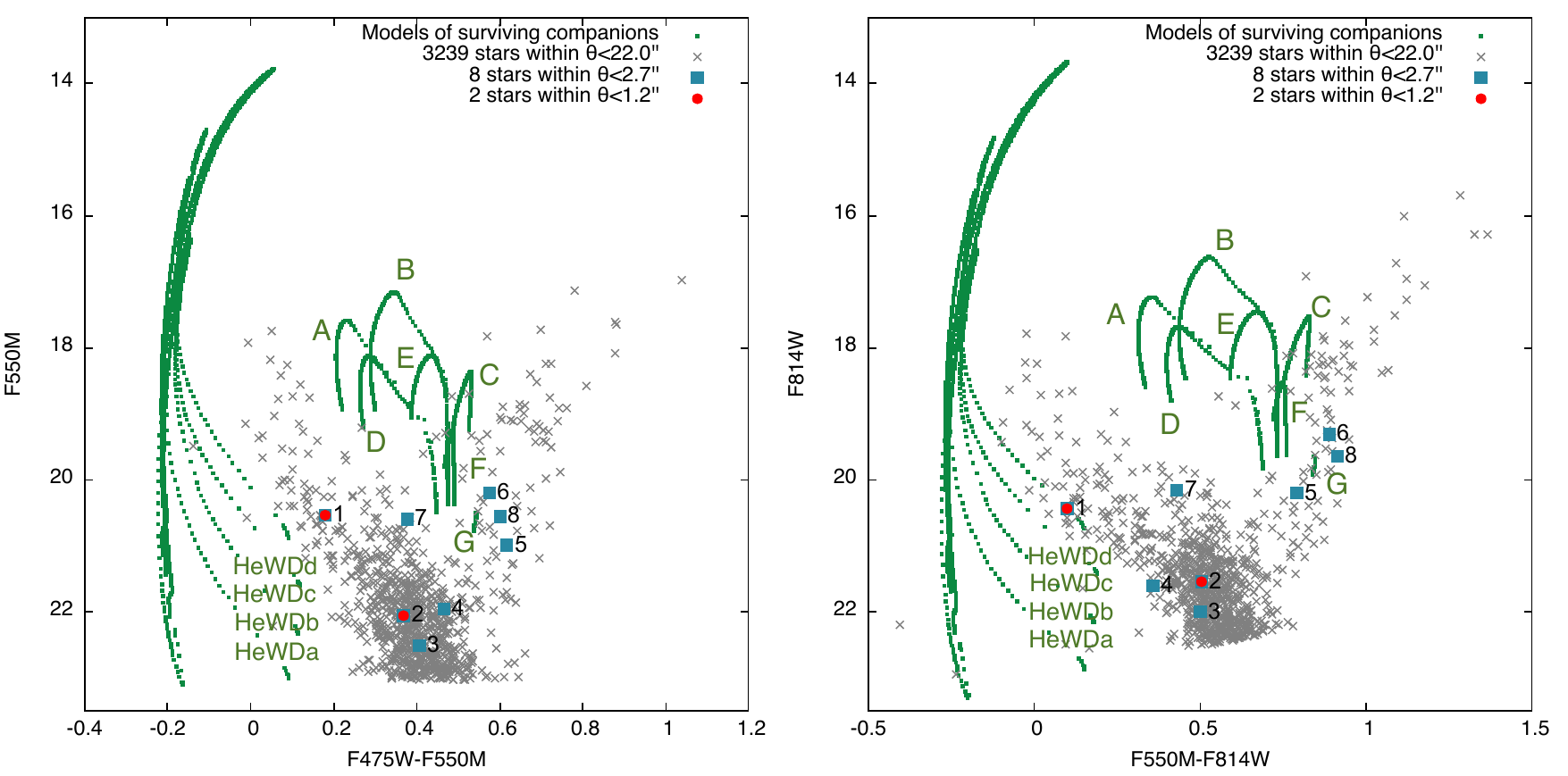}
\caption{
Left: The $V$ versus $B$ -- $V$ CMD of stars projected in and near the SNR 0519--69.0. Right: The $I$ versus $V - I$ CMD of the same stars.
Stars that are found within the runaway distances of helium star and main-sequence (MS) surviving companions from the center are plotted in blue filled squares and red filled circles, respectively. Stars that are superposed on and near the remnant are plotted as gray crosses to illustrate the general background stellar population. The post-impact evolutionary tracks are plotted in small green squares, and those of surviving helium star and MS companions are to the left and above the MS, respectively.
Different tracks of helium star and MS companions correspond to different companion mass in a range of 0.70 to 1.21 \msun\ and 1.17 to 1.88 \msun, respectively. The details of these helium star and MS companions can be found in models of \citet{pan2014}.
}
\label{figure:CMD0519}
\end{figure*}

%--------------------------------------------
%             Figure 

\begin{figure}
\epsscale{1.4}
\hspace*{-1.2cm}
\plotone{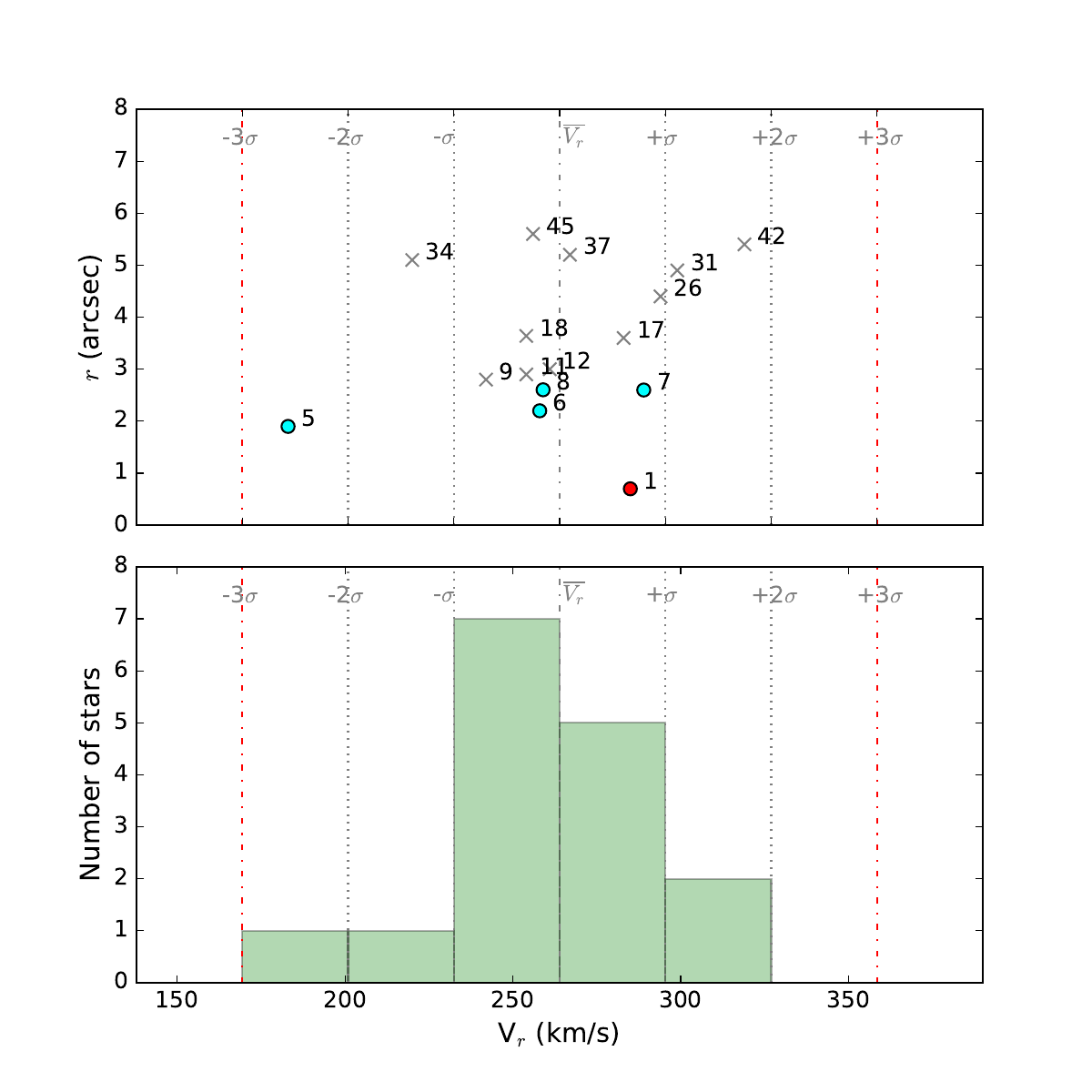}
\caption{ 
Top: a plot of the radial velocity ($V_r$) versus the distance to 
the site of SN explosion ($r$) for stars with $V < 21.6$ mag within a 
6\farcs0 radius in the SNR 0519--69.0. The stars within 1\farcs2 and 2\farcs7 from the site of explosion are marked as red and cyan solid circles, respectively. The background stars within 6\farcs0 from the explosion site are marked as grey crosses.
Bottom: the cumulative number of stars within standard deviations ($\sigma$) from 
the mean of radial velocity ($\overline{V_r}$).
}
\label{figure:fig_0519_V_r}
\end{figure}

%--------------------------------------------
%             Figure 

\begin{figure*}
%\epsscale{2.5}
\hspace*{-1.5cm}
\includegraphics[width=21cm]{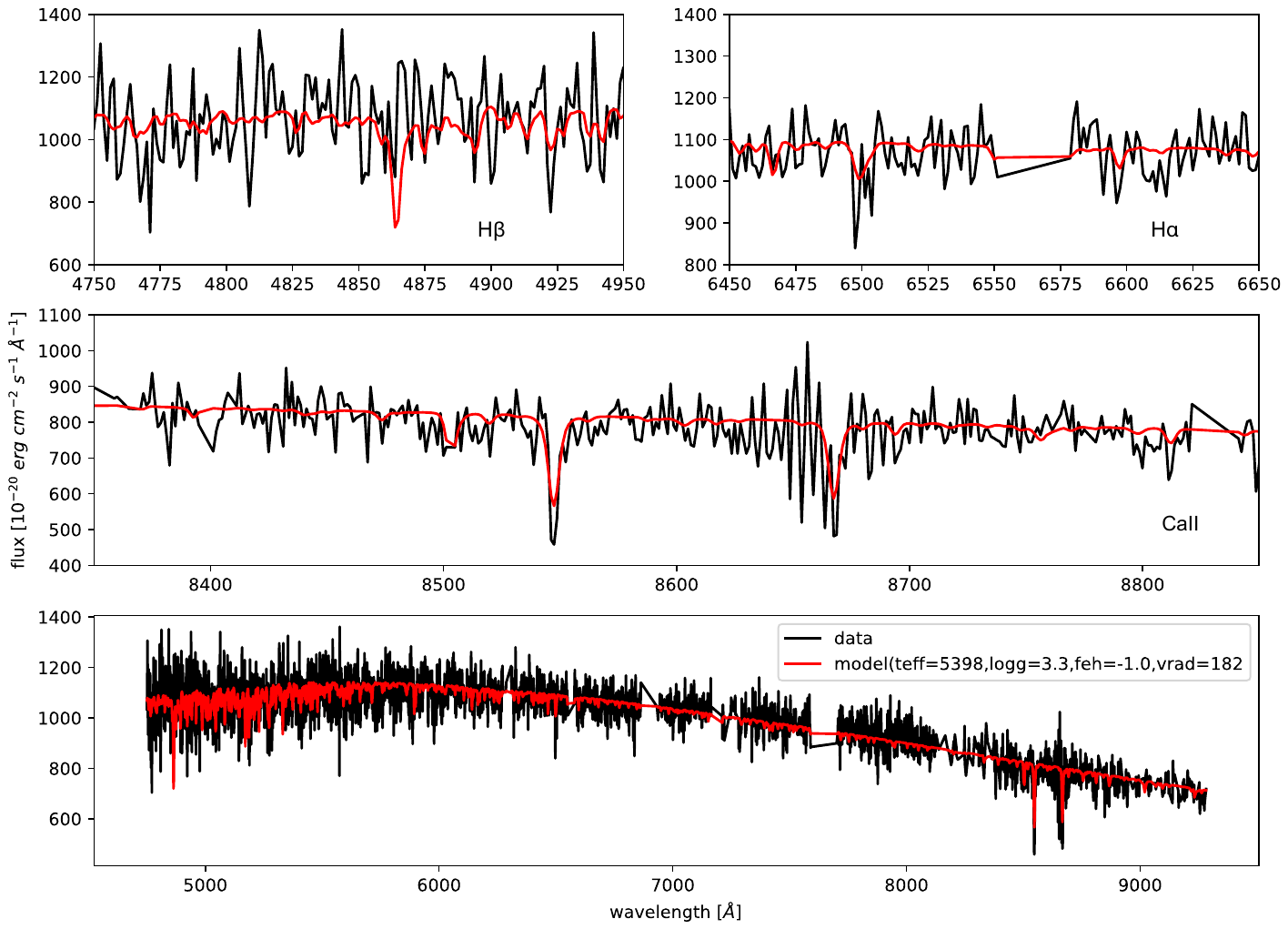}
\caption{ The VLT MUSE spectrum of star 5 in SNR 0519--69.0. Top: close-up wavelength windows around \hb and \ha lines.
Middle: close-up Ca II lines.
Bottom: the stellar spectrum and its model fit. A systematic uncertainty term has been added in quadrature to statistical uncertainties of the fit, as shown in Table \ref{table:starkit0519}.
}
\label{figure:fig_0519_star_5}
\end{figure*}

%--------------------------------------------
%             Figure 
 
\begin{figure}  
\epsscale{1.2}
\hspace*{-0.2cm}\plotone{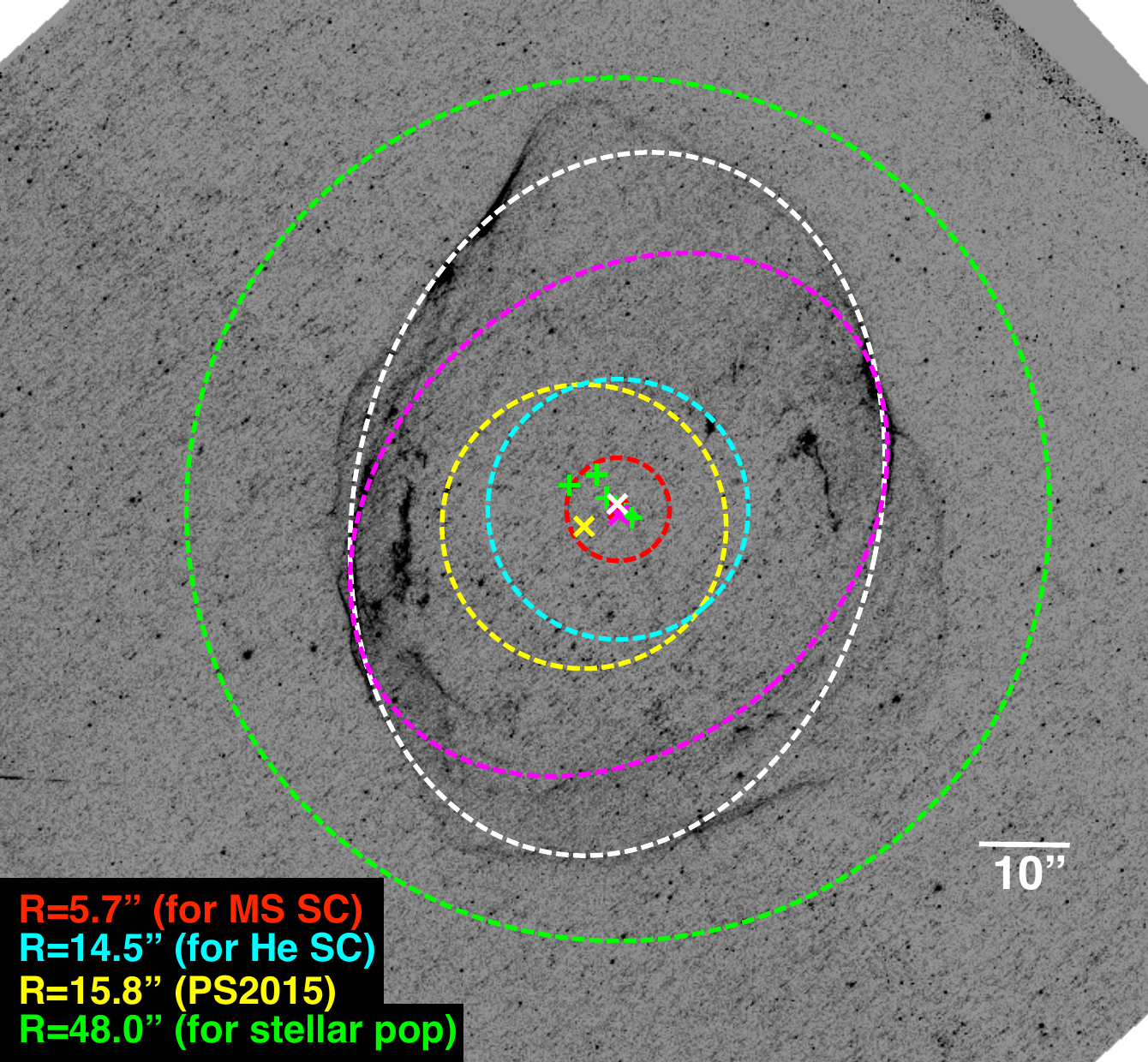}
\caption{
Same as Figure \ref{figure:center0519}, but for DEM L71. We have visually fitted two ellipses (white and magenta) to the shell and mark their centers and their average for comparison. The detail is described in the Section \ref{subsection:investigationofdeml71}.
\iffalse
\ha image of SNR DEM L71, with our 
center marked by a red cross, and the center of 
\citetalias{pagnotta2015} marked by a yellow cross.
The dashed red circle over the \ha image illustrates the runaway distance for the MS surviving companion, and the dash Cyan circle represent the runaway distance for the helium surviving companion; the dashed yellow circle marks the search area of \citetalias{pagnotta2015}. The green dash circle is the radius for the study of the underlying stellar population.
\fi
}
\label{figure:centerdeml71}
\end{figure}

%--------------------------------------------
%             Figure 

\begin{figure}  
\epsscale{1.2}
\hspace*{-0.2cm}\plotone{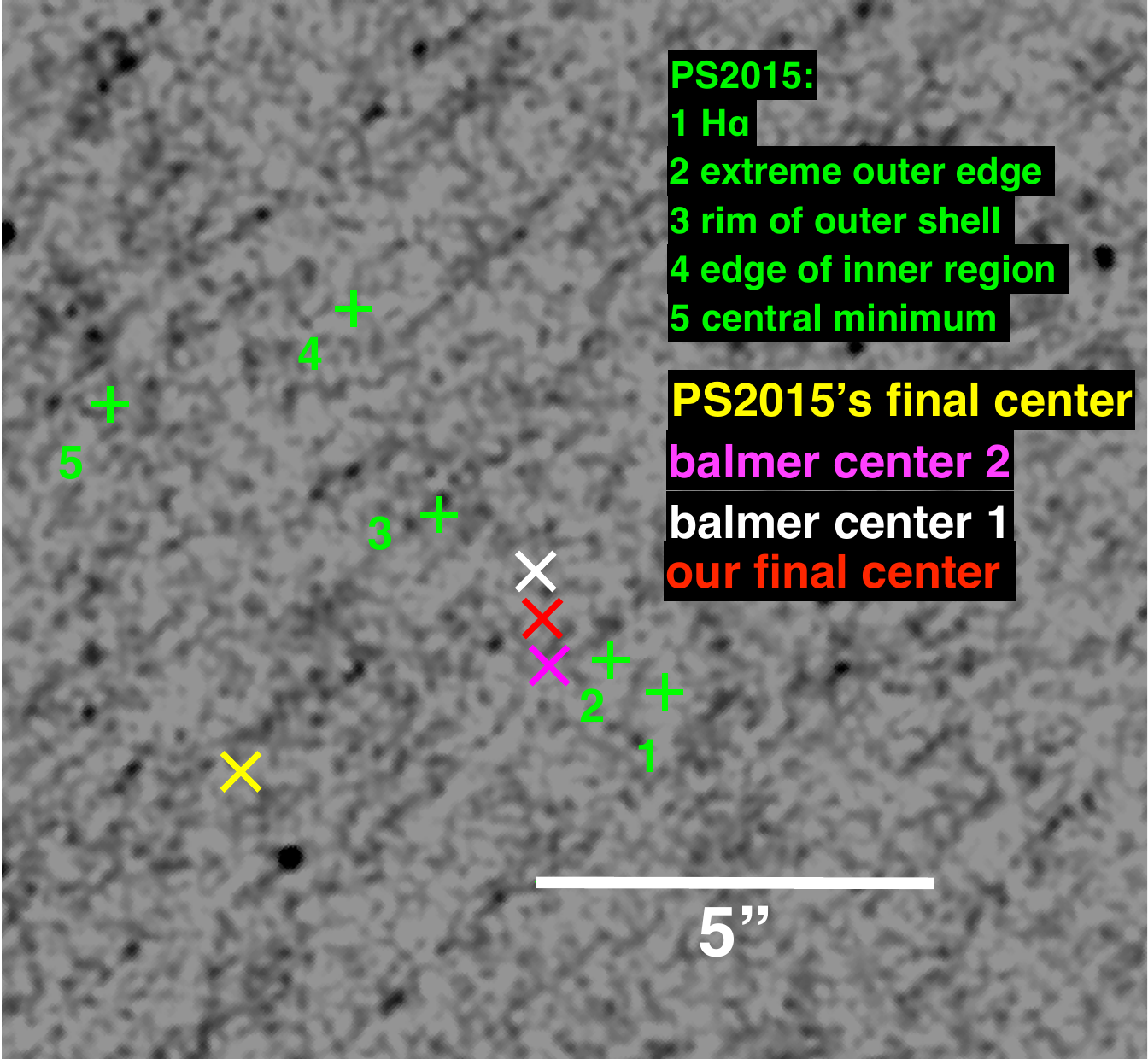}
\caption{
Close-up \ha image of different centers of DEM L71.
}
\label{figure:centerdeml71closeup}
\end{figure}

%%%%%%%%  Stars in central region of DEM L71, Ha + RGB  %%%%%%%%

%--------------------------------------------
%             Figure 

\begin{figure*}   
\epsscale{1.2}
%\vspace*{-1.5cm}
%\hspace*{-0.5cm}\plottwo{central_stars_deml71_ha.pdf}{central_stars_deml71_rgb.pdf}
\plotone{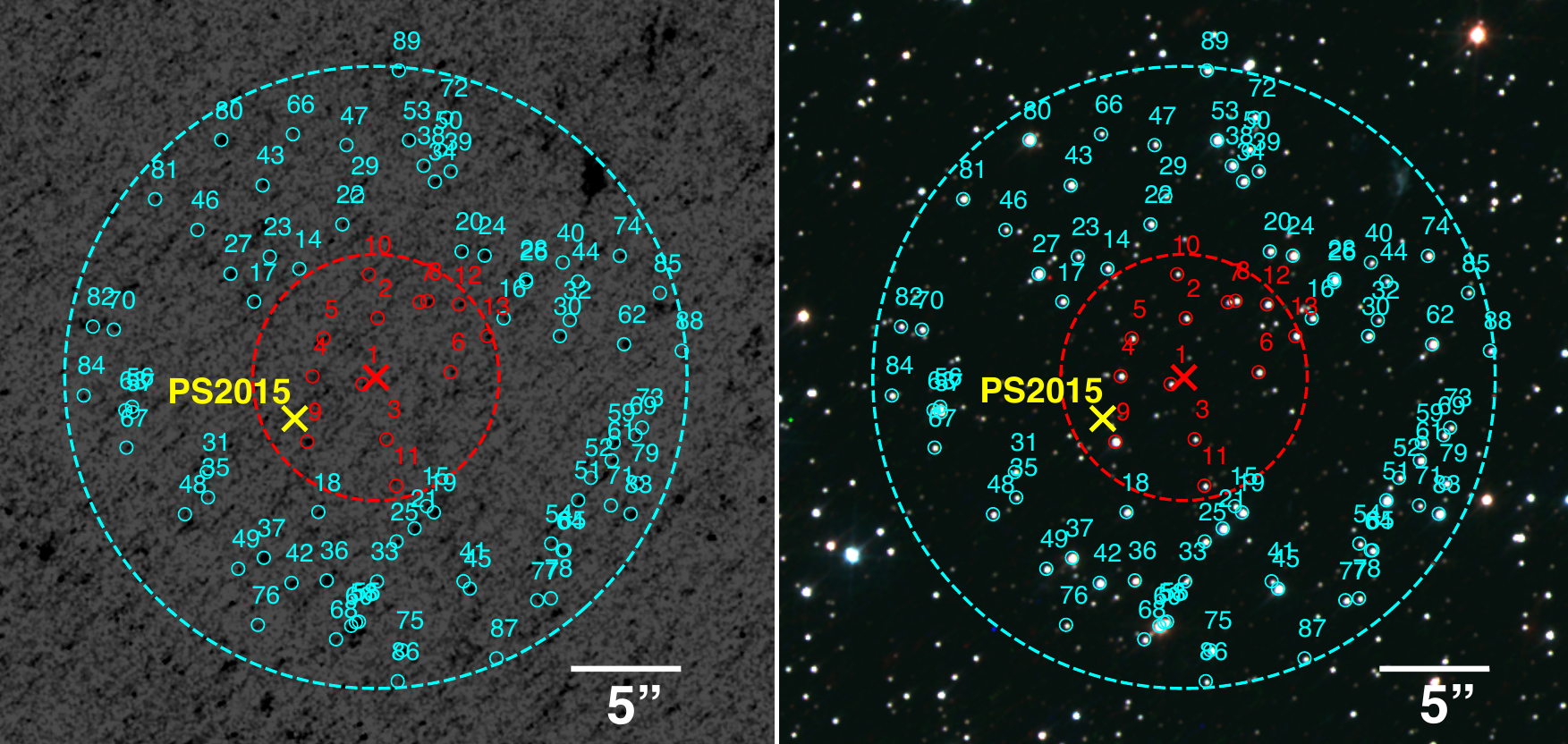}
\caption{
Same as Figure \ref{figure:stars0519}, but for DEM L71.
}
\label{figure:starsdeml71}
\end{figure*}

%%%%%%%%  CMDs of DEM L71 %%%%%%%%

%--------------------------------------------
%             Figure 

\begin{figure*}   
\epsscale{1.2}
\hspace*{-0.5cm}\plotone{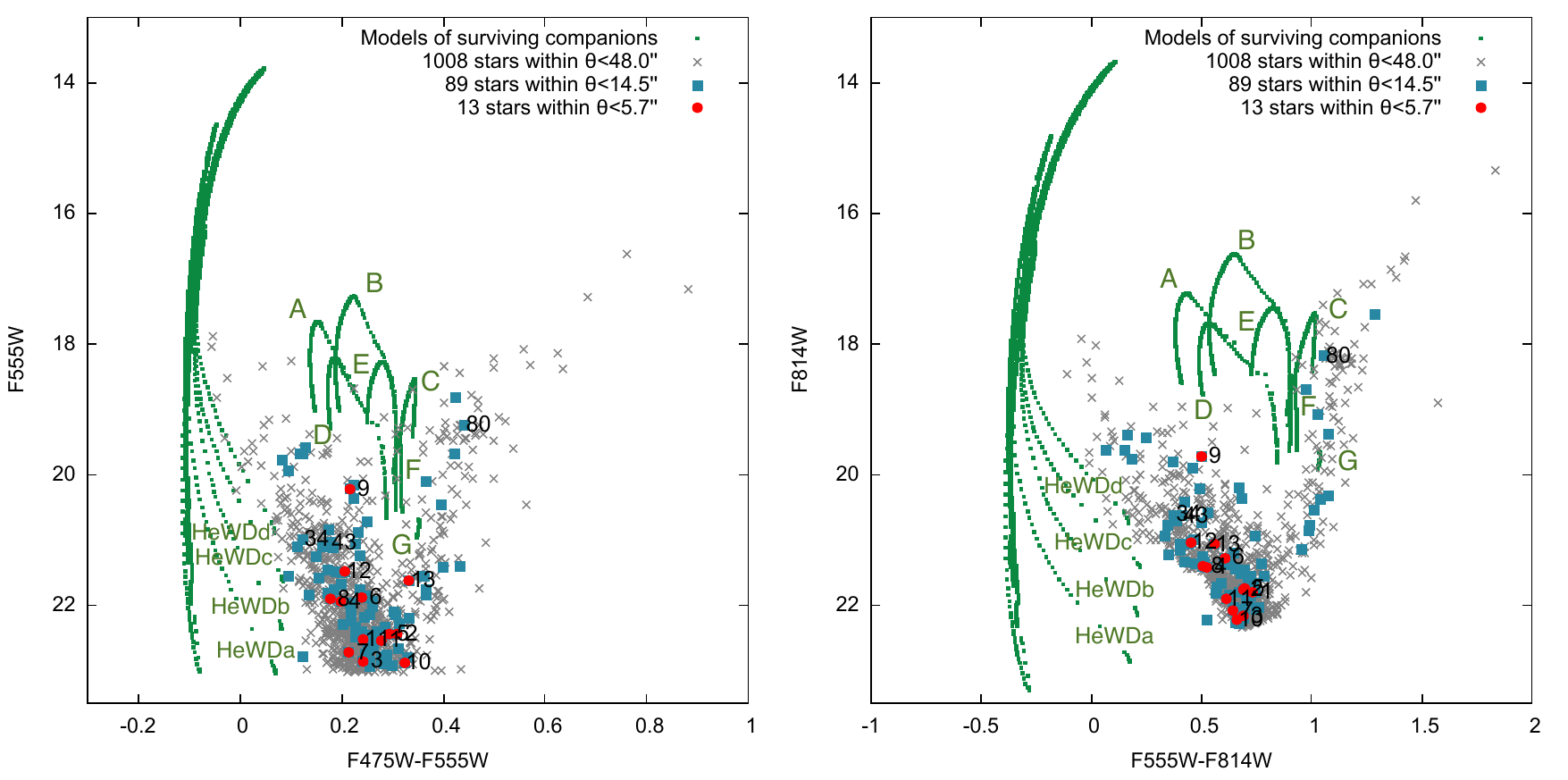}
\caption{Same as Figure \ref{figure:CMD0519}, but for DEM L71.
Left: V versus B -- V CMD.  Right: I versus V -- I CMD. 
The photometry of 89 candidates of MS surviving companions within 14\farcs5
from the center of DEM L71 is summarized in Table \ref{table:photometrydeml71}.
} 
\label{figure:CMDdeml71}
\end{figure*}

%--------------------------------------------
%             Figure 

\begin{figure*}  
\epsscale{1.2}
\hspace*{-0.5cm}\plotone{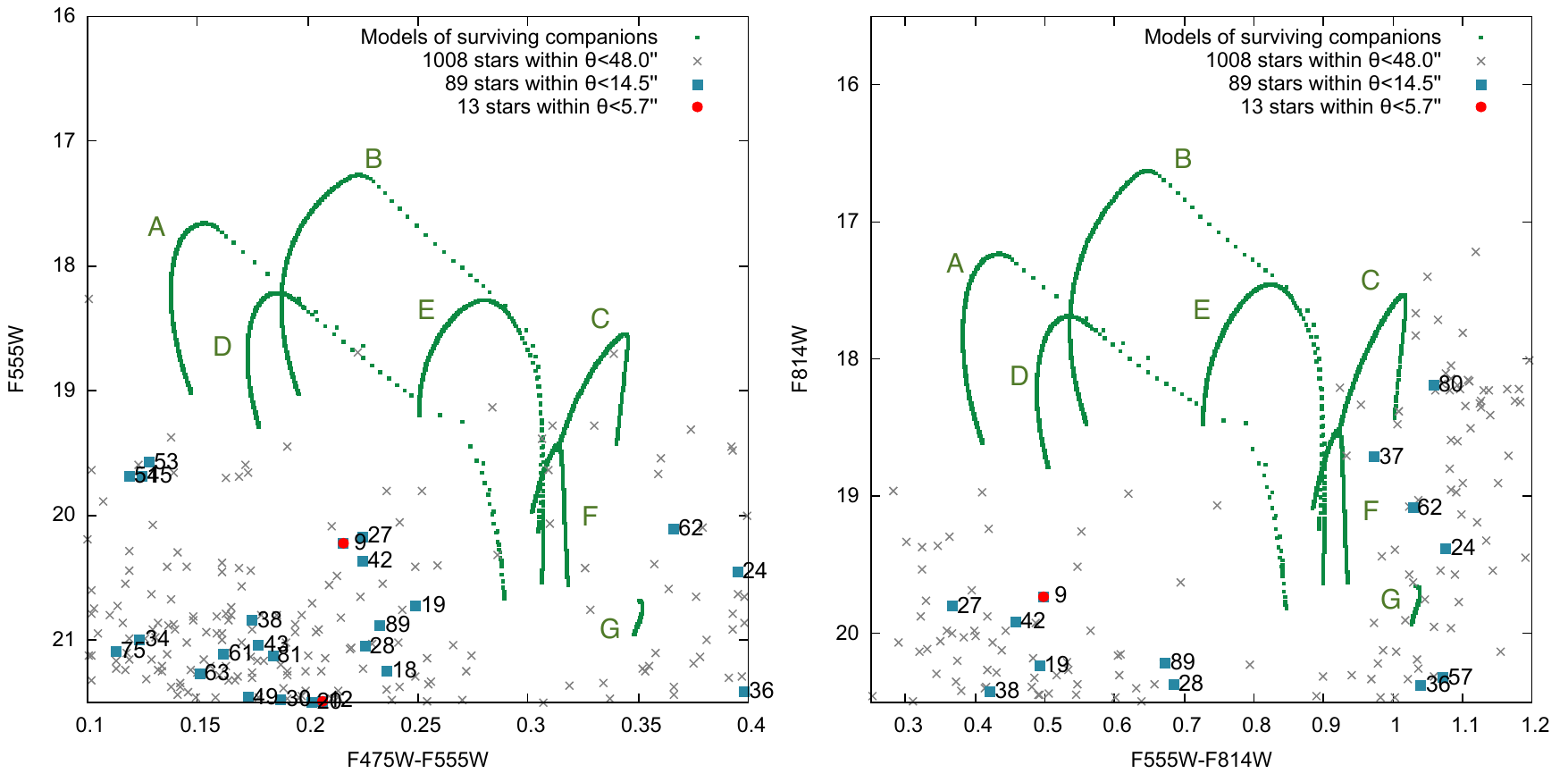}
\caption{ Close-up (Left) V versus B -- V and (Right) I versus V -- I CMDs near the 
evolutionary tracks for the case of surviving MS stars of DEML71.}
\label{figure:CMDdeml71closeup}
\end{figure*}

%--------------------------------------------
%             Figure 

\begin{figure*}  
\epsscale{1.2}
\hspace*{-0.5cm}\plotone{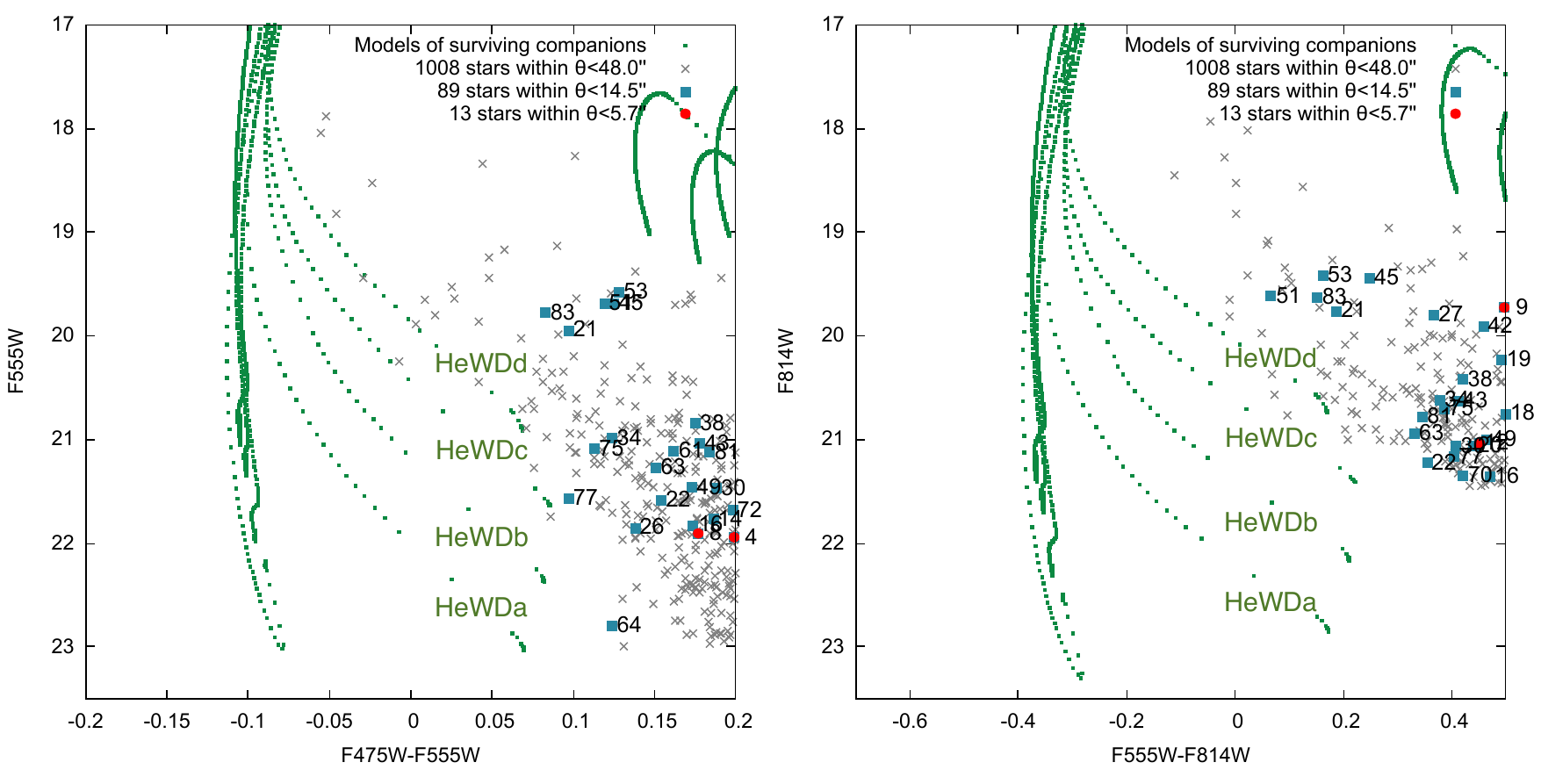}
\caption{A close-up I versus V -- I CMD near the evolutionary tracks 
for the case of surviving helium stars of DEML71. }
\label{figure:CMDdeml71closeup2}
\end{figure*}

%--------------------------------------------
%             Figure 

\begin{figure}
\epsscale{1.4}
\hspace*{-1cm}
\plotone{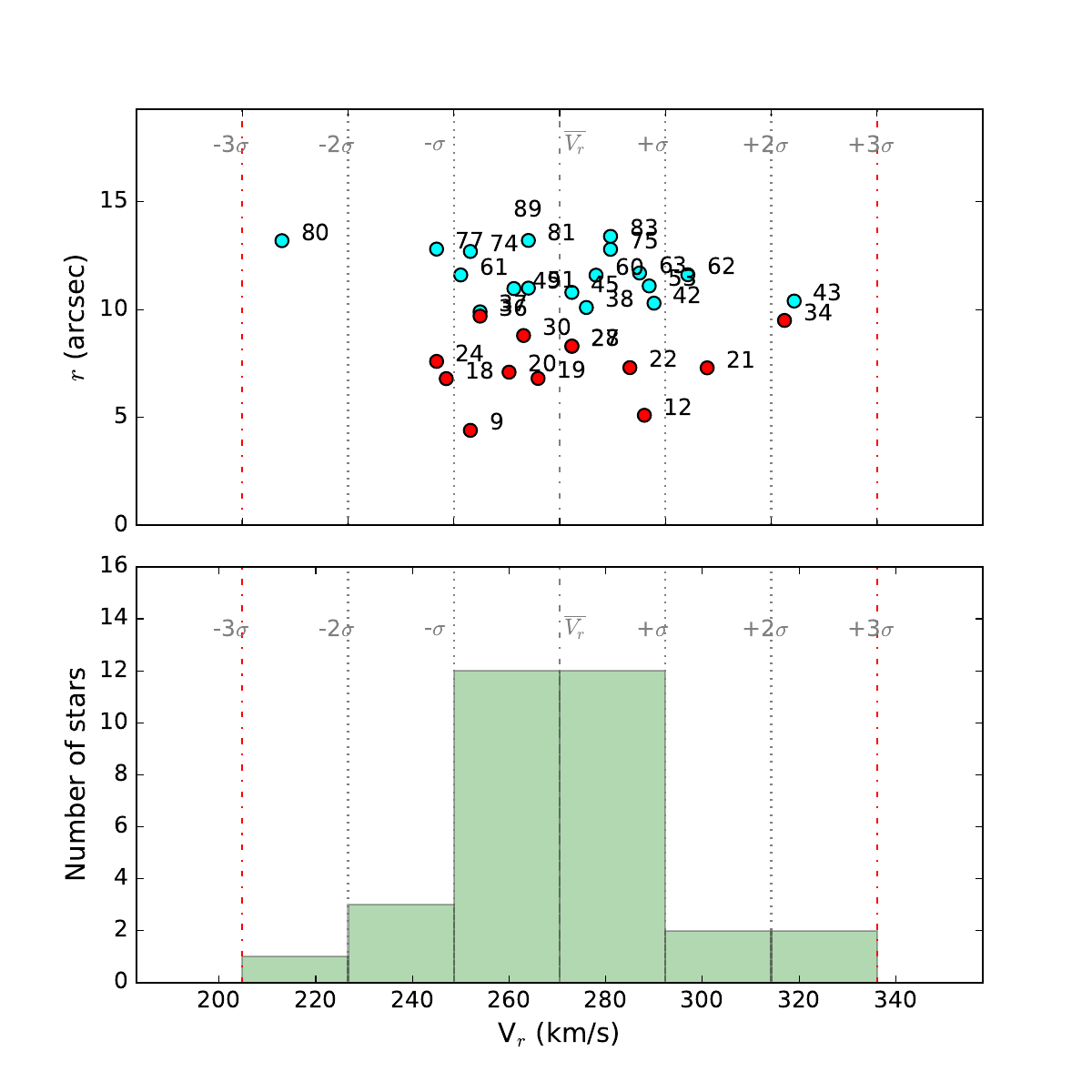}
\caption{
Top: a plot of radial velocity ($V_r$) versus the distance to 
the site of SN explosion ($r$) for stars with $V < 21.6$ mag within a 
14\farcs5 radius in DEM\,L71. The stars within 5\farcs7 and 14\farcs5 from the site of explosion are marked as red and cyan solid circles, respectively. 
Bottom: the cumulative number of stars within standard deviations ($\sigma$) from 
the mean of radial velocity ($\overline{V_r}$).
}
\label{figure:fig_deml71_V_r}
\end{figure}

%--------------------------------------------
%             Figure 

\begin{figure*}
%\epsscale{2.5}
\hspace*{-1.5cm}
\includegraphics[width=21cm]{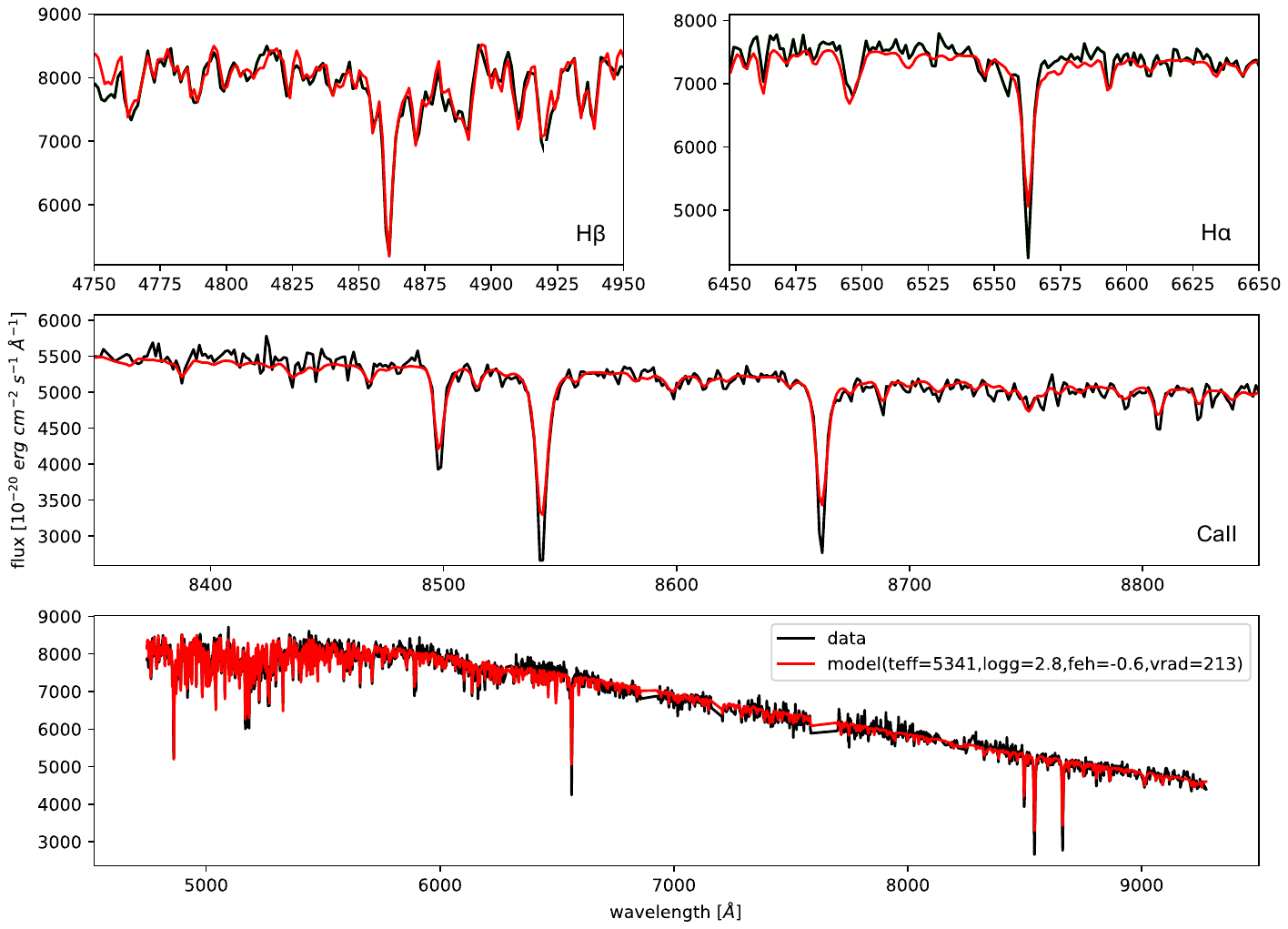}
\caption{ 
The VLT MUSE spectrum of star 80 in DEM L71. Top: close-up \hb and \ha lines. Middle: close-up Ca II lines. Bottom: the stellar spectrum and its model fit. A systematic uncertainty term has been added in quadrature to statistical uncertainties of the fit, as shown in Table \ref{table:starkitdeml71}.
}
\label{figure:fig_deml71_star_80}
\end{figure*}

%--------------------------------------------
%     Figure 

\begin{figure}   
\epsscale{1.2}
%\vspace*{-1.5cm}
\hspace*{-0.5cm}\plotone{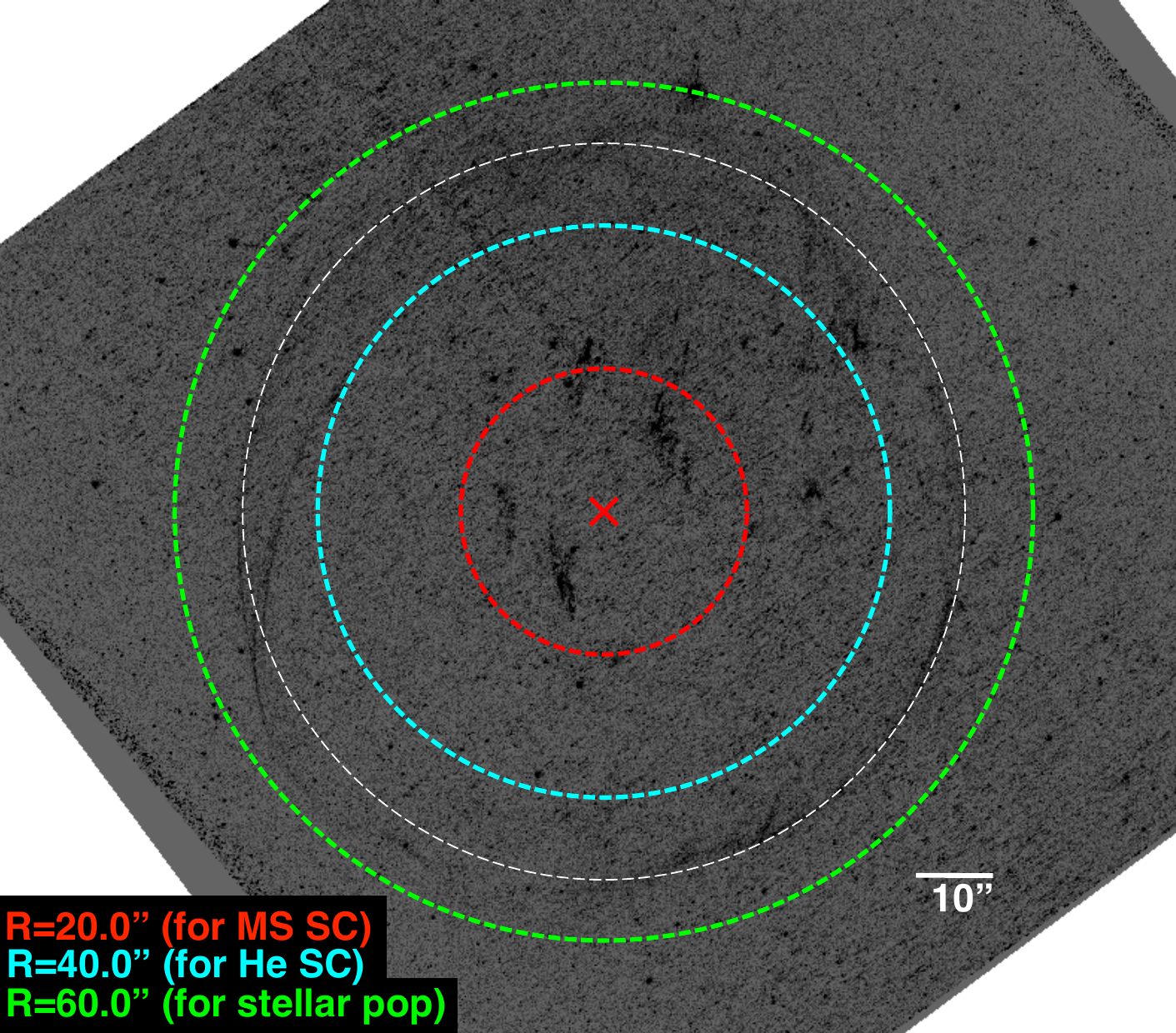}
\caption{
Same as Figure \ref{figure:centerdeml71}, but for SNR 0548--70.4. 
\iffalse
\ha image of SNR 0548--70.4, with our 
center marked by a white cross.
The dashed red circle over the \ha image illustrates our search radius for the MS surviving companion, and the dash Cyan circle represent our search radius for the helium surviving companion; the green dash circle is the radius for the study of the underlying stellar population.
\fi
}
\label{figure:center0548}
\end{figure}

The Large Magellanic Cloud (LMC), on the other hand, is an ideal galaxy
where we can study SNRs \citep{ou2018}
and search for surviving companions of SN progenitors
because it has a large sample of SNRs all 
at a known moderate distance, $\sim$
50 kpc \citep{pietrzynski2013, pietrzynski2019}, 
near enough for stars and SNRs to be 
easily resolved by the {\it{Hubble Space Telescope}} ({\it{HST}}). 
Furthermore, confusion and extinction along the lines of sight are 
minimized by the LMC disk's nearly face-on orientation \citep{nikolaev2004, 
olsen2002, vandermarel2001, zaritsky2004}.
Searches for surviving companions have been conducted for 4 young Type Ia 
SNRs in the LMC: 0509--67.5, 0519--69.0, 0505--67.9 (DEM\,L71), and 
0509--68.7 (N103B).  As summarized below, no surviving companion has
been unambiguously identified and confirmed.

{\it SNR 0509--67.5}: This remnant is 400 $\pm$ 50 yr old based on analyses 
of the light echoes of its SN \citep{rest2005, rest2008}.
 \hst images reveal a patch of diffuse emission with a point-like central source
 near the explosion center of this SNR more prominently in the red and near-IR
 bands than in the blue bands \citep{litke2017}.  The Gemini GMOS long-slit
 spectrum of this patch shows one emission line and it is identified as \ha line at 
 z $\sim$ 0.03 \citep{pagnotta2014}; however, based on the spectral energy distribution
 of this diffuse patch and absence of other nebular lines, 
 \citet{litke2017} conclude that the diffuse patch is a background galaxy at z $\sim$ 0.8
 and that the central point-like source is the nucleus/bulge. Based on comparisons of stars with post-impact evolution models of \citet{pan2014}, \citet{litke2017} concur with the conclusion of \citet{pagnotta2014} and \citet{schaefer2012} that the SN progenitor of 0509--67.5 has no surviving companion.
 
{\it SNR 0519--69.0}:  This remnant is 600 $\pm$ 200 yr old based on the light echoes
of its SN \citep{rest2005}.  \citet[][hereafter EPS2012]{edwards2012} use \hst F550M ($V$ band) and 
F656N (\ha line) images of this SNR to exclude post-main sequence companion, but not MS companion, for the SN progenitor, and further suggest that the SN progenitor was 
either a supersoft source \citep{hachisu1999, langer2000, han2004} or a DD system. 

{\it SNR 0505--67.9 (DEM\,L71)}: This remnant is more commonly called DEM\,L71
\citep{davies1976}.  Its age estimated from size and shock velocity is 
4360 $\pm$ 290 yr
\citep{ghavamian2003}.   \citet[][hereafter PS2015]{pagnotta2015} use {\it{Chandra}} observations 
\citep{hughes2003, rakowski2003} to assess the SNR boundary and explosion center 
and use Gemini GMOS $g'$, $r'$, $i'$, and \ha images to select a large number of 
possible candidates for the SN's surviving companion; however, none has been confirmed.

{\it SNR 0509-68.7 (N103B)}:  
This remnant's age estimated from the SN light echoes is $\sim$860 yr  \citep{rest2005}.
A SD origin has been suggested because a dense CSM is detected in the 
remnant  \citep{williamsbf2014,li2017}.
\citetalias{pagnotta2015} used radio \citep{dickel1995} and \xray observations \citep{lewis2003} 
to determine the site of the SN explosion, then obtained Gemini GMOS images to examine
the stars and found eight candidates for the surviving companion.  Alternatively, \citet{li2017} 
used the Balmer-dominated filamentary shell, revealed by \hst \ha 
image, as the fronts of the collisionless shocks to determine the explosion center, and found 
fifteen stars within 1 pc from the explosion center.  They further suggest that the star
closest to the explosion center is the most likely candidate for the surviving companion 
of the SN progenitor, as its location in the CMDs is similar to that expected from a post-impact
$\sim$1 \msun\ subgiant modeled by \citet{podsiadlowski2003}.

A detailed, up-to-date compilation of methods and results of searches (including this work) for surviving companions in Galactic and LMC Type Ia SNRs is given in Appendix 
 \ref{appendix:summary}.

To investigate surviving companions of Type Ia SNe in the LMC, we have obtained new 
\hst images in continuum and \ha bands for nine Type Ia SNRs in the LMC.  
We have reported our study of the SNR 0509--67.5 \citep{litke2017} and 
N103B \citep{li2017}.  In this paper, we report our investigation of surviving companions
in the other three Balmer-dominated young Type Ia SNRs:
0519--69.0, 0505--67.9 (DEM\,L71), and 0548--70.4 
(see Figures \ref{figure:0519}, \ref{figure:deml71} and  \ref{figure:0548}). 
The observations used in this study are described in Section 2.  The methodology 
is detailed in Section 3.  Sections 4-6 report 
our investigation of the three SNRs, respectively. The results are discussed 
in Section 7, and a summary is given in Section 8.

%--------------------------------------------  
%             Table 2

\begin{deluxetable}{ccccc}
%\tabletypesize{\scriptsize}
%\rotate
%\tablewidth{0pc}
\tablecaption{MUSE Observations}
\tablehead{SNR & PI & Program ID &   Date & t$_{exp}$ (s)}
\startdata 
0519--69.0 &  Leibundgut & 096.D-0352(A) & 2016 Jan 17 & 900\\
DEM L71  &  Leibundgut & 096.D-0352(A) & 2015 Nov 16 & 900              
\enddata
\label{table:muse}
\end{deluxetable}

%======================================================================
\section{Observations and Data Reduction} \label{sec:obs}  
%======================================================================

\subsection{Hubble Space Telescope Observations}

We have obtained new \hst images of SNR 0519--69.0, DEM\,L71, and 
SNR 0548--70.4, using the UVIS channel of Wide Field Camera 3 (WFC3) 
in Program 13282 (PI: Chu). For DEM\,L71 and SNR 0548--70.4, the 
images were obtained with the F475W ($B$ band), F555W ($V$ band), 
F814W ($I$ band), and F656N (\ha line) filters. For SNR 0519--69.0, new 
images were obtained with only the F475W and F814W filters, because \hst
F550M and F658N images taken with the Advanced Camera for Surveys 
(ACS)  are available in the Hubble Legacy Archive. The \hst images used in 
this paper are listed in Table \ref{table:hst}. The continuum band images  
can be used to search for surviving companions and study the underlying 
stellar population,  while the \ha images can be used to analyze physical 
structures of the remnants and to determine their explosion centers.

The UVIS channel of WFC3 has a 162\arcsec\ $\times $162\arcsec\ field of view. 
The 0\farcs04 pixel size corresponds to 0.01 pc in the LMC, at a distance of $\sim$ 50 kpc. 
The observations of DEM\,L71 and SNR 0548--70.4 were dithered with the 
WFC3-UVIS-GAP-LINE pattern for 3 points and point spacings of 2\farcs414. 
Each observation had a minimum total exposure time of 1050 s. 
In order to mitigate the charge transfer efficiency (CTE) issues in WFC3, 
FLASH = 5, 11 and 4 options were used for the F475W, F656N and F814W 
observations, respectively. The observations of SNR 0519--69.0 in the F475W 
and F814W passbands were dithered using POSTARG. The F475W observation 
had a total exposure time of 1070 s, and FLASH = 5; the F814W observation 
had a total exposure time of 1174 s, and FLASH = 4.

To examine the stars projected within or near the three Balmer-dominated 
SNRs, we have used the stellar photometry package \texttt{DOLPHOT}, 
adapted from \texttt{HSTPHOT} with {\it HST}-specific modules \citep{dolphin2000}, 
to perform point-spread function photometry on the \hst images. 
Following \cite{williamsbf2014}, we adopt specific criteria on photometric 
parameters, such as signal-to-noise ratio (S/N) $>$ 5, $sharp^2 < 0.1$, and 
$crowd < 1.0$, to filter out  non-stellar sources \citep{dolphin2000}. All stars 
have been measured in the Vega magnitude system for the $B$, $V$, and $I$ 
passbands. 
The \hst photometry allows us to compare the locations of stars in the CMDs 
with the expected evolutionary tracks of the post-impact surviving companions.

\subsection{Chandra \xray Observations}

SNR 0519--69.0, DEML71 and SNR 0548--70.4 were observed with the Advanced 
CCD Imaging Spectrometer (ACIS) of the \chandra \xray Observatory 
for Program 01500024 (PI: Holt; 39.2 ks), Program 01500900 (PI: Hughes; 45.6 ks), 
and Program 02500872 (PI: Borkowski; 59.3 ks), respectively. 
The event files of these observations are available in the Chandra Data Archive,
and images in various energy bands are conveniently available from the Chandra 
Supernova Remnant 
Catalog\footnote{https://hea-www.harvard.edu/ChandraSNR/index.html}. 
While these SNRs are detected from 0.3 keV to 4--7 keV, the 0.3--2.1 keV 
band contains the bulk of the \xray emission and the 0.3--2.1 keV band 
images are used to locate the boundaries of SNRs 
in order to assess their explosion centers.

\subsection{VLT MUSE Observations}
 
Multi-Unit Spectroscopic Explorer (MUSE) observations of SNR 0519--69.0 
and DEM\,L71 were obtained with the Very Large Telescope (VLT) UT4 of the 
European Southern Observatory (ESO) on 2015 November 16 and 2016 
January 17 (see Table \ref{table:muse}) for Program 096.D-0352(A) (PI: Leibundgut). VLT MUSE is an
integral-field Unit (IFU), providing optical spectrum for every position 
in the field of view (FOV).  The wide-field mode used for the observations 
has a 60\arcsec $\times$\,60\arcsec\ FOV.  This FOV is large enough to  
cover the entire SNR 0519--69.0 and adequate surroundings for background, 
but only about 80$\%$ of DEM\,L71 and a small region for background.
The spectral coverage is 4750 -- 9350 \AA, which includes nebular emission 
lines  such as \ha, H$\beta$, [\ion{O}{3}]$\lambda$$\lambda$4595, 5007, 
[\ion{N}{2}]$\lambda\lambda$6548, 6583, and [\ion{S}{2}]$\lambda\lambda$6716, 6731, 
as well as stellar absorption lines such as H$\alpha$, H$\beta$, \ion{He}{1} $\lambda$5875,
\ion{He}{2} $\lambda$5411 and $\lambda$4686, and \ion{Na}{1} D-lines 
$\lambda\lambda$5890, 5896. 
The spatial and spectral samplings are 0\farcs2 spaxel$^{-1}$ and 1.25 \AA\, pixel$^{-1}$,
respectively. The VLT MUSE observations of SNR 0519--69.0 and DEM L71 each had 
an exposure time of 900 s. 

We use the VLT MUSE data reduction pipeline \citep{weilbacher2014} 
to process bias subtraction, flat fielding, and wavelength and geometrical
calibrations. 
The fields are crowded and large parts of the FOV are covered by 
either stars or nebular emission. To subtract the sky background, 
%we tweak the number of eigenvectors manually and did this for one pointing to blank sky. 
we use one pointing to blank sky and assume that the result would 
be viable for all pointings, and make use of algorithms from the Zurich 
Atmospheric Package \citep{soto2016};
however, artificial flux fluctuations are found in the wavelength range 
7700 -- 9000 \AA, and they are likely caused by the 
imperfect subtraction of sky background. 
To perform the flux calibration, we use existing photometry of stars 
in the field and synthetic photometry from the VLT MUSE data cube to
make sure that differences of magnitudes of stars are within 0.05  
mag.
The detailed procedures of data reduction are described in \citet{Kruhler2017}.
The VLT MUSE observations can be used to carry out spectroscopic 
analyses to search for stars that have large peculiar radial velocities.

%--------------------------------------------  
%             Table 3

\begin{deluxetable*}{c|ccc|cl|cc|cc}
\tabletypesize{\scriptsize}
%\rotate
\tablewidth{0pc}
\tablecaption{Explosion Centers and Search Radius}
\tablehead{ &  & Previous$^{*}$ & & $\hspace{30pt}$ Center of ellipse 
$\hspace{-30pt}$ & & $\hspace{30pt}$Adopted center$\hspace{-30pt}$ & & $\hspace{30pt}$Search radius$\hspace{-30pt}$ & \\
\hline
SNR & R.A. & Decl. & R$_{\mathrm{search}}$ & R.A. & Decl. & R.A. & Decl.  & R$_{\mathrm{MS}}$ & R$_{\mathrm{helium}}$\\
(name) & (J2000) & (J2000) & (arcsec) & (J2000) & (J2000) & (J2000) & (J2000) & (arcsec) & (arcsec) }
\startdata 
0519--69.0 & 5:19:34.83 & -69:02:06.92 &  4.7 & 5:19:34.72 & -69:02:07.57 & 5:19:34.77 & -69:02:07.25 & 1.2 & 2.7   \\ 
DEM L71 & 5:05:42.71 & -67:52:43.50 & 15.8 & 5:05:42.06 & -67:52:41.00$^{a}$ & 5:05:42.04 &  -67:52:41.58 & 5.7 & 14.5  \\  
DEM L71 & - & - & - & 5:05:42.03 & -67:52:42.17$^{b}$ & - & - & - & - \\ 
0548--70.4 & - & - & - & 5:47:48.46 & -70:24:52.43 & 5:47:48.46 & -70:24:52.43  & 20.0 & 40.0 
\enddata
\tablenotetext{*}{The previous search of surviving companion in 
SNR 0519--69.0 and DEM L71 has been reported by
\citetalias{edwards2012}  and \citetalias{pagnotta2015} respectively.}
\tablenotetext{a, b}{For DEM L71, we visually fit two ellipses to the shell with and without considering the blowout-like structures, respectively. We adopt the average of centers of the ellipses as the explosion center.}
\label{table:regions}
\end{deluxetable*}

\subsubsection{Extracting Spectra from VLT MUSE Data}

We use the \texttt{PampelMUSE} package \citep{kamann2013} to 
extract spectra of all stars with $V<23.0$ mag within search radii 
from the centers of SNR 0519--69.0 and DEM L71. This software is 
well-designed with the purpose of improving the analysis of crowded stellar fields in IFU observations, and its graphical user interface (GUI) 
allows us to check the results visually in an interactive mode.
To identify the sources that are feasible to extract the spectra,
\texttt{PampelMUSE} requires an accurate reference catalogue 
which covers relative positions and magnitudes of the stars in the 
MUSE field of view. In this study, we use the high-resolution \hst 
observations in continuum bands as the reference. From the \hst 
catalogue, \texttt{PampelMUSE} creates a mock image of stellar 
field from the existing photometry of stars. This mock image is 
used to cross-correlate with the VLT MUSE data in the routine 
INITFIT, to estimate positions of the sources that allow us to
extract spectra in the VLT MUSE observations. From their positions, 
INITFIT identifies the sources as the resolved stars if they pass 
the criteria for the measured and modeled parameters, such as a 
local density of brighter sources $<$ 0.4 per resolution element, 
a S/N $>$ 3, and a distance to the nearest brighter source $>$ 0.3 
full width at half maximum (FWHM) of the point spread function (PSF). 

Once the resolved stars in the VLT MUSE observations are identified, 
the routine CUBEFIT extracts all stellar spectra by fitting the PSF, 
the positions and the flux of these resolved stars layer by layer for 
the VLT MUSE data cube. In this step, we used the Moffat function to 
be the PSF profile, avoiding underestimation of the PSF wings 
for the Gaussian function. Afterwards, the routine POLYFIT fits the 
wavelength dependence of the PSF parameters and the VLT MUSE 
coordinates to smooth the fitted variations caused by the atmosphere 
refraction. In our analysis, the routine POLYFIT was performed with 
a fifth order Legendre polynominal. With these polynominal fits, 
we run the routine CUBEFIT again to extract the final spectra of the 
stars. In the end, we used the routine GETSPECTRA to save all 
stellar spectra into individual FITS file. 

In the extracted stellar spectra, the presence of telluric lines would 
bias the spectral fits. To mitigate this problem, we identify the telluric 
absorption lines from the measurements of the VLT Ultraviolet and 
Visual Echelle Spectrograph (UVES) \citep{dekker2000}, and 
manually remove the prominent telluric lines over the spectral 
coverage of 4750 -- 9350 \AA. 
Moreover, we find the Balmer absorption lines
in some extracted stellar spectra are contaminated by Balmer line emission from nearby 
collisionless shocks. To make the spectral fits more reliable,
we manually remove the contaminated Balmer
emission lines, such as 
\hb and \ha lines, from those stellar spectra.
The resulting spectra are 
used in the stellar spectra fitting to determine radial velocities. 

% being under or over estimated easily during the background subtraction, especially in \ha and \hb lines. This kind of contamination would be a challenge to the kinematic search for the surviving companions within the Balmer-dominated SNRs.

\subsubsection{Stellar Parameter Inference}

Spectral template fitting with the PHOENIX grid was performed using the method described in \citet{do2013} and \citet{stostad2015} with the software package \textit{StarKit} \citep{kerzendorf2015}. \textit{StarKit} is a modular spectral fitting framework using Bayesian inference to determine the best fit parameters and their uncertainties. \textit{StarKit} simultaneously fits the physical parameters of stars, the spectral continuum, the radial velocity, and the rotational velocity. The set of physical parameters available for fitting is determined by the parameters sampled by the spectral grid. For the PHOENIX grid, the parameters are stellar effective temperature $T_{eff}$, surface gravity $\log g$, metallicity [M/H], and $\alpha$-element to iron abundance ratio [$\alpha$/Fe]. We use a linear interpolator to interpolate between the synthetic spectral grid points. We then convolve the spectral resolution of the grid to R = 3,000 in order to match the instrumental resolution.

Statistically, the fit is done by computing the posterior distribution in Bayes' Theorem:
\begin{equation}
P(\theta| D) = \frac{P(D|\theta)P(\theta)}{P(D)}
\end{equation}
where $D$ is the observed spectrum, and the model parameters $ \theta = (T_{eff}, ~\log g, ~\textrm{[M/H]}, \textrm{[}\alpha\textrm{/Fe]}, V_r,V_{rot})$, where $V_r$ is the radial velocity and $V_{rot}$ is the rotational velocity. The priors on the model parameters are $P(\theta)$ and $P(D)$ is the evidence, which acts as the normalization. The combined likelihood for an observed spectrum is:
\begin{equation}
P(D|\theta) = \prod^{\lambda_n}_{\lambda=\lambda_0}\frac{1}{\epsilon_{\lambda,obs}\sqrt{2\pi}}\exp{(-(F_{\lambda,obs} - F_\lambda(\theta))^2/2\epsilon_{\lambda,obs}^{2})},
\end{equation}
where $F_{\lambda,obs}$ is the observed spectrum, $F_{\lambda}(\theta)$ is the model spectrum evaluated with a given set of model parameters, and $\epsilon_{\lambda,obs}$ is the 1 $\sigma$ uncertainty for each observed flux point. This likelihood assumes that the uncertainty for each flux point is approximately Gaussian. For computational efficiency, we use the log-likelihood in place of the likelihood:
\begin{equation}
\ln P(D|\theta) \propto -\frac{1}{2} \sum^{\lambda_n}_{\lambda=\lambda_0}((F_{\lambda,obs} - F_\lambda(\theta))^2/\epsilon_{\lambda,obs}^{2}).
\end{equation}

We use flat priors in all the fit parameters and sample the posterior using Nestle, a nested sampling implementation \citep{feroz2009}. We use the peak posterior value to be the best fit values and the marginalized 68\% central confidence intervals for each fit parameter to be its uncertainty. Based on the tests against empirical references described in \citet{feldmeierKrause2017}, we include a systematic uncertainty term added in quadrature to the statistical uncertainties of each fit of $\sigma_{T_{eff}} = 200$ K, $\sigma_{[M/H]} = 0.2$, $\sigma_{[\alpha/Fe]} = 0.2$, and $\sigma_{\log g} = 1.0$.

%%%%%%%%  Stars in central region of SNR 0548, Ha + RGB  %%%%%%%%

\begin{figure*}    
\epsscale{1.0}
\vspace*{-1cm}
\hspace*{-0.5cm}\plotone{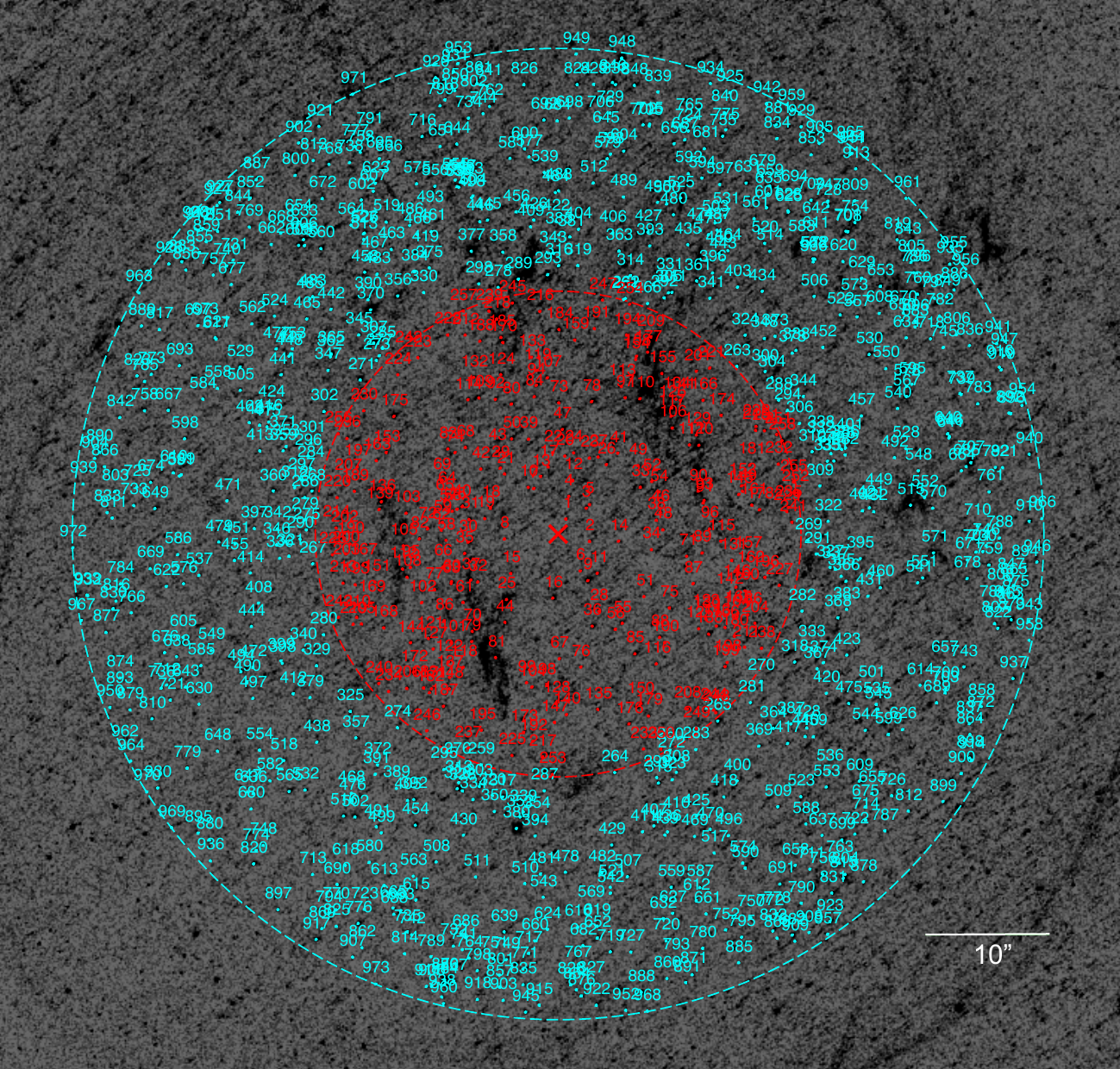}
\caption{
Same as the left panel in Figure \ref{figure:stars0519}, but for SNR 0548--70.4. 
}
\label{figure:central_stars_0548_ha}
\end{figure*}

\begin{figure*}    
\epsscale{1.0}
%\vspace*{-1.5cm}
\hspace*{-0.5cm}\plotone{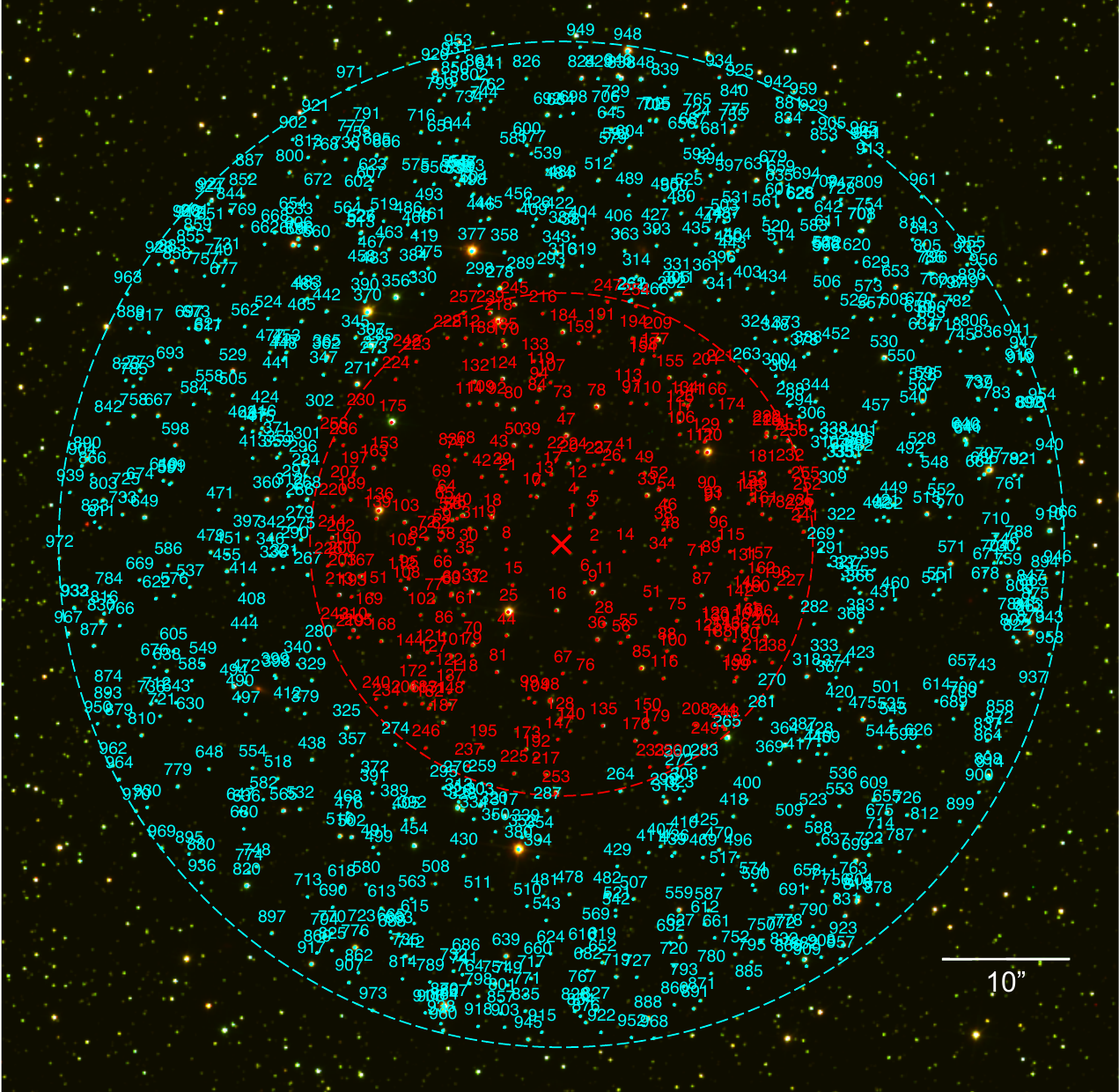}
\caption{
Same as the right panel in Figure \ref{figure:stars0519}, but for SNR 0548--70.4. 
}
\label{figure:central_stars_0548_rgb}
\end{figure*}
%======================================================================
\section{Methodology} 
%======================================================================

To search for the surviving companion of a Type Ia SN progenitor,
we first locate the site of SN explosion, estimate possible range
of velocity for a surviving companion, and determine the search 
radius.  We then use two methods to search for surviving 
companions within this radius.  In the first method, we 
compare \hst photometric measurements of stars with those  
expected from models of post-impact evolution of surviving 
companions \citep{pan2014}. In the second method, we use 
MUSE observations to carry out spectroscopic analyses of
stars and search for large peculiar radial velocities as diagnostics of surviving companions.

\subsection{Site of SN Explosion and Search Radius}

After the explosion of a Type Ia SN, the SN ejecta expands and interacts 
with the partially neutral ambient medium to produce the Balmer-dominated 
shell \citep{chevalier1980}.  If its progenitor had a non-degenerate stellar 
companion, as in the SD scenario, the stellar companion would survive the 
explosion and run away at  a velocity comparable to its final orbital velocity. 
If the binary progenitor had a translational motion through the ISM
\citep[e.g., N103B,][]{li2017}, the 
Balmer dominated shell would still be centered at the site of SN explosion 
as the expansion velocity of the SN ejecta is much higher than the progenitor's 
translational velocity.  The surviving companion will run away from the 
site of SN explosion at a velocity that is the sum of its orbital velocity 
and the progenitor's translational velocity.

With the above understandings, we first estimate the site of SN explosion 
using the Balmer-dominated shell.  If an SNR's Balmer shell is regular and 
exhibits reflection symmetry about two orthogonal lines of symmetry, the shell
can be fitted by an ellipse, and its geometric center can be easily 
identified and measured with small uncertainties. We then adopt the geometric
center to be the SN explosion site \citep[e.g.,][]{litke2017}. 

If an SNR's Balmer shell is irregular, we still visually fit its general 
shape with an ellipse. Usually, over 75\% of the shell periphery 
can be fitted well by an ellipse, and regions showing large deviations 
are faint and may be caused by a particularly low ambient density; 
thus, it is reasonable to adopt the center of the ellipse as the site of 
SN explosion. This is the case for SNR 0519--69.0 and 
SNR 0548--70.4.

If the ellipse fitted to an SNR's Balmer shell is visibly unsatisfactory,
for example, no ellipse can describe more than 70\% of DEM L71's 
Balmer shell periphery, then X-ray images of the reverse-shocked 
SN ejecta are used to make independent estimates of site 
of SN explosion (e.g., \citealt{schaefer2012}, \citetalias{edwards2012}, \citetalias{pagnotta2015}).  
The differences among these different estimates contribute to  
the uncertainty of the SN explosion site.

A surviving companion runs away from the site of SN explosion
with a velocity that is the sum of its orbital velocity and the
progenitor's translational velocity.  Assuming $\sim$2--3 $M_\odot$
for a SN Ia progenitor's total mass and a period of $\le$1.0 day,
the orbital velocity of the companion can reach as high as 400 km s$^{-1}$
\citep{schaefer2012}.  The translational velocity of the SN progenitor is
unknown, but it is most likely within the range of peculiar velocities of
stars and much lower than velocities of runaway stars. 
In the post-impact models of surviving companion \citep{pan2014}, the largest runaway velocity is 730 km s$^{-1}$ for a helium star companion.
To be on the safe
side, we adopt a translational velocity of $<$100 km s$^{-1}$ and a
``runaway velocity'' of $<$730 km s$^{-1}$ for the surviving companion. 
Note that in a thermonuclear supernova model with a WD and a sub-dwarf helium star donor going through double detonation may eject the companion at velocities up to 900 \kms \citep{Bauer2019,Geier2013}; 
however, these surviving sub-dwarfs are too faint to be detected in the LMC.

The ages of the young Balmer-dominated SNRs have been determined
from SN light echo observations \citep{rest2005} or sizes and shock 
velocities \citep{smith1991, ghavamian2003}. The ``runaway distance'' of a 
surviving companion from the site of SN explosion is equal to the runaway
velocity times the SNR's age $t$. We thus adopt a search radius of 
$7.5\times10^{-4}~t_{\rm yr}$ pc for the surviving companions. 
The younger the SNR is, the easier it is to search for a surviving companion within, as the surviving companion has not moved too far.

\subsection{Photometric Search}

%\subsection{Photometric Comparison with Models of Post-Shock Evolution }

%The group of evolutionary tracks on the left are for the case of He stars, and the group of evolutionary tracks near the center of the diagram are for the case of MS stars.

The impact of SN ejecta on a non-degenerate stellar companion 
has been studied with numerous hydrodynamic simulations 
\citep{marietta2000, pakmor2008, pan2010, pan2012a, liu2012, liu2013, Bauer2019}. 
Among these, \cite{pan2014} have calculated the post-impact evolution of 
luminosity and effective temperature of the surviving companion and 
plotted them in the theoretical Hertzsprung-Russell (H-R) diagram, i.e., 
luminosity versus stellar effective temperature. For direct comparisons 
with observations, a blackbody model with the surviving companion's
temperature and luminosity is used to calculate its magnitudes in
different passbands.  The post-impact evolution of a surviving companion
can then be plotted in an observational H-R diagram, i.e., color-magnitude 
diagrams (CMDs).

\citet{schaefer2012} have considered various published SD models
for Type Ia SN progenirots, such as recurrent novae, symbiotic stars,
supersoft sources, helium donor companions, etc., and suggested
that a surviving companion would have $V < 22.7$ mag at the 
distance of the LMC.  Thus, in this study we consider only stars
with $V < 23.0$ in the search for surviving companions.

We plot both stars in the search radius and stars within a large radius 
encompassing the entire SNR and some surrounding regions in 
$V$ versus $B$--$V$ and $I$ versus $V$--$I$ CMDs.  Stars within the 
large radius allow us to establish a background stellar population for 
comparison.  We then compare the stars in the search radius with the
post-impact evolutionary tracks of surviving companions from
\citet{pan2012b, pan2013, pan2014}.  A star falling on an evolutionary
track with consistent age would be a candidate for surviving companion.
Only few stars appear double and too blended for reliable photometric 
measurements, and most of these are outside the search radii; therefore,
we choose to ignore them.

\subsection{Stars with Peculiar Radial Velocities}

We use VLT MUSE observations of SNR 0519--69.0 
and DEM L71 to carry out spectral analyses of 
stars within the search radius from the site 
of SN explosion, and use stellar atmosphere 
model fits to determine physical parameters 
and radial velocities of stars. 
We find that stellar atmosphere models show 
more reliable fits for stars with $V$ $<$ 21.6 mag. 
%while brighter stars have higher signal to noise ratio. 
This turns out a safe limiting magnitude for our spectroscopic
search for surviving MS companions, as the post-impact 
evolution of surviving MS companions modeled by
\citet{pan2014} show $V$ well brighter than $\sim$ 
21.0 mag within the Balmer-dominated phase of Type Ia
SNRs, less than $\sim$10,000 yr after the SN blast.
We thus spectroscopically examine the photometric 
candidates of surviving companions that have
$V$ $<$ 21.6 and are located within the search radii,
using large radial velocities as diagnostics of 
surviving companions. 

The distributions of radial velocities
of these stars are plotted and 
examined. The standard deviation of the 
distribution, $\sigma$, is computed. Assuming 
a Gaussian distribution of the radial 
velocities, we expect only 0.3\% of the 
population to have velocities deviating by
more than 3$\sigma$  from the mean ($\overline{V_r}$).  
For a population of $<$200 stars, fewer than 
1 star is expected to be $>$3$\sigma$ from 
the mean. Therefore, stars with radial 
velocities more than 3$\sigma$ from the mean 
are carefully examined to 
determine whether they are viable candidates 
for surviving companions of the SN progenitors.

%--------------------------------------------
%             Table 4

\begin{deluxetable*}{ccccccccc}
\tablecolumns{9}
\tabletypesize{\scriptsize}
%\rotate
\tablewidth{0pc}
\tablecaption{Stars Brighter Than $V = 23.0$ mag near Central Region in the SNR 0519--69.0}
\tablehead{Star & R.A. (J2000) & Decl. (J2000) & B & V & I& B-V & V-I & $r$}
\startdata
1 & 05:19:34.81 & -69:02:7.85 & 20.72 $\pm$ 0.00 & 20.54 $\pm$ 0.00 & 20.44 $\pm$ 0.01 & 0.18 $\pm$ 0.00 & 0.10 $\pm$ 0.01 & 0.63\arcsec \\ 
2 & 05:19:34.97 & -69:02:7.36 & 22.42 $\pm$ 0.01 & 22.05 $\pm$ 0.01 & 21.54 $\pm$ 0.01 & 0.37 $\pm$ 0.01 & 0.50 $\pm$ 0.01 & 1.06\arcsec \\ 
3 & 05:19:34.98 & -69:02:6.15 & 22.91 $\pm$ 0.01 & 22.50 $\pm$ 0.01 & 22.00 $\pm$ 0.01 & 0.41 $\pm$ 0.01 & 0.50 $\pm$ 0.01 & 1.57\arcsec \\ 
4 & 05:19:34.84 & -69:02:5.45 & 22.42 $\pm$ 0.01 & 21.95 $\pm$ 0.01 & 21.60 $\pm$ 0.01 & 0.47 $\pm$ 0.01 & 0.36 $\pm$ 0.01 & 1.83\arcsec \\ 
5 & 05:19:34.99 & -69:02:8.72 & 21.61 $\pm$ 0.01 & 21.00 $\pm$ 0.01 & 20.20 $\pm$ 0.01 & 0.62 $\pm$ 0.01 & 0.79 $\pm$ 0.01 & 1.87\arcsec \\ 
6 & 05:19:34.54 & -69:02:5.33 & 20.77 $\pm$ 0.00 & 20.19 $\pm$ 0.00 & 19.30 $\pm$ 0.00 & 0.57 $\pm$ 0.00 & 0.89 $\pm$ 0.00 & 2.29\arcsec \\ 
7 & 05:19:35.16 & -69:02:8.79 & 20.97 $\pm$ 0.00 & 20.59 $\pm$ 0.01 & 20.16 $\pm$ 0.01 & 0.38 $\pm$ 0.01 & 0.43 $\pm$ 0.01 & 2.60\arcsec \\ 
8 & 05:19:34.68 & -69:02:9.80 & 21.15 $\pm$ 0.01 & 20.55 $\pm$ 0.00 & 19.64 $\pm$ 0.00 & 0.60 $\pm$ 0.01 & 0.91 $\pm$ 0.00 & 2.60\arcsec 
\enddata
\label{table:photometry0519}
\end{deluxetable*}

% V_tan cross the sky calculated from the theta and the age of SN
%1, V_tan = (2.7+/-1.3)e+02km/s , V_overall = (3.9+/-0.9)e+02 km/s
%2, V_tan = (4.7+/-1.8)e+02km/s , V_overall = ? km/s => confidence level of its Vr is D 
%3, V_tan = (7.5+/-2.7)e+02km/s , V_overall = (8.2+/-2.5)e+02 km/s
%4, V_tan = (7.9+/-2.8)e+02km/s , V_overall = (8.1+/-2.7)e+02 km/s
%5, V_tan = (9.1+/-3.2)e+02km/s , V_overall = (9.4+/-3.0)e+02 km/s
%6, V_tan = (1.1+/-0.4)e+03km/s , V_overall = (1.1+/-0.4)e+03 km/s
%7, V_tan = (1.1+/-0.4)e+03km/s , V_overall = (1.1+/-0.4)e+03 km/s

%--------------------------------------------
%             Table 5

\begin{deluxetable*}{cccccccccc}
\tablecolumns{9}
\tabletypesize{\scriptsize}
%\rotate
\tablewidth{0pc}
\tablecaption{%Fitting 
Stellar Parameters of the Stars Brighter Than $V = 21.6$ mag in Table \ref{table:photometry0519}} 
\tablehead{Star & T$_{eff}$ (K) & log $g$ (dex) & [Fe/H] (dex) & $V_{r}$ (km s$^{-1}$) &  }
\startdata
1 & 7873 $\pm$ 201 & 4.6 $\pm$ 1.0 & -0.5 $\pm$ 0.2 & 285 $\pm$ 0  \\
5 & 5398 $\pm$ 201 & 3.3 $\pm$ 1.0 & -1.0 $\pm$ 0.2 & 182 $\pm$ 0 \\
6 & 5263 $\pm$ 200 & 3.6 $\pm$ 1.0 & -1.0 $\pm$ 0.2 & 258 $\pm$ 0  \\
7 & 7000 $\pm$ 200 & 4.2 $\pm$ 1.0 & -0.4 $\pm$ 0.2 & 289 $\pm$ 0 \\
8 & 5042 $\pm$ 200 & 3.3 $\pm$ 1.0 & -0.9 $\pm$ 0.2 & 259 $\pm$ 0 
\enddata
\tablenotetext{}{We have included a systematic uncertainty term added in quadrature to statistical uncertainties of each fit, based on the tests against empirical references described in \citet{feldmeierKrause2017}. The details about the spectral fitting can be found in Section 2.3.2.}
\label{table:starkit0519}
\end{deluxetable*}

\section{Investigation of SNR 0519--69.0}

The previous search for surviving companion in SNR 0519--69.0
by \citetalias{edwards2012} used only \hst $V$ and \ha band images.
We have obtained additional $B$ and $I$ band images. This full 
set of data, in $B$, $V$, $I$, and \ha bands, allow us to carry out 
more accurate analyses of the underlying stellar population.

\subsection{Site of SN Explosion and Search Radius}

\citetalias{edwards2012} argued that the SNR shell of 0519--69.0
does not suggest highly asymmetric explosion and that its
geometric center is close to the site of SN explosion.  They 
used 9 sets of perpendicular bisectors from edge to edge 
for the \ha shell and X-ray shell, and adopted their average 
center, 05$^{\mathrm{h}}$19$^{\mathrm{m}}$34$^{\mathrm{s}}$.83, 
$-$69$^\circ$02$'$06\farcs92 (J2000), to be the site of
SN explosion in SNR 0519--69.0. In our study, we fit an 
ellipse to the bright rim of the Balmer-dominated shell and 
adopt its center as the site of SN explosion, 
05$^{\mathrm{h}}$19$^{\mathrm{m}}$34$^{\mathrm{s}}$.72, 
$-$69$^\circ$02$'$07\farcs57 (J2000). As shown in Figure \ref{figure:center0519}, there is an offset of 0\farcs45 
between our center and that of \citetalias{edwards2012}. 
This offset is caused by our different treatment of 
the faint northeast arc -- we consider it an anomaly due to 
a lower density in this quadrant and ignore it when fitting 
an ellipse to the \ha shell rim, while \citetalias{edwards2012} 
included the faint arcs as the shell edge. Without knowing 
details of the explosion geometry, we can only treat the 
difference in these two determinations of SN explosion site 
as additional uncertainty. We adopt the average of centers 
determined by \citetalias{edwards2012} and our method as the 
site of SN explosion,
05$^{\mathrm{h}}$19$^{\mathrm{m}}$34$^{\mathrm{s}}$.77, 
$-$69$^\circ$02$'$07\farcs25 (J2000), see Figure 
\ref{figure:center0519}. The coordinates of these centers 
are listed in Table \ref{table:regions}.

SNR 0519--69.0 has an age of 600 $\pm$ 200 yrs \citep{rest2005}. 
Adopting the largest runaway velocity of a surviving MS 
companion, 270 km s$^{-1}$, and a surviving helium star 
companion, 730 km s$^{-1}$ \citep{pan2014}, the runaway distance for 
a surviving MS companion can be up to 0.2 pc (0\farcs9), and 
a surviving helium star companion up to 0.6 pc (2\farcs5).  The search 
radii for MS and helium star companions listed in Table \ref{table:regions} 
are determined by adding the uncertainty of explosion center, 
%0\farcs23, 
0\farcs2, to the runaway distance. All stars with $V < 23.0$ mag 
within the helium star search radius, 2\farcs7, are examined for plausible candidates of surviving companion.  %The search radii for MS and He companions are given in Table \ref{table:regions}.

%2.39+0.23=2.62
%the two radii correspond to the runaway distances of a MS and a helium surviving companion with a post-impact velocity of 270 s$^{-1}$ and 730 km s$^{-1}$ respectively  
%50000*1/3600*0.05*3.1416/180=0.012
%270*(600+200)*365*24*3600/3.08567758e13*0.05/0.012=0.92"
%730*(600+200)*365*24*3600/3.08567758e13*0.05/0.012=2.49''
%1200*(600+200)*365*24*3600/3.08567758e13*0.05/0.012=4.09"

%0.92+0.225=1.15
%2.49+0.225=2.72
%The offset may indicate a small proper motion of the SN progenitor system through the ISM across the sky, and relatively uniform densities in the ambient ISM. 

\subsection{Photometric Search}

Within the search radius 2\farcs7 in SNR 0519--69.0, 
only eight stars are brighter than $V = 23.0$ mag. These 
eight stars, numbered in the order 
of increasing distance from the explosion center, are listed 
in Table \ref{table:photometry0519} and marked in 
the close-up images in Figure \ref{figure:stars0519}
and the $V$ versus $B$--$V$ and $I$ versus $V$--$I$ 
CMDs in Figure \ref{figure:CMD0519}.
These stars are too few to show clearly 
the locations of the MS and the RG branch in the CMDs.
We have thus plotted stars with $V < 23.0$ mag within 
22\farcs0 from the site of SN explosion in the CMDs in 
Figure \ref{figure:CMD0519}. 
% remove stars 632, 785 that have weird colors, both numbers are very close binaries.
The greatest majority of these stars 
belong to the background stellar population, and the MS 
and RG branch are clearly visible in the CMDs, providing
convenient references for candidate stars within the search 
radius. %{\color{red}\uline{ In the CMDs, few binaries that are too close to have reliable photometry near the boundary of the search area are ignored.}}
% They are unlikely the surviving companion candidates.

In these CMDs we have also plotted the post-impact evolutionary 
tracks of MS (the curved tracks above the MS) and helium star 
(the vertical tracks to the left of the MS) surviving companions 
\citep{pan2014}. None of the eight stars within 2\farcs7 from 
the site of explosion fall on the two sets of tracks. Thus we 
conclude that no viable candidates of surviving MS or helium star 
companion are present based on the $V$ versus $B$--$V$ 
and $I$ versus $V$--$I$ CMDs, consistent with but more 
stringent than the conclusion of \citetalias{edwards2012}.
%The magnitudes of the stars with $V < 23.0$ mag within 2\farcs7 from the explosion site are listed in Table \ref{table:photometry0519} in order of increasing angular distance from the site of SN explosion. 

\subsection{Stars with Peculiar Radial Velocities }

Only five stars are brighter than $V = 21.6$ mag 
within the search radius 2\farcs7 of SNR 0519--69.0. 
The physical parameters and radial velocities
of these stars are derived from model fits and  
are listed in Table \ref{table:starkit0519}. We 
have plotted the radial velocities of these five 
stars versus their distance to the SN explosion 
site in Figure \ref{figure:fig_0519_V_r}.  These 
five stars are too few to illustrate a statistically 
meaningful radial velocity distribution; therefore, 
we have added all stars with $V < 21.6$ mag 
within 6\farcs0 from the SN explosion site in
Figure \ref{figure:fig_0519_V_r} to show the 
radial velocity distribution in this neighborhood. 
It is noted that this larger radius, 6\farcs0, 
encompasses candidate stars with $V < 21.6$ 
mag that are selected in \citetalias{edwards2012}. 
From Figure \ref{figure:fig_0519_V_r}, we find that
star 5 has a radial velocity, 182 $\pm$ 0 km~s$^{-1}$, that  
deviates more than 2.5$\sigma$ from the mean velocity, 
264 km~s$^{-1}$; furthermore, its radial 
velocity is not well populated by the LMC or Galactic 
stars. To examine whether the radial velocity from model 
fit is reliable, we display the spectrum of star 5 with 
its model fit in Figure \ref{figure:fig_0519_star_5}. 
As shown in this figure,
the \ha and \hb lines of star 5 are contaminated 
by the nearby Balmer-dominated shocks and are 
not possible for the determination of radial velocity, 
but the calcium line 8542.1\AA, 
%8498.0, 8542.1, and 8662.1\AA\, 
with the model fit looks viable. 
This peculiar radial velocity indicates that star 5 
may be a promising candidate of surviving companion. 
Furthermore, assuming that star 5 has a mass of $\sim$1 $M_\odot$, the stellar effective temperature and surface gravity from the spectral fit implies an $M_V$ of 2.54 mag, and the observed $V$ = 21.0 mag requires the distance to be $\sim$49 kpc, consistent 
with the LMC distance.
We will further discuss this star in Section 7.

%remove blended stars

%star 5, 10^((21-1.5)/5)*10 ~ 80kpc
%  1.87/4 parsec to km=1.44e13km
%  1.44e13/800/365/24/3600~570km/s

%1, V_tan = (2.7+/-1.3)e+02km/s , V_overall = (3.9+/-0.9)e+02 km/s
%2, V_tan = (4.7+/-1.8)e+02km/s , V_overall = ? km/s => confidence level of its Vr is D 
%3, V_tan = (7.5+/-2.7)e+02km/s , V_overall = (8.2+/-2.5)e+02 km/s
%4, V_tan = (7.9+/-2.8)e+02km/s , V_overall = (8.1+/-2.7)e+02 km/s
%5, V_tan = (9.1+/-3.2)e+02km/s , V_overall = (9.4+/-3.0)e+02 km/s
%6, V_tan = (1.1+/-0.4)e+03km/s , V_overall = (1.1+/-0.4)e+03 km/s

%%%%%%%%  CMDs  of SNR 0548 %%%%%%%%

%--------------------------------------------
%             Figure 

\begin{figure*}   
\epsscale{1.2}
\hspace*{-0.5cm}\plotone{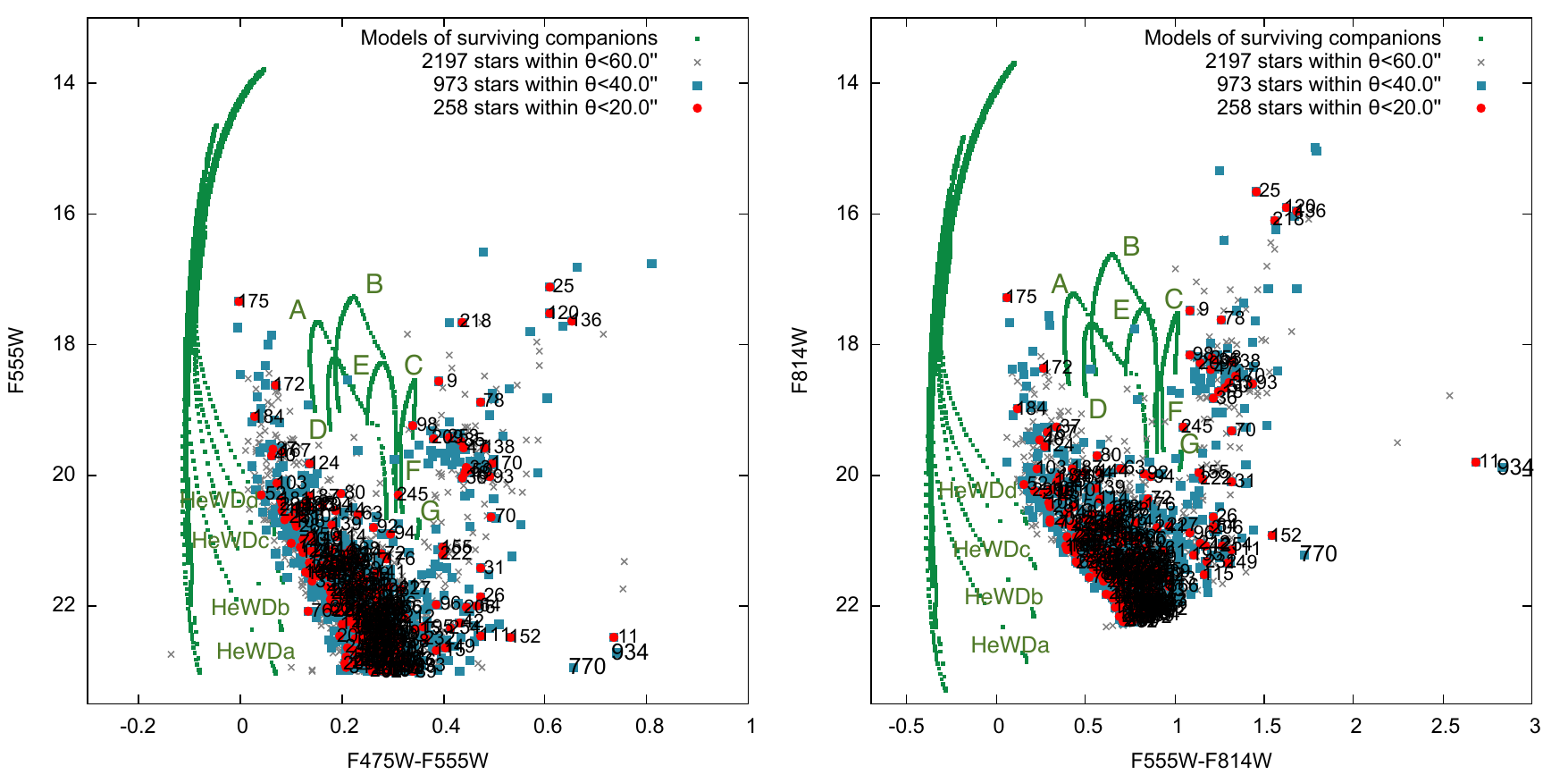}
\caption{ 
Same as Figure \ref{figure:CMD0519}, but for SNR 0548--70.4.
(Left) V versus B -- V CMD (Right) I versus V -- I CMD for the underlying stars of SNR0548--70.4. }
\label{figure:CMD0548}
\end{figure*}

%--------------------------------------------
%             Figure 

\begin{figure*}   
\epsscale{1.0}
\hspace*{-0.5cm}\plotone{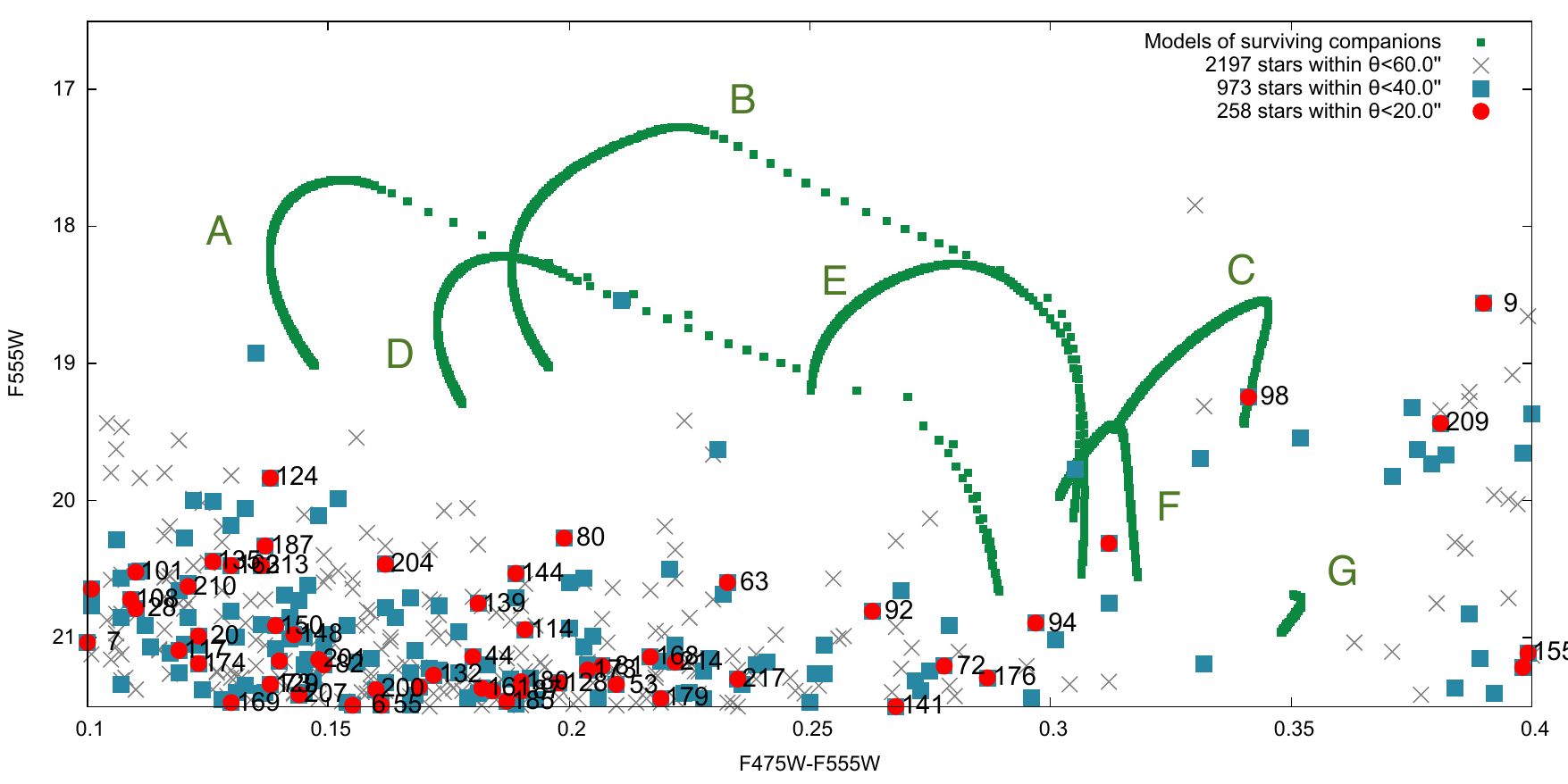}
\caption{ Close-up V versus B -- V CMD for the case of MS stars of SNR 0548--70.4 .}
\label{figure:CMD0548closeup}
\end{figure*}

\begin{figure*}   
\epsscale{1.0}
\hspace*{-0.5cm}\plotone{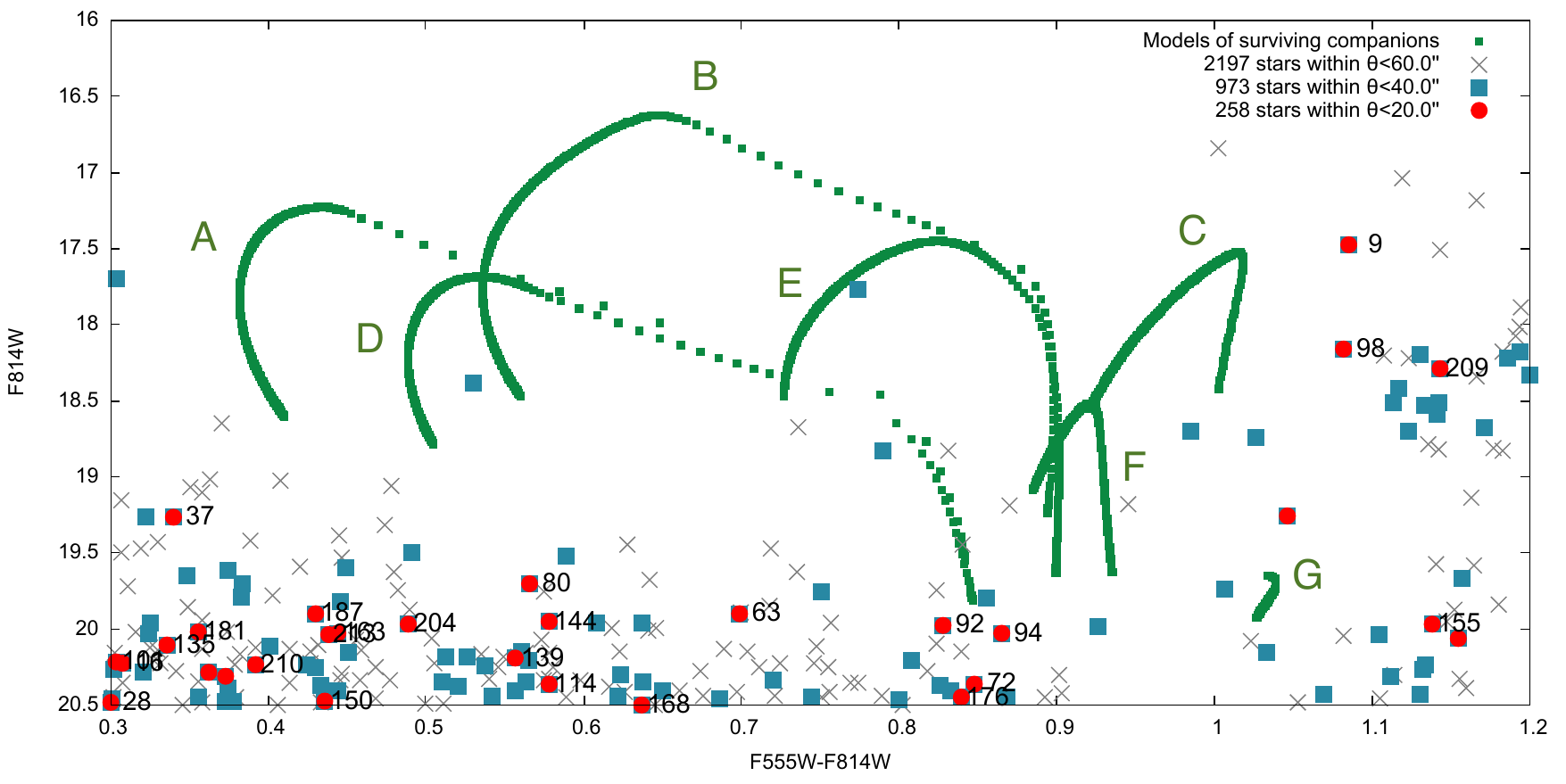}
\caption{ Close-up I versus V -- I CMD for the case of MS stars of SNR 0548--70.4 .}
\label{figure:CMD0548closeup2}
\end{figure*}

\section{Investigation of DEM L71}

%270*(4650)*365*24*3600*3.24078e-14*4=5.13
%730*(4650)*365*24*3600*3.24078e-14*4=13.88
%0.6+5.13 = 5.73 
%0.6+13.88 = 14.48

The previous search for surviving companion in DEM L71 by 
\citetalias{pagnotta2015} used Gemini GMOS $g'$, $r'$, $i'$, and 
\ha images. 
%Parallel to their work, 
We have obtained \hst $B$, $V$, $I$, and H$\alpha$ 
images of DEM L71. These new \hst images and 
VLT MUSE observations are used in our investigation.

\subsection{Site of SN Explosion and Search Radius}
\label{subsection:investigationofdeml71}

\citetalias{pagnotta2015} applied the perpendicular bisector method \citep{schaefer2012} to the GMOS \ha image and \chandra \xray images \citep{hughes2003, rakowski2003} of DEM L71 to assess an average geometric center of the remnant, and
adopted this center as the explosion site. For the
\xray images, the authors used four gas regions to locate the
\xray geometric center: the extreme faint outer edge, the bright rim of the outer shell, the inner region, 
and the central minimum. 
Their final SN explosion site is adopted to be 
05$^{\mathrm{h}}$05$^{\mathrm{m}}$42$^{\mathrm{s}}$.71, 
$-$67$^\circ$52$'$43\farcs50 (J2000). 
These centers are marked in Figure
\ref{figure:centerdeml71closeup}.

The physical origins of the above four gas regions are different and need to be noted.  
DEM L71 presents an ideal example of the double-shock structure, with forward shocks expanding into the ISM and reverse shocks moving into and heating the ejecta \citep{hughes2003, rakowski2003}. 
The outer shell in X-rays is thus associated with the forward (collisionless) shock into the partially neutral ISM. The bright X-ray shell rims correspond to sightlines parallel to the shock fronts and include the longest emitting path lengths, while the faint emission extending beyond the bright shell rim is associated with blowout-like structures into a lower-density medium.  The inner X-ray emission region corresponds to SN ejecta that has been heated by the reverse shock, and the central minimum originates from the central cavity.  

In our previous study of N103B \citep{li2017}, the geometric center of its Balmer-dominated shell was identified as the site of SN explosion because the Balmer shell follows closely the shock fronts into the ISM.  The Balmer-dominated shell of DEM L71 is somewhat irregular.  We have visually fitted two ellipses to the shell (as shown in Figure 9) and mark their centers and their average in Figure \ref{figure:centerdeml71closeup} for comparison. 
Our centers are offset to the northeast of the H$\alpha$ center of \citetalias{pagnotta2015} because of our different treatment of the faint southwest protrusion of the Balmer shell.  We assume the faint protrusion is caused by a local anomalously low density in the ISM and fit ellipses to only the bright rims of the Balmer shell, while \citetalias{pagnotta2015} include the outer edge of the protrusion in their derivation of H$\alpha$ center.  

Figure \ref{figure:centerdeml71closeup} also shows that the centers \citetalias{pagnotta2015}
determined from the central minimum and the inner X-ray emission are both offset significantly to the northeast of the center we derived from the Balmer shell.
As noted above, the inner \xray emission is associated with the SN ejecta. This offset is expected because the SN ejecta carried the same orbital velocity of the WD while the Balmer shell interacts and expands in the ISM. The offset between the center of ejecta and the center of SNR is expected to increase with age.

The detailed analysis of physical structure of DEM L71 will be reported in a future paper.  Here we adopt our average Balmer center as the site of SN explosion and use the difference between our centers as uncertainty.  All centers are listed in Table \ref{table:regions}. 
  
DEM L71 has an age of 4360 $\pm$ 290 yr \citep{ghavamian2003}.
Adopting the largest runaway velocity,
270 km s$^{-1}$ and 730 km s$^{-1}$, for surviving 
MS and helium star companions, respectively, 
the runaway distance for a surviving MS companion can be up
to 1.3 pc (5\farcs1), and a surviving helium star companion up to 
3.5 pc (13\farcs9). We determine the search radii for surviving 
MS and helium star companions by adding the uncertainty of the 
SN explosion site, %2\farcs24, 
0\farcs6, to the runaway distance. These search radii are listed 
in Table \ref{table:regions}.  In this study, 
all stars with $V < 23.0$ mag within the helium star search radius, 
14\farcs5, are examined as potential candidates of surviving companion.
%This helium star search radius, 17\farcs2, is $\sim$ 10\% larger than \citetalias{pagnotta2015}'s  search radius, 15\farcs8.

%The ejecta should in an opposite direction. 
%\xray (0.7--1.1 keV) 

\subsection{Photometric Search}

A total of 89 stars are brighter than $V = 23.0 $
mag within the search radius 14\farcs5 in DEM L71.
These 89 stars are numbered 
in the order of increasing distance to the site 
of explosion and marked in the close-up images 
(Figure \ref{figure:starsdeml71}) and $V$ versus
$B$--$V$ and $I$ versus $V$--$I$ CMDs (Figure
\ref{figure:CMDdeml71}). Their photometric measurements
are listed in Table \ref{table:photometrydeml71}.  
In the CMDs, these 89 stars 
do not show clearly the locations of the MS and 
RG branch.  We have thus plotted all stars 
with $V < 23.0$ mag within 48\farcs0 from the site 
of explosion in the CMDs (Figure \ref{figure:CMDdeml71}) 
in a different symbol (grey cross) to illustrate the MS 
and RG branch in the CMDs. 

In the CMDs (Figure \ref{figure:CMDdeml71}), we have
over-plotted the post-impact evolutionary tracks of the
MS (the curved tracks above the MS) and helium star (the 
vertical tracks to the left of the MS) surviving 
companions \citep{pan2014}. 
From locations of stars in the CMDs, we find 
no star matches the evolutionary tracks for 
MS and helium star surviving companion candidates 
within the search radius of 14\farcs5, as shown in 
the close-up in Figure \ref{figure:CMDdeml71closeup} 
and Figure \ref{figure:CMDdeml71closeup2}.
Our study has different search area and candidates 
from \citetalias{pagnotta2015}, but our photometric 
search has not found any obvious candidate of surviving 
companion, neither.

\subsection{Stars with Peculiar Radial Velocities}

There are 32 stars with $V < 21.6$ mag within the 
14\farcs5 search radius in DEM L71. 
The stellar parameters and radial velocities 
of these stars from model fits are listed in 
Table \ref{table:starkitdeml71}. We have plotted
the distribution of 
radial velocities versus distances to the explosion 
site of these stars %with standard deviations of the distribution, $\sigma$s, 
in Figure \ref{figure:fig_deml71_V_r}. While the 
overall distribution roughly reveals a Gaussian
profile, we find that star 80 has a radial velocity, 213 $\pm$ 0 km~s$^{-1}$, that are deviated more than 2.5$\sigma$ 
from the mean, 270 \kms (Figure \ref{figure:fig_deml71_V_r});
furthermore, its radial velocity is not well
populated by the LMC or Galactic stars. The model
fit of the spectrum of star 80, displayed in 
Figure \ref{figure:fig_deml71_star_80}, looks reasonable. 
The peculiar radial velocity of star 80 makes it
an intriguing candidate for surviving companion.
Assuming that star 80 has a mass of $\sim$1 $M_\odot$, the stellar effective temperature and surface gravity from the spectral fit implies an $M_V$ of 1.33 mag, and the star would be at a distance of $\sim$38 kpc as it has an observed $V$ = 19.25 mag; if a mass of $\sim$1.69 \msun\ is assumed, the star would be at a distance of $\sim$50 kpc.
Follow-up spectroscopic observations of star 80 in 
blue wavelengths are needed for better determination 
of its physical properties. We will further discuss 
this star in Section 7.

%remove blended stars

\section{Investigation of SNR 0548--70.4}
  
No search for surviving companion in the SNR 
0548--70.4 has been reported in the literature. 
In this study, we use our \hst $B$, $V$, $I$,
and \ha images of SNR 0548--70.4 to 
analyze the underlying stellar population, 
and search for the surviving companion candidates.
Unfortunately, no VLT MUSE observation of SNR
0548--70.4 is available for radial velocity 
analyses.

\subsection{Site of SN Explosion and Search Radius} 

The Balmer shell of SNR 0548--70.4 shows a 
regular shape, similar 
to the case of SNR 0509--67.5 \citep{litke2017}. 
We find that over 75\% of the outer shell periphery
can be well fitted by an ellipse.
%, centered at 05$^{\mathrm{h}}$47$^{\mathrm{m}}$48$^{\mathrm{s}}$.47, $-$70$^\circ$24$'$52\farcs20 (J2000)
In the northwest and southeast quadrants, the faint
extended emission deviates from the fitted ellipse, 
and appears to be associated with blowout structures. 
Thus we identify the center of the ellipse as the 
site of SN explosion.  We will use a larger search 
radius to take into account the uncertainty due 
to the irregularity in the Balmer shell periphery.

We choose the MS and helium star search radii of 
SNR 0548--70.4 that are sufficiently larger than the 
respective runaway distances of the MS and 
helium star companions from the explosion site. SNR
0548--70.4 has an age of 10,000 yr 
\citep{hendrick2003}. Adopting the largest runaway
velocity of 270 km s$^{-1}$ and 730 km s$^{-1}$ 
for surviving MS and helium star companions, we
calculate their runaway distance up to 2.8 pc
(11\farcs0) and 7.5 pc (29\farcs8), respectively.
We then adopt 5.0 pc (20\farcs0) and 10.0 pc (40\farcs0) 
as the search radii for MS and helium star companions, 
respectively. 
These limits correspond to a kick velocity of 500 
km s$^{-1}$ and 1000 km s$^{-1}$, respectively.
All stars with $V < 23.0$ mag within the search 
radius of helium star, 40\farcs0, are compiled for 
surviving companion candidates.

%270*(10000)*365*24*3600/3.08567758e13*4~11.0377312979"
%250*(10000)*365*24*3600/3.08567758e13*4~10.22"
%730*(10000)*365*24*3600/3.08567758e13*4~29.84"
%1200*(10000)*365*24*3600/3.08567758e13*4~49.057"
%whole SNR 56'' = 1413.4312 pixels

  %radius+uncertainty
%11.0377+uncertainty ~ 20"
%29.84+uncertainty ~ 40"

\subsection{Photometric Search}

Within the search radius 40\farcs0 in SNR 
0548--70.4, 973 stars are brighter than 
$V < 23.0$ mag. These 973 
stars are numbered in the order of increasing
distance from the explosion site and listed in 
Table \ref{table:photometry0548} with their 
photometric measurements. We have marked these
stars in the \ha (Figure
\ref{figure:central_stars_0548_ha}) and 
color-composite images (Figure \ref{figure:central_stars_0548_rgb}),
as well as in the $V$ versus $B$--$V$ and $I$ 
versus $V$--$I$ CMDs (Figure \ref{figure:CMD0548}). 
In the CMDs, these stars show hints of the 
MS and the RG branch. 
To better illustrate the locations of the MS
and RG branch in the CMDs, we 
have also plotted all stars with $V < 23.0$ mag 
within 60\farcs0 from the explosion site.  

In the CMDs (Figure \ref{figure:CMD0548}), we have 
over-plotted the post-impact evolutionary tracks of 
surviving MS (the tracks above the MS) and helium star 
(the vertical tracks to the left of the MS) companions
\citep{pan2014}. We find that star 98 lies on a 
track of a surviving MS star model in the $V$ versus 
$B$--$V$ CMD (Figure \ref{figure:CMD0548closeup}), 
but not in the $I$ versus $V$--$I$ CMD (Figure \ref{figure:CMD0548closeup2}); furthermore, its 
location in the $V$ versus $B$--$V$ CMD indicates 
a MS surviving companion going though $\sim$ 110 
years after the SN explosion. It is not consistent 
with the age of SNR 0548--70.4, 10,000 yr 
\citep{hendrick2003}. 
%, and has luminosity $ \sim 17 L_\sun$ , radius $\sim 8 R_\sun$, and effective temperature $T_{eff} =$ 5400 K. 
Assuming negligible extinction, star 98's $B$ -- $V$ 
$\sim$ 0.34 suggests an early F spectral type, 
and its $V$ -- $I$ $\sim$ 1.08 suggests an
early K spectral type.

While photometric comparisons with post-impact evolution
models in CMDs do not yield promising candidates for surviving
companions, we notice puzzling photometric properties of star 11.
In the CMDs, star 11 is located in regions not well populated by 
LMC or Galactic stars, as shown in Figure \ref{figure:CMD0548}. 
It is far away from the post-impact evolution tracks of stars;
furthermore, its colors are inconsistent with any spectral types.
Assuming negligible extinction, star 11's $B$ -- $V$ $\sim$ 0.74
suggests a spectral type of G type and its $V$ 
-- $I$ $\sim$ 2.68 suggests a spectral type of M3 type. In 
the $V$ band, star 11 is twice as bright as the Sun, while 
in the $I$ band star 11 is 13 times as bright as the Sun. 
This particularly red $V-I$ color suggests infrared excess.
Interestingly, star 770 and 934 appear puzzling photometric properties in the CMDs as well. These three stars will be discussed further in Section 7.

\section{Discussions}
To differentiate between the SD and DD origins of a Type Ia SN, we use the strategy to 
search for a surviving companion within its SNR.\citep{marietta2000, canal2001, ruiz-lapuente2004, gonzalez2009, gonzalez2012, kerzendorf2014, pan2014}.
%--------------------------------------------  
%             Table 6 
\clearpage
\LongTables
\begin{deluxetable*}{ccccccccc}
\tablecolumns{9}
\tabletypesize{\scriptsize}
%\rotate
\tablewidth{0pc}
\tablecaption{Stars Brighter Than $V = 23.0$ mag near Central Region in the DEML71}
\tablehead{Star & R.A. (J2000) & Decl. (J2000) & B & V & I & B-V & V-I & $r$}
\startdata
1 & 05:05:42.15 & -67:52:41.89 & 22.82 $\pm$ 0.01 & 22.54 $\pm$ 0.01 & 21.80 $\pm$ 0.01 & 0.28 $\pm$ 0.01 & 0.73 $\pm$ 0.01 & 0.70\arcsec \\ 
2 & 05:05:42.03 & -67:52:38.82 & 22.75 $\pm$ 0.01 & 22.44 $\pm$ 0.01 & 21.75 $\pm$ 0.01 & 0.31 $\pm$ 0.01 & 0.69 $\pm$ 0.01 & 2.76\arcsec \\ 
3 & 05:05:41.95 & -67:52:44.45 & 23.10 $\pm$ 0.01 & 22.86 $\pm$ 0.01 & 22.17 $\pm$ 0.01 & 0.24 $\pm$ 0.01 & 0.69 $\pm$ 0.01 & 2.92\arcsec \\ 
4 & 05:05:42.56 & -67:52:41.53 & 22.14 $\pm$ 0.01 & 21.95 $\pm$ 0.01 & 21.42 $\pm$ 0.01 & 0.20 $\pm$ 0.01 & 0.52 $\pm$ 0.01 & 2.95\arcsec \\ 
5 & 05:05:42.47 & -67:52:39.76 & 22.73 $\pm$ 0.01 & 22.44 $\pm$ 0.01 & 21.75 $\pm$ 0.01 & 0.29 $\pm$ 0.01 & 0.69 $\pm$ 0.01 & 3.04\arcsec \\ 
6 & 05:05:41.43 & -67:52:41.34 & 22.12 $\pm$ 0.01 & 21.88 $\pm$ 0.01 & 21.27 $\pm$ 0.01 & 0.24 $\pm$ 0.01 & 0.60 $\pm$ 0.01 & 3.47\arcsec \\ 
7 & 05:05:41.68 & -67:52:38.08 & 22.94 $\pm$ 0.01 & 22.73 $\pm$ 0.01 & 22.08 $\pm$ 0.01 & 0.21 $\pm$ 0.01 & 0.64 $\pm$ 0.01 & 4.04\arcsec \\ 
8 & 05:05:41.62 & -67:52:38.02 & 22.09 $\pm$ 0.01 & 21.91 $\pm$ 0.01 & 21.40 $\pm$ 0.01 & 0.18 $\pm$ 0.01 & 0.51 $\pm$ 0.01 & 4.29\arcsec \\ 
9 & 05:05:42.61 & -67:52:44.58 & 20.44 $\pm$ 0.00 & 20.22 $\pm$ 0.00 & 19.73 $\pm$ 0.00 & 0.22 $\pm$ 0.00 & 0.50 $\pm$ 0.00 & 4.39\arcsec \\ 
10 & 05:05:42.10 & -67:52:36.78 & 23.21 $\pm$ 0.01 & 22.88 $\pm$ 0.01 & 22.22 $\pm$ 0.01 & 0.32 $\pm$ 0.01 & 0.66 $\pm$ 0.01 & 4.81\arcsec \\ 
11 & 05:05:41.87 & -67:52:46.62 & 22.76 $\pm$ 0.01 & 22.52 $\pm$ 0.01 & 21.91 $\pm$ 0.01 & 0.24 $\pm$ 0.01 & 0.61 $\pm$ 0.01 & 5.13\arcsec \\ 
12 & 05:05:41.36 & -67:52:38.18 & 21.69 $\pm$ 0.01 & 21.49 $\pm$ 0.01 & 21.04 $\pm$ 0.01 & 0.21 $\pm$ 0.01 & 0.45 $\pm$ 0.01 & 5.14\arcsec \\ 
13 & 05:05:41.12 & -67:52:39.65 & 21.96 $\pm$ 0.01 & 21.63 $\pm$ 0.01 & 21.07 $\pm$ 0.01 & 0.33 $\pm$ 0.01 & 0.56 $\pm$ 0.01 & 5.52\arcsec \\ 
14 & 05:05:42.67 & -67:52:36.52 & 21.96 $\pm$ 0.01 & 21.77 $\pm$ 0.01 & 21.27 $\pm$ 0.01 & 0.19 $\pm$ 0.01 & 0.50 $\pm$ 0.01 & 6.19\arcsec \\ 
15 & 05:05:41.62 & -67:52:47.57 & 22.60 $\pm$ 0.01 & 22.28 $\pm$ 0.01 & 21.56 $\pm$ 0.01 & 0.31 $\pm$ 0.01 & 0.72 $\pm$ 0.01 & 6.44\arcsec \\ 
16 & 05:05:40.99 & -67:52:38.83 & 22.00 $\pm$ 0.01 & 21.82 $\pm$ 0.01 & 21.35 $\pm$ 0.01 & 0.17 $\pm$ 0.01 & 0.47 $\pm$ 0.01 & 6.54\arcsec \\ 
17 & 05:05:43.04 & -67:52:38.05 & 22.10 $\pm$ 0.01 & 21.85 $\pm$ 0.01 & 21.21 $\pm$ 0.01 & 0.25 $\pm$ 0.01 & 0.65 $\pm$ 0.01 & 6.68\arcsec \\ 
18 & 05:05:42.51 & -67:52:47.85 & 21.48 $\pm$ 0.01 & 21.25 $\pm$ 0.01 & 20.75 $\pm$ 0.01 & 0.24 $\pm$ 0.01 & 0.50 $\pm$ 0.01 & 6.82\arcsec \\ 
19 & 05:05:41.56 & -67:52:47.86 & 20.97 $\pm$ 0.01 & 20.72 $\pm$ 0.00 & 20.23 $\pm$ 0.01 & 0.25 $\pm$ 0.01 & 0.49 $\pm$ 0.01 & 6.84\arcsec \\ 
20 & 05:05:41.33 & -67:52:35.72 & 21.70 $\pm$ 0.01 & 21.50 $\pm$ 0.01 & 21.06 $\pm$ 0.01 & 0.20 $\pm$ 0.01 & 0.44 $\pm$ 0.01 & 7.09\arcsec \\ 
21 & 05:05:41.72 & -67:52:48.61 & 20.04 $\pm$ 0.00 & 19.95 $\pm$ 0.00 & 19.76 $\pm$ 0.00 & 0.10 $\pm$ 0.00 & 0.19 $\pm$ 0.00 & 7.25\arcsec \\ 
22 & 05:05:42.32 & -67:52:34.46 & 21.73 $\pm$ 0.01 & 21.58 $\pm$ 0.01 & 21.23 $\pm$ 0.01 & 0.15 $\pm$ 0.01 & 0.35 $\pm$ 0.01 & 7.29\arcsec \\ 
23 & 05:05:42.92 & -67:52:35.96 & 22.89 $\pm$ 0.01 & 22.63 $\pm$ 0.01 & 21.93 $\pm$ 0.02 & 0.26 $\pm$ 0.01 & 0.70 $\pm$ 0.02 & 7.49\arcsec \\ 
24 & 05:05:41.15 & -67:52:35.91 & 20.85 $\pm$ 0.01 & 20.45 $\pm$ 0.00 & 19.38 $\pm$ 0.00 & 0.40 $\pm$ 0.01 & 1.08 $\pm$ 0.00 & 7.59\arcsec \\ 
25 & 05:05:41.87 & -67:52:49.22 & 22.21 $\pm$ 0.01 & 21.84 $\pm$ 0.01 & 20.86 $\pm$ 0.01 & 0.37 $\pm$ 0.01 & 0.98 $\pm$ 0.01 & 7.70\arcsec \\ 
26 & 05:05:40.80 & -67:52:37.14 & 21.99 $\pm$ 0.01 & 21.85 $\pm$ 0.01 & 21.34 $\pm$ 0.01 & 0.14 $\pm$ 0.01 & 0.51 $\pm$ 0.01 & 8.28\arcsec \\ 
27 & 05:05:43.24 & -67:52:36.76 & 20.39 $\pm$ 0.00 & 20.17 $\pm$ 0.00 & 19.80 $\pm$ 0.00 & 0.23 $\pm$ 0.00 & 0.37 $\pm$ 0.00 & 8.31\arcsec \\ 
28 & 05:05:40.81 & -67:52:37.01 & 21.28 $\pm$ 0.01 & 21.05 $\pm$ 0.00 & 20.37 $\pm$ 0.01 & 0.23 $\pm$ 0.01 & 0.69 $\pm$ 0.01 & 8.34\arcsec \\ 
29 & 05:05:42.20 & -67:52:33.15 & 22.80 $\pm$ 0.01 & 22.54 $\pm$ 0.01 & 21.80 $\pm$ 0.01 & 0.26 $\pm$ 0.01 & 0.74 $\pm$ 0.01 & 8.48\arcsec \\ 
30 & 05:05:40.52 & -67:52:39.66 & 21.66 $\pm$ 0.01 & 21.48 $\pm$ 0.01 & 21.07 $\pm$ 0.01 & 0.19 $\pm$ 0.01 & 0.41 $\pm$ 0.01 & 8.77\arcsec \\ 
31 & 05:05:43.43 & -67:52:45.98 & 22.62 $\pm$ 0.01 & 22.34 $\pm$ 0.01 & 21.55 $\pm$ 0.01 & 0.28 $\pm$ 0.01 & 0.78 $\pm$ 0.01 & 9.02\arcsec \\ 
32 & 05:05:40.44 & -67:52:38.91 & 23.11 $\pm$ 0.01 & 22.85 $\pm$ 0.01 & 22.19 $\pm$ 0.01 & 0.26 $\pm$ 0.01 & 0.67 $\pm$ 0.01 & 9.40\arcsec \\ 
33 & 05:05:42.03 & -67:52:51.08 & 23.06 $\pm$ 0.01 & 22.81 $\pm$ 0.01 & 22.07 $\pm$ 0.01 & 0.25 $\pm$ 0.01 & 0.75 $\pm$ 0.01 & 9.50\arcsec \\ 
34 & 05:05:41.55 & -67:52:32.47 & 21.12 $\pm$ 0.01 & 20.99 $\pm$ 0.01 & 20.62 $\pm$ 0.01 & 0.12 $\pm$ 0.01 & 0.38 $\pm$ 0.01 & 9.52\arcsec \\ 
35 & 05:05:43.42 & -67:52:47.17 & 23.05 $\pm$ 0.01 & 22.77 $\pm$ 0.01 & 22.11 $\pm$ 0.01 & 0.28 $\pm$ 0.01 & 0.66 $\pm$ 0.01 & 9.61\arcsec \\ 
36 & 05:05:42.45 & -67:52:51.03 & 21.81 $\pm$ 0.01 & 21.41 $\pm$ 0.01 & 20.37 $\pm$ 0.01 & 0.40 $\pm$ 0.01 & 1.04 $\pm$ 0.01 & 9.73\arcsec \\ 
37 & 05:05:42.96 & -67:52:49.98 & 20.11 $\pm$ 0.00 & 19.68 $\pm$ 0.00 & 18.71 $\pm$ 0.00 & 0.42 $\pm$ 0.00 & 0.97 $\pm$ 0.00 & 9.89\arcsec \\ 
38 & 05:05:41.65 & -67:52:31.73 & 21.02 $\pm$ 0.01 & 20.84 $\pm$ 0.00 & 20.42 $\pm$ 0.01 & 0.17 $\pm$ 0.01 & 0.42 $\pm$ 0.01 & 10.10\arcsec \\ 
39 & 05:05:41.43 & -67:52:31.98 & 22.41 $\pm$ 0.01 & 22.11 $\pm$ 0.01 & 21.15 $\pm$ 0.01 & 0.30 $\pm$ 0.01 & 0.96 $\pm$ 0.01 & 10.20\arcsec \\ 
40 & 05:05:40.50 & -67:52:36.23 & 23.14 $\pm$ 0.01 & 22.86 $\pm$ 0.01 & 22.19 $\pm$ 0.01 & 0.28 $\pm$ 0.01 & 0.67 $\pm$ 0.01 & 10.20\arcsec \\ 
41 & 05:05:41.32 & -67:52:51.07 & 23.20 $\pm$ 0.01 & 22.95 $\pm$ 0.01 & 22.28 $\pm$ 0.01 & 0.26 $\pm$ 0.01 & 0.67 $\pm$ 0.01 & 10.33\arcsec \\ 
42 & 05:05:42.74 & -67:52:51.15 & 20.59 $\pm$ 0.00 & 20.37 $\pm$ 0.00 & 19.91 $\pm$ 0.01 & 0.23 $\pm$ 0.00 & 0.46 $\pm$ 0.01 & 10.34\arcsec \\ 
43 & 05:05:42.97 & -67:52:32.64 & 21.22 $\pm$ 0.01 & 21.04 $\pm$ 0.00 & 20.63 $\pm$ 0.01 & 0.18 $\pm$ 0.01 & 0.41 $\pm$ 0.01 & 10.38\arcsec \\ 
44 & 05:05:40.38 & -67:52:37.11 & 22.95 $\pm$ 0.01 & 22.70 $\pm$ 0.01 & 22.05 $\pm$ 0.01 & 0.25 $\pm$ 0.01 & 0.65 $\pm$ 0.01 & 10.41\arcsec \\ 
45 & 05:05:41.27 & -67:52:51.41 & 19.81 $\pm$ 0.00 & 19.69 $\pm$ 0.00 & 19.44 $\pm$ 0.00 & 0.13 $\pm$ 0.00 & 0.25 $\pm$ 0.00 & 10.76\arcsec \\ 
46 & 05:05:43.51 & -67:52:34.72 & 23.10 $\pm$ 0.01 & 22.85 $\pm$ 0.01 & 22.18 $\pm$ 0.01 & 0.25 $\pm$ 0.01 & 0.67 $\pm$ 0.01 & 10.78\arcsec \\ 
47 & 05:05:42.28 & -67:52:30.77 & 22.79 $\pm$ 0.01 & 22.54 $\pm$ 0.01 & 21.80 $\pm$ 0.01 & 0.25 $\pm$ 0.01 & 0.74 $\pm$ 0.01 & 10.90\arcsec \\ 
48 & 05:05:43.61 & -67:52:47.95 & 22.21 $\pm$ 0.01 & 21.95 $\pm$ 0.01 & 21.34 $\pm$ 0.01 & 0.26 $\pm$ 0.01 & 0.60 $\pm$ 0.01 & 10.94\arcsec \\ 
49 & 05:05:43.17 & -67:52:50.49 & 21.64 $\pm$ 0.01 & 21.46 $\pm$ 0.01 & 21.00 $\pm$ 0.01 & 0.17 $\pm$ 0.01 & 0.47 $\pm$ 0.01 & 10.98\arcsec \\ 
50 & 05:05:41.51 & -67:52:31.00 & 22.32 $\pm$ 0.01 & 22.10 $\pm$ 0.01 & 21.44 $\pm$ 0.01 & 0.22 $\pm$ 0.01 & 0.66 $\pm$ 0.01 & 11.00\arcsec \\ 
51 & 05:05:40.37 & -67:52:47.30 & 19.80 $\pm$ 0.00 & 19.68 $\pm$ 0.00 & 19.62 $\pm$ 0.00 & 0.12 $\pm$ 0.00 & 0.07 $\pm$ 0.00 & 11.01\arcsec \\ 
52 & 05:05:40.27 & -67:52:46.25 & 22.67 $\pm$ 0.01 & 22.40 $\pm$ 0.01 & 21.66 $\pm$ 0.01 & 0.27 $\pm$ 0.01 & 0.74 $\pm$ 0.01 & 11.04\arcsec \\ 
53 & 05:05:41.77 & -67:52:30.54 & 19.70 $\pm$ 0.00 & 19.57 $\pm$ 0.00 & 19.41 $\pm$ 0.00 & 0.13 $\pm$ 0.00 & 0.16 $\pm$ 0.00 & 11.14\arcsec \\ 
54 & 05:05:40.60 & -67:52:49.31 & 22.84 $\pm$ 0.01 & 22.58 $\pm$ 0.01 & 21.86 $\pm$ 0.01 & 0.25 $\pm$ 0.01 & 0.73 $\pm$ 0.01 & 11.24\arcsec \\ 
55 & 05:05:42.18 & -67:52:52.95 & 22.46 $\pm$ 0.01 & 22.24 $\pm$ 0.01 & 21.66 $\pm$ 0.01 & 0.22 $\pm$ 0.01 & 0.59 $\pm$ 0.01 & 11.40\arcsec \\ 
56 & 05:05:44.05 & -67:52:42.95 & 22.45 $\pm$ 0.01 & 22.14 $\pm$ 0.01 & 21.37 $\pm$ 0.01 & 0.31 $\pm$ 0.01 & 0.77 $\pm$ 0.01 & 11.43\arcsec \\ 
57 & 05:05:44.04 & -67:52:43.20 & 21.83 $\pm$ 0.01 & 21.39 $\pm$ 0.01 & 20.32 $\pm$ 0.01 & 0.43 $\pm$ 0.01 & 1.07 $\pm$ 0.01 & 11.44\arcsec \\ 
58 & 05:05:42.21 & -67:52:52.98 & 22.13 $\pm$ 0.01 & 21.76 $\pm$ 0.01 & 20.77 $\pm$ 0.01 & 0.36 $\pm$ 0.01 & 0.99 $\pm$ 0.01 & 11.44\arcsec \\ 
59 & 05:05:40.08 & -67:52:44.62 & 22.41 $\pm$ 0.01 & 22.15 $\pm$ 0.01 & 21.46 $\pm$ 0.01 & 0.26 $\pm$ 0.01 & 0.69 $\pm$ 0.01 & 11.47\arcsec \\ 
60 & 05:05:42.24 & -67:52:53.14 & 19.25 $\pm$ 0.00 & 18.82 $\pm$ 0.00 & 17.54 $\pm$ 0.00 & 0.42 $\pm$ 0.00 & 1.29 $\pm$ 0.00 & 11.62\arcsec \\ 
61 & 05:05:40.10 & -67:52:45.44 & 21.27 $\pm$ 0.01 & 21.11 $\pm$ 0.01 & 20.58 $\pm$ 0.01 & 0.16 $\pm$ 0.01 & 0.53 $\pm$ 0.01 & 11.63\arcsec \\ 
62 & 05:05:40.00 & -67:52:40.04 & 20.48 $\pm$ 0.00 & 20.11 $\pm$ 0.00 & 19.08 $\pm$ 0.00 & 0.37 $\pm$ 0.00 & 1.03 $\pm$ 0.00 & 11.64\arcsec \\ 
63 & 05:05:44.10 & -67:52:43.11 & 21.42 $\pm$ 0.01 & 21.27 $\pm$ 0.01 & 20.94 $\pm$ 0.01 & 0.15 $\pm$ 0.01 & 0.33 $\pm$ 0.01 & 11.76\arcsec \\ 
64 & 05:05:40.50 & -67:52:49.64 & 22.91 $\pm$ 0.01 & 22.79 $\pm$ 0.01 & 22.10 $\pm$ 0.01 & 0.12 $\pm$ 0.01 & 0.69 $\pm$ 0.01 & 11.85\arcsec \\ 
65 & 05:05:40.49 & -67:52:49.62 & 22.97 $\pm$ 0.01 & 22.66 $\pm$ 0.01 & 21.94 $\pm$ 0.01 & 0.31 $\pm$ 0.01 & 0.72 $\pm$ 0.01 & 11.90\arcsec \\ 
66 & 05:05:42.72 & -67:52:30.26 & 23.00 $\pm$ 0.01 & 22.75 $\pm$ 0.01 & 22.23 $\pm$ 0.01 & 0.25 $\pm$ 0.01 & 0.52 $\pm$ 0.01 & 11.96\arcsec \\ 
67 & 05:05:44.10 & -67:52:44.85 & 22.54 $\pm$ 0.01 & 22.21 $\pm$ 0.01 & 21.53 $\pm$ 0.01 & 0.33 $\pm$ 0.01 & 0.68 $\pm$ 0.01 & 12.07\arcsec \\ 
68 & 05:05:42.37 & -67:52:53.77 & 22.43 $\pm$ 0.01 & 22.19 $\pm$ 0.01 & 21.51 $\pm$ 0.01 & 0.24 $\pm$ 0.01 & 0.68 $\pm$ 0.01 & 12.33\arcsec \\ 
69 & 05:05:39.90 & -67:52:44.27 & 22.57 $\pm$ 0.01 & 22.34 $\pm$ 0.01 & 21.62 $\pm$ 0.01 & 0.23 $\pm$ 0.01 & 0.73 $\pm$ 0.01 & 12.37\arcsec \\ 
70 & 05:05:44.20 & -67:52:39.36 & 22.00 $\pm$ 0.01 & 21.77 $\pm$ 0.01 & 21.34 $\pm$ 0.01 & 0.24 $\pm$ 0.01 & 0.42 $\pm$ 0.01 & 12.40\arcsec \\ 
71 & 05:05:40.10 & -67:52:47.54 & 23.19 $\pm$ 0.01 & 22.90 $\pm$ 0.01 & 22.21 $\pm$ 0.01 & 0.28 $\pm$ 0.01 & 0.69 $\pm$ 0.01 & 12.45\arcsec \\ 
72 & 05:05:41.46 & -67:52:29.49 & 21.88 $\pm$ 0.01 & 21.68 $\pm$ 0.01 & 20.94 $\pm$ 0.01 & 0.20 $\pm$ 0.01 & 0.74 $\pm$ 0.01 & 12.53\arcsec \\ 
73 & 05:05:39.85 & -67:52:43.92 & 22.71 $\pm$ 0.01 & 22.48 $\pm$ 0.01 & 21.81 $\pm$ 0.01 & 0.23 $\pm$ 0.01 & 0.67 $\pm$ 0.01 & 12.60\arcsec \\ 
74 & 05:05:40.03 & -67:52:35.92 & 21.92 $\pm$ 0.01 & 21.56 $\pm$ 0.01 & 20.55 $\pm$ 0.01 & 0.36 $\pm$ 0.01 & 1.01 $\pm$ 0.01 & 12.68\arcsec \\ 
75 & 05:05:41.82 & -67:52:54.28 & 21.21 $\pm$ 0.01 & 21.09 $\pm$ 0.00 & 20.71 $\pm$ 0.01 & 0.11 $\pm$ 0.01 & 0.39 $\pm$ 0.01 & 12.76\arcsec \\ 
76 & 05:05:43.01 & -67:52:53.10 & 23.13 $\pm$ 0.01 & 22.80 $\pm$ 0.01 & 22.04 $\pm$ 0.01 & 0.33 $\pm$ 0.01 & 0.76 $\pm$ 0.01 & 12.76\arcsec \\ 
77 & 05:05:40.71 & -67:52:51.95 & 21.67 $\pm$ 0.01 & 21.57 $\pm$ 0.01 & 21.17 $\pm$ 0.01 & 0.10 $\pm$ 0.01 & 0.40 $\pm$ 0.01 & 12.80\arcsec \\ 
78 & 05:05:40.60 & -67:52:51.86 & 22.63 $\pm$ 0.01 & 22.39 $\pm$ 0.01 & 21.83 $\pm$ 0.01 & 0.23 $\pm$ 0.01 & 0.57 $\pm$ 0.01 & 13.11\arcsec \\ 
79 & 05:05:39.89 & -67:52:46.50 & 22.18 $\pm$ 0.01 & 21.96 $\pm$ 0.01 & 21.35 $\pm$ 0.01 & 0.22 $\pm$ 0.01 & 0.61 $\pm$ 0.01 & 13.12\arcsec \\ 
80 & 05:05:43.32 & -67:52:30.53 & 19.69 $\pm$ 0.00 & 19.25 $\pm$ 0.00 & 18.19 $\pm$ 0.00 & 0.44 $\pm$ 0.00 & 1.06 $\pm$ 0.00 & 13.19\arcsec \\ 
81 & 05:05:43.86 & -67:52:33.28 & 21.31 $\pm$ 0.01 & 21.12 $\pm$ 0.00 & 20.78 $\pm$ 0.01 & 0.18 $\pm$ 0.01 & 0.35 $\pm$ 0.01 & 13.21\arcsec \\ 
82 & 05:05:44.37 & -67:52:39.22 & 22.50 $\pm$ 0.01 & 22.29 $\pm$ 0.01 & 21.72 $\pm$ 0.01 & 0.20 $\pm$ 0.01 & 0.57 $\pm$ 0.01 & 13.39\arcsec \\ 
83 & 05:05:39.94 & -67:52:47.93 & 19.86 $\pm$ 0.00 & 19.78 $\pm$ 0.00 & 19.63 $\pm$ 0.00 & 0.08 $\pm$ 0.00 & 0.15 $\pm$ 0.00 & 13.44\arcsec \\ 
84 & 05:05:44.45 & -67:52:42.42 & 22.28 $\pm$ 0.01 & 22.02 $\pm$ 0.01 & 21.39 $\pm$ 0.01 & 0.26 $\pm$ 0.01 & 0.63 $\pm$ 0.01 & 13.63\arcsec \\ 
85 & 05:05:39.70 & -67:52:37.64 & 23.22 $\pm$ 0.01 & 22.92 $\pm$ 0.01 & 22.27 $\pm$ 0.01 & 0.30 $\pm$ 0.01 & 0.65 $\pm$ 0.01 & 13.78\arcsec \\ 
86 & 05:05:41.86 & -67:52:55.72 & 23.05 $\pm$ 0.01 & 22.76 $\pm$ 0.01 & 22.06 $\pm$ 0.01 & 0.29 $\pm$ 0.01 & 0.70 $\pm$ 0.01 & 14.18\arcsec \\ 
87 & 05:05:41.05 & -67:52:54.66 & 22.72 $\pm$ 0.01 & 22.46 $\pm$ 0.01 & 21.84 $\pm$ 0.01 & 0.26 $\pm$ 0.01 & 0.63 $\pm$ 0.01 & 14.23\arcsec \\ 
88 & 05:05:39.52 & -67:52:40.34 & 22.65 $\pm$ 0.01 & 22.41 $\pm$ 0.01 & 21.69 $\pm$ 0.01 & 0.24 $\pm$ 0.01 & 0.72 $\pm$ 0.01 & 14.28\arcsec \\ 
89 & 05:05:41.85 & -67:52:27.29 & 21.12 $\pm$ 0.01 & 20.89 $\pm$ 0.00 & 20.21 $\pm$ 0.01 & 0.23 $\pm$ 0.01 & 0.67 $\pm$ 0.01 & 14.33\arcsec 
\enddata
\label{table:photometrydeml71}
\end{deluxetable*}

%--------------------------------------------
%             Table 7
\begin{deluxetable*}{ccccccccc}
\tablecolumns{9}
\tabletypesize{\scriptsize}
%\rotate
\tablewidth{0pc}
\tablecaption{
Fitting Stellar Parameters of the Stars Brighter Than $V = 21.6$ mag in Table \ref{table:photometrydeml71} }
\tablehead{Star & T$_{eff}$ (K) & log $g$ (dex) & [Fe/H] (dex) & $V_{r}$ (km s$^{-1}$) &  }
\startdata
9 & 7222 $\pm$ 200 & 4.1 $\pm$ 1.0 & -0.7 $\pm$ 0.2 & 252 $\pm$ 0 \\
12 & 7202 $\pm$ 201 & 3.7 $\pm$ 1.0 & -0.8 $\pm$ 0.2 & 288 $\pm$ 0 \\
18 & 7090 $\pm$ 201 & 3.8 $\pm$ 1.0 & -0.9 $\pm$ 0.2 & 247 $\pm$ 0 \\
19 & 6710 $\pm$ 201 & 4.2 $\pm$ 1.0 & -0.5 $\pm$ 0.2 & 266 $\pm$ 0 \\
20 & 7418 $\pm$ 201 & 4.5 $\pm$ 1.0 & -0.6 $\pm$ 0.2 & 260 $\pm$ 0 \\
21 & 9181 $\pm$ 200 & 3.9 $\pm$ 1.0 & -0.8 $\pm$ 0.2 & 301 $\pm$ 0 \\
22 & 7698 $\pm$ 201 & 4.3 $\pm$ 1.0 & -0.4 $\pm$ 0.2 & 285 $\pm$ 0 \\
24 & 5783 $\pm$ 200 & 4.3 $\pm$ 1.0 & -0.4 $\pm$ 0.2 & 245 $\pm$ 0 \\
27 & 7549 $\pm$ 200 & 4.4 $\pm$ 1.0 & -0.3 $\pm$ 0.2 & 273 $\pm$ 0 \\
28 & 7044 $\pm$ 200 & 4.1 $\pm$ 1.0 & -0.7 $\pm$ 0.2 & 273 $\pm$ 0 \\
30 & 7411 $\pm$ 201 & 4.3 $\pm$ 1.0 & -0.2 $\pm$ 0.2 & 263 $\pm$ 0 \\
34 & 7733 $\pm$ 201 & 4.3 $\pm$ 1.0 & -0.7 $\pm$ 0.2 & 317 $\pm$ 0 \\
36 & 5612 $\pm$ 202 & 4.2 $\pm$ 1.0 & -0.7 $\pm$ 0.2 & 254 $\pm$ 0 \\
37 & 5613 $\pm$ 200 & 3.8 $\pm$ 1.0 & -0.3 $\pm$ 0.2 & 254 $\pm$ 0 \\
38 & 7692 $\pm$ 201 & 4.0 $\pm$ 1.0 & -0.7 $\pm$ 0.2 & 276 $\pm$ 0 \\
42 & 7338 $\pm$ 200 & 3.9 $\pm$ 1.0 & -0.7 $\pm$ 0.2 & 290 $\pm$ 0 \\
43 & 7201 $\pm$ 201 & 3.1 $\pm$ 1.0 & -0.7 $\pm$ 0.2 & 319 $\pm$ 0 \\
45 & 8051 $\pm$ 200 & 4.3 $\pm$ 1.0 & -0.4 $\pm$ 0.2 & 273 $\pm$ 0 \\
49 & 7145 $\pm$ 201 & 3.8 $\pm$ 1.0 & -1.0 $\pm$ 0.2 & 261 $\pm$ 0 \\
51 & 12780 $\pm$ 200 & 4.7 $\pm$ 1.0 & -1.0 $\pm$ 0.2 & 264 $\pm$ 0 \\
53 & 8699 $\pm$ 201 & 3.9 $\pm$ 1.0 & -0.9 $\pm$ 0.2 & 289 $\pm$ 0 \\
60 & 4846 $\pm$ 200 & 3.0 $\pm$ 1.0 & -0.6 $\pm$ 0.2 & 278 $\pm$ 0 \\
61 & 7032 $\pm$ 201 & 3.9 $\pm$ 1.0 & -0.8 $\pm$ 0.2 & 250 $\pm$ 0 \\
62 & 5862 $\pm$ 200 & 3.8 $\pm$ 1.0 & -1.0 $\pm$ 0.2 & 297 $\pm$ 0 \\
63 & 7883 $\pm$ 201 & 4.7 $\pm$ 1.0 & 0.2 $\pm$ 0.2 & 287 $\pm$ 0 \\
74 & 6203 $\pm$ 201 & 5.0 $\pm$ 1.0 & -0.2 $\pm$ 0.2 & 252 $\pm$ 0 \\
75 & 7961 $\pm$ 201 & 4.3 $\pm$ 1.0 & -0.4 $\pm$ 0.2 & 281 $\pm$ 0 \\
77 & 7688 $\pm$ 201 & 4.5 $\pm$ 1.0 & -0.5 $\pm$ 0.2 & 245 $\pm$ 0 \\
80 & 5341 $\pm$ 200 & 2.8 $\pm$ 1.0 & -0.6 $\pm$ 0.2 & 213 $\pm$ 0 \\
81 & 7944 $\pm$ 201 & 4.1 $\pm$ 1.0 & -0.4 $\pm$ 0.2 & 264 $\pm$ 0 \\
83 & 10782 $\pm$ 201 & 4.4 $\pm$ 1.0 & 0.7 $\pm$ 0.2 & 281 $\pm$ 0 \\
89 & 6273 $\pm$ 200 & 3.5 $\pm$ 1.0 & -0.9 $\pm$ 0.2 & 257 $\pm$ 0
\enddata
\tablenotetext{}{See the note in Table \ref{table:starkit0519}.}
\label{table:starkitdeml71}
\end{deluxetable*}

\noindent  We have used
photometric and spectroscopic methods to search for surviving companions in three young 
LMC Type Ia SNRs: 0519--69.0, DEM L71 and 0548--70.4. Below we discuss the pros and cons 
of these two methods in Subsections 7.1 and 7.2, and further discuss the inadequacy of 
Gaia Data in the search for surviving companions in Subsection 7.3. Finally, we discuss 
implications of our results on progenitor systems of Type Ia SNe.

\subsection{Photometric Search}
 
To probe the parameter space of a close non-degenerate companion of a WD that leads to a 
Type Ia SN of SD origin, detailed models of binary stellar evolution for a range of
separations, mass ratios, and WD masses have been calculated \citep{hachisu1999,han2004,
meng2007,hachisu2008,han2008,wang2009a,wang2009b,meng2009,wang2010}; however, most of
these models end at the SN explosion without extending to the SN impact on the stellar
companion and afterwards. On the other hand, some hydrodynamic model calculations start from 
the Type Ia SN explosion and focus on the impact of SN ejecta on the surviving stellar
companion and its subsequent evolution, taking into consideration of explosion geometry 
and impact angle \citep{marietta2000,pakmor2008,liu2012,pan2014}. In addition, less
sophisticated calculations assuming ad hoc energy input and mass stripping have been 
made for the post-impact evolution of a 1 $M_\odot$ subgiant or MS surviving companions 
\citep{podsiadlowski2003,shappee2013}. 

Among the above models, \cite{pan2014} have carried out comprehensive calculations 
that detail the post-impact evolution of a surviving companion's effective temperature 
and luminosity. They considered only MS and helium star companions.  For a given temperature 
and luminosity of a surviving companion, its stellar emission was approximated by 
a blackbody model and convolved with a filter response curve to compute the photometric 
magnitude in that filter. Thus, the post-impact evolution of a surviving companion in 
the temperature-luminosity domain could be easily plotted in CMDs for direct comparisons 
with observations.

Using \hst photometric measurements of stars in SNRs 0519--69.0, DEM L71, and 0548--70.4,
we have constructed CMDs in $V$ versus $B-V$ and $I$ versus $V-I$, and compared the stars with
\citet{pan2014} evolutionary tracks of surviving MS and helium star companions 
(Figures \ref{figure:CMD0519}, \ref{figure:CMDdeml71}, and \ref{figure:CMD0548}).  
We do not find any MS or helium star candidates with $V <$ 23.0 for surviving 
companion in these three Type Ia SNRs.  As \citet{pan2014} did not model cases with RG
companions, we cannot assess the existence of surviving RG companions in these SNRs, 
although there are a number of stars within our search radii lie on the RG branch in 
the CMDs. As shown by spectroscopic observations and discussed in the next subsection, 
some RG stars appear to show large peculiar radial velocities. Future models of 
surviving RG companions are needed for comparison.  

Interestingly, in the SNR 0548--70.4, we find star 11 within the MS search radius 
in a strange location in the CMDs (Figure \ref{figure:CMD0548}). Its colors cannot be 
reproduced by stars of any temperature with a normal extinction law.  As there are
additional sources with similar colors and location in the CMDs
(e.g., star 770 and star 934)
, and the colors are
similar to some cataloged galaxies, we think star 11 is most likely a background galaxy, instead of a star, as well as source 934.  
Unfortunately, star 11 is below the detection limit of the 
2MASS catalog; future sensitive $JHK$ photometric measurements are needed to extend
its spectral energy distribution (SED) to confirm its nature as a background galaxy.

\subsection{Peculiar Radial Velocity Search}
We have also used VLT MUSE observations to carry out spectroscopic analyses, and used 
stellar spectra to search for high radial velocities as diagnostics of surviving 
companions. In the SNRs 0519--69.0 and DEM L71, we find each has a star with radial 
velocity that deviate by more than 2.5$\sigma$ (75 and 50 km~s$^{-1}$) from the mean of the underlying 
stellar population, 264 \kms and 270 km~s$^{-1}$, respectively
(see Figure \ref{figure:fig_0519_V_r} 
and Figure \ref{figure:fig_deml71_V_r}).
%; {\color{red}\uline{Star 5 in 0519--69.0 with a radial velocity of 182 \kms and Star 80 in DEM L71 with a radial velocity of 213 \kms).}}
These peculiar radial velocities are intriguing and may suggest that these stars are surviving companions.  In the case of 0519$-69.0$, star 5 is half way between the MS and helium star search radii, while in the case of DEM L71, star 80 is close to the helium star search radius; however, both stars with peculiar radial velocities are located in the RG branch in the CMDs. Because of the uncertainties in the exact site of SN explosion and
our inadequate understanding of the Type Ia SN progenitors, we can neither confirm nor exclude these two stars as candidates for surviving companions in SNRs 0519--69.0 and DEM L71.  Note that DEM L71 has two other stars with radial velocities at $>2\sigma$ but $<2.5\sigma$ from the mean radial velocity of the underlying population, and both their locations appear to be in the MS in the CMDs. 
% assume 1 msun, star 34 is 39 kpc? 43 is 125kpc.
Follow-up observations to measure chemical abundance and stellar rotation of these stars are needed to determine whether they are indeed surviving companions of Type Ia SN progenitors.

\subsection{Peculiar Proper Motion Search}
Beside the aforementioned methods, we have also looked into Gaia Data Release 2 (DR2) to investigate proper motions of stars near the explosion sites of the three SNRs, and search for large proper motions to diagnose surviving companions.  A transverse velocity of 300 km s$^{-1}$ at the LMC distance of $\sim$ 50 kpc corresponds to a proper motion of 1.26 mas yr$^{-1}$. Most stars observed in these Type Ia SNRs are fainter than $V$ = 18-19.  As the quoted uncertainty in proper motion in Gaia DR2 is 1.20 mas yr$^{-1}$ for a star with $G = 20$, it would be impossible to use Gaia DR2 to conduct a meaningful investigation of peculiar proper motions of intermediate- and low-mass stars.  Stars near the tip of the RG branch are bright enough to have more accurate proper motion measurements, but no such RGs are seen in these Type Ia SNRs.

\subsection{Implications for Type Ia SNe Progenitors}

In the SD scenario, the companion of the WD progenitor is expected to survive the SN explosion and be identifiable \citep{marietta2000,pakmor2008,liu2013,pan2014}; 
however, recent searches of surviving companion in the Milky Way (MW)
\citep{ruiz-lapuente2004, kerzendorf2009, 
gonzalez2009, gonzalez2012, kerzendorf2012, kerzendorf2013, kerzendorf2014, kerzendorf2018a, kerzendorf2018b, ruiz-lapuente2018b, ruiz-lapuente2019} and the LMC
(\citealt{schaefer2012}, \citetalias{edwards2012}, \citetalias{pagnotta2015}, \citealt{li2017}, \citealt{litke2017}) have not unambiguously identified and confirmed any surviving companion. In this study, we have used photometric and spectroscopic methods to search for candidates of surviving companion within the three young LMC Type Ia SNRs, but are still unable to make unambiguous identification and confirmation. See Appendix  \ref{appendix:summary} for a detailed compilation of these searches and their results.

Nevertheless, the lack of an obvious surviving companion cannot exclude the SD scenario. In the SD scenario, the modeled post-impact properties of surviving companion are usually calculated under reasonable assumptions for a simplified condition. It is conceivable that the assumed conditions are not appropriate and lead to discrepancies between the observations and model predictions of surviving companions. For instance, models of \cite{pan2014} adopt a classical SD scenario, in which a normal Chandrasekhar mass model \citep{whelan1973, nomoto1982} with an initial spherically symmetric explosion is assumed. 
If the WD in the SD case has a sub-Chandrasekhar mass instead, and explodes through double detonation \citep{nomoto1982}, the explosion energy will be lower than that assumed by \citet{pan2014}.  Furthermore, the SN explosion may not be spherically symmetric, as a result of the binary orbit and stellar rotation
\citep{kashi2011}.  
These different explosion energy and geometry will certainly affect the post-impact temperature and luminosity of the surviving companions and change their locations in the CMDs.
Future post-impact evolution models of surviving companions need to consider different types of SN explosions and, more importantly, include RG companions.

It should be noted that the conventional SD and DD classifications of Type Ia SNe may be an over-simplification.  Many alternative models for Type Ia SNe have been proposed (see \citealt{wang2012,maoz2014,ruiz-lapuente2014,wang2018,ruiz-lapuente2018a} for reviews), such as the sub-Chandrasekhar model \citep{nomoto1982,woosley1986,bildsten2007,shen2009,fink2010,guillochon2010,dan2011,raskin2012,pakmor2013,shen2018},
the super-Chandrasekhar model \citep{yoon2004, hachisu2012a, hachisu2012b, Boshkayev2013, wang2014},
the spin-up/spin-down model \citep{distefano2011,justham2011}, 
the single-star model \citep{iben1983,tout2008},
the violent DD merger model \citep{Pakmor2010, Pakmor2011, Pakmor2012},
the collisional DD model \citep{chomiuk2008, raskin2009,kushnir2013}, the core-degenerate (CD) model \citep{kashi2011}, and the M-dwarf model \citep{wheeler2012}, etc.
%The angular momentum from the rotation of the WD can not be ignored.
These models predict different outcomes for Type Ia SNe apart from the classical SD scenario, resulting in different geometric distribution and abundance of the SN ejecta whose impact on a surviving companion would manifest differently and need further exploration.

Finally, from studies of the five young Balmer-dominated Type Ia SNRs in the LMC, we find: no surviving stellar companion exists in 0509--67.5 \citep{schaefer2012, litke2017}; Star 1 in N103B is a promising candidate for surviving companion \citep{li2017}; 0548--70.4 has no obvious candidate for surviving companion, while DEM L71 and 0519--69.0 each has a possible candidate for surviving companion based on anomalous radial velocities (this paper).  These results imply 20\% -- 60\% of them may originate from SD Type Ia SNe.  This is not inconsistent with the 20\% suggested by \citet{gonzalez2012}; however, these results are derived with small number statistics. Future surveys of young Balmer-dominated Type Ia SNRs in nearby galaxies, such as M31 and M33, should be made in order to expand the Type Ia SNR sample size and the 30m-class telescopes under construction can be used to search for surviving companions of Type Ia SNe in the future.

\section{Summary}

We have been searching for surviving companions of Type Ia SN progenitors in five SNRs in the LMC.  This paper reports our study of three Type Ia SNRs in the LMC: 0519--69.0, DEM\,L71, and 0548--70.4.

We used our and archival \hst images in \ha and $BVI$
continuum bands to examine structures of the SNRs and their
underlying stellar populations, respectively.
The continuum-band images were used to make photometric measurements and construct CMDs for stars projected in the SNRs and their vicinities, in order to be compared with model predictions of post-impact evolution of surviving companions.

All stars within a search radius from an estimated SN explosion site were considered initially.
To find the sites of SN explosion, we fitted ellipses to the SNRs' Balmer shells and adopted the centers of the ellipses as the SN explosion sites.  The search radii were adopted from the runaway distances for surviving MS and He star companions, estimated as SNR age times possible kick velocities.

The faintest post-impact surviving companions in the models of  \citet{pan2014} have $V \sim 23.0$ mag for He star companions; thus we only include stars with $V < 23.0$ in the CMDs for comparisons with models.  In these comparisons We do not find any star located on post-impact evolutionary tracks consistently between the $V$ versus $B-V$ and $I$ versus $V-I$ CMDs. Note that the explosion mechanism and geometry maybe different from those assumed by
 \citet{pan2014}; thus, our search cannot exclude faint companions that were not considered by
\citet{pan2014}.

We have also used VLT MUSE observations of 0519--69.0 and DEM\,L71 to perform spectroscopic analyses of the stars with $V < 21.6$ mag in each SNR to search for peculiar radial velocities as diagnostic for candidate surviving companions.
We find one star in each SNR with radial velocity offset from the mean velocity of the underlying stellar population, 264--270 km~s$^{-1}$, by more than 2.5$\sigma$ (75--50 
km~s$^{-1}$).  Both stars with large peculiar radial velocities are RG stars. These stars need to be investigated further for abundance anomaly to confirm their surviving companion status.

As compiled in Appendix \ref{appendix:summary},
the number of Type Ia SNRs in which surviving companions have been searched is too small for statistically significant conclusions.  More young Type Ia SNRs in nearby galaxies, such as M31 and M33, need to be surveyed to enlarge the sample for searches of surviving companions using 30m-class telescopes in the future.

\acknowledgments
We thank Dr.\ Thomas Krühler for his help with the VLT MUSE data reduction.  This project is supported by the
NASA grant HST-GO-13282.01-A. Y.-H.C. and C.-J.L. are
supported by Taiwanese Ministry of Science and Technology
grant MOST 108-2811-M-001-587. K.-C.P. is supported by the MOST grant MOST 107-2112-M-007-032-MY3. T.-W.C. acknowledgments the funding provided by the Alexander von Humboldt Foundation. The Programme IDs of VLT MUSE data acquired at ESO are 096.D-0352(A) and 096.D-0352(A).

% old MOST: 104-2112-M-001-044-MY3 (Y.-H.C. and C.-J.L.)

\appendix

%--------------------------------------------  
%             Table 

\section{Appendix A}
\label{appendix:summary}
\clearpage
\LongTables
\begin{landscape}
\tabletypesize{\tiny}
\begin{deluxetable*}{c|c|c|c|c|c|c|c}
\tablecaption{A detailed, up-to-date compilation of searches for Type Ia SN progenitors's surviving companions in Galactic and LMC SNRs} 
\tablehead{ Galaxy/\specialcell{SNR} & Reference & Observations & \specialcell{ Companion\\ Candidate} & Diagnostics & Result$^{\text{*}}$ & \specialcell{ Progenitors\\ (WD + ?)} }
\startdata 
MW/Tycho & \specialcell{\citet{ruiz-lapuente2004} \\ (RLCM04)} & \specialcell{ {\it{Photometry and astrometry}}: \hst WFPC2\\ F555W, F675W.  {\it{Spectroscopy}}: WHT 4.2 m\\ UES and ISIS, NOT 2.5 m ALFOSC,  Keck I\\ 10 m ESI and LRIS, Keck II 10 m LRIS }  & Tycho G & Radial velocity and proper motion & \scriptsize{+} & SD (WD+Tycho G). RG excluded.\\ %(WD+subgiant/MS)\\  
%\tabledashline  
\rule{0pt}{5ex}. & \citet{fuhrmann2005} & \specialcell{Radial velocity, proper motion and distance\\ from RLCM04} & Tycho G & The kinematics in Toomre diagram & \scriptsize{--} & . \\
%\tabledashline 
\rule{0pt}{6ex}. & \citet{ihara2007} & \specialcell{ {\it{Photometry}}:  Subaru 8.2 m Suprime-Cam\\ $V$, $R_{c}$, $I_{c}$. {\it{Spectroscopy}}: Subaru 8.2 m\\ FOCAS/MOS } & Tycho G & \specialcell{Blueshifted Fe I absorption by the\\ SN ejecta} &  \scriptsize{--} & SD (WD+Tycho E) \\
%\tabledashline 
\rule{0pt}{3ex}. & \citet{gonzalez2009} & \specialcell{ {\it{Spectroscopy}}: Keck I 10 m HIRES and LRIS } & Tycho G & Nickel (Ni) and cobalt (Co) abundances & \scriptsize{+} & SD (WD+Tycho G) \\
%\tabledashline 
\rule{0pt}{6ex}. & \citet{kerzendorf2009} & \specialcell{ {\it{Astrometry}}: Palomar 5.1 m, INT 2.5 m\\ {\it{Spectroscopy}}: Subaru 8.2 m HDS } & Tycho G & \specialcell{Radial velocity, distance, stellar\\ parameters, rotational velocity\\ and proper motion} & \scriptsize{--}  & . \\
%\tabledashline 
\rule{0pt}{3ex}. & \citet{lu2011} & \specialcell{ {\it{X-ray imaging}}: \chandra ACIS } & Tycho G & Proper motion and X-ray morphology & \scriptsize{+}  & SD \\
%\tabledashline 
\rule{0pt}{5ex}. & \citet{kerzendorf2013} & \specialcell{ {\it{Astrometry}}: \hst ACS/WFC F555W\\ {\it{Spectroscopy}}: Keck I 10 m HIRES and LRIS} & Tycho G & \specialcell{Ni abundance, radial velocity, distance,\\ rotational velocity and proper motion} & \scriptsize{--}  & Non-classical SD or DD (WD+WD) \\
%\tabledashline 
\rule{0pt}{6ex}. & \specialcell{\citet{bedin2014} \\ (BRLGH14)}   & \specialcell{ {\it{Photometry and astrometry}}: \hst ACS/WFC\\ F555W, WFC3/UVIS F555W.  {\it{Spectroscopy}}:\\Keck I 10 m HIRES, La Silla 3.6 m HARPS} & Tycho G & \specialcell{Proper motion, metallicity and Ni\\ abundance} & \scriptsize{+} & SD (WD+Tycho G) or DD (WD+WD)\\
%\tabledashline 
\rule{0pt}{6ex}. & \citet{xue2015} & \specialcell{ {\it{Astrometry}}: 42 observations from\\ seven astronomers. } & Tycho G & \specialcell{Position in the SNR} & \,\,\,\scriptsize{--}$^\dagger$ & .\\
%\tabledashline 
%\rule{0pt}{4ex}. & \citet{williams2016} & \textbf{\specialcell{ {\it{SNR imaging}}: \chandra ACIS, VLA  }} & \textbf{.} & \textbf{\specialcell{Position in the SNR}} & \textbf{\scriptsize{?}} & \textbf{.}\\
%\tabledashline 
\rule{0pt}{7ex}. & \citet{ruiz-lapuente2019} & \specialcell{ {\it{Photometry and astrometry}}: Gaia DR2.\\Photometry, astrometry and radial velocity\\ from BRLGH14. } & Tycho G & \specialcell{Galactic orbital kinematics\\ and metallicity} & \scriptsize{+} & \specialcell{SD (WD+Tycho G) or DD (WD+WD)\\ or CD (WD+AGB)}\\
%\tabledashlinegap 
\rule{0pt}{6ex}MW/Tycho & \citet{ihara2007} & \specialcell{ {\it{Photometry}}:  Subaru 8.2 m Suprime-Cam\\ $V$, $R_{c}$, $I_{c}$.  {\it{Spectroscopy}}: Subaru 8.2 m\\ FOCAS/MOS  } & Tycho E & \specialcell{Blueshifted Fe I absorption by the\\ SN ejecta}  & \scriptsize{+} & \specialcell{SD (WD+Tycho E) }\\
%\tabledashline 
\rule{0pt}{3ex}. & \citet{gonzalez2009} & \specialcell{ {\it{Spectroscopy}}: Keck I 10 m HIRES and LRIS  } & Tycho E & A double-lined spectroscopic binary  & \scriptsize{--} & \specialcell{SD (WD+Tycho G)}\\
%\tabledashline  
\rule{0pt}{5ex}. & \citet{kerzendorf2009} & \specialcell{ {\it{Astrometry}}: Palomar 5.1 m, INT 2.5 m\\ {\it{Spectroscopy}}: Subaru 8.2 m HDS  } & Tycho E & Radial velocity and proper motion & \scriptsize{--} & \specialcell{.}\\
%\tabledashline   
\rule{0pt}{5ex}. & \citet{kerzendorf2013} & \specialcell{ {\it{Astrometry}}: \hst ACS/WFC F555W\\ {\it{Spectroscopy}}: Keck I 10 m HIRES and LRIS } & Tycho E & \specialcell{Radial velocity, rotational velocity,\\ proper motion and distance}  & \scriptsize{--} & \specialcell{Non-classical SD or DD (WD+WD) }\\
%\tabledashline 
\rule{0pt}{6ex}. & \citet{xue2015} & \specialcell{ {\it{Astrometry}}: 42 observations from\\ seven astronomers. }& Tycho E & \specialcell{Position in the SNR} & \,\,\,\scriptsize{--}$^\dagger$ & .\\
%\tabledashline  
\rule{0pt}{7ex}. & \citet{ruiz-lapuente2019} & \specialcell{ {\it{Photometry and astrometry}}: Gaia DR2.\\Photometry, astrometry and radial velocity\\ from BRLGH14  } & Tycho E & Distance & \scriptsize{--} & \specialcell{SD (WD+Tycho G) or DD (WD+WD)\\ or CD (WD+AGB)}\\
%\tabledashlinegap  
MW/Tycho & \citet{kerzendorf2013} & \specialcell{ {\it{Astrometry}}: \hst ACS/WFC F555W\\ {\it{Spectroscopy}}: Keck I 10 m HIRES and LRIS } & Tycho B & \specialcell{ \rule{0pt}{3ex} Radial velocity and proper motion\\ ........................................................\\ \rule{0pt}{3ex} Temperature, rotational velocity,\\ abundance, and position in the SNR } & \specialcell{\scriptsize{\,--}\\\,\,........\\ \rule{0pt}{4ex} \scriptsize{+}} & \specialcell{Non-classical SD or DD (WD+WD) }\\
%\tabledashline  
\rule{0pt}{6ex}. & \specialcell{\citet{bedin2014} \\ (BRLGH14)}  & \specialcell{ {\it{Photometry and astrometry}}: \hst ACS/WFC\\ F555W, WFC3/UVIS F555W.  {\it{Spectroscopy}}:\\Keck I 10 m HIRES, La Silla 3.6 m HARPS} & Tycho B & \specialcell{Rotational velocity, radial velocity,\\ and proper motion} & \scriptsize{--} & \specialcell{SD (WD+Tycho G) or DD (WD+WD)}\\
%\tabledashline 
\rule{0pt}{6ex}. & \citet{xue2015} & \specialcell{ {\it{Astrometry}}: 42 observations from\\ seven astronomers. }& Tycho B & \specialcell{Position in the SNR} & \,\,\,\scriptsize{--}$^\dagger$ & .\\
%%\tabledashline 
\rule{0pt}{7ex}. & \citet{kerzendorf2018b} & \specialcell{ {\it{Spectroscopy}}: \hst STIS  } & Tycho B & \specialcell{Fe II absorption by cold SN ejecta\\ in the UV spectra and the luminosity\\ distance} & \scriptsize{--} & \specialcell{Non-classical SD}\\
%\tabledashline  
. & \citet{ruiz-lapuente2019} & \specialcell{ {\it{Photometry and astrometry}}: Gaia DR2.\\Photometry, astrometry and radial velocity\\ from BRLGH14  }  & Tycho B & Radial velocity and proper motion & \scriptsize{--} & \specialcell{SD (WD+Tycho G) or DD (WD+WD)\\ or CD (WD+AGB)}\\
%\tabledashline 
%\rule{0pt}{6ex} MW/Tycho & \citet{zhou2016} & \textbf{\specialcell{ {\it{SNR imaging}}:  IRAM 30 m }} & \textbf{.} & \textbf{\specialcell{}} & \textbf{\scriptsize{.}} & \textbf{SD (WD+)}\\
%\tabledashline  
%MW/SN 1006 & \citet{schweizer1980} & \textbf{\specialcell{  {\it{Photometry}}:  \\ {\it{Spectroscopy}}: }} & \textbf{SM-star} & \textbf{\specialcell{ }} & \textbf{\scriptsize{+}} & \textbf{\specialcell{SD (WD+) }}\\
%\tabledashline 
%. & \citet{wu1983} & \textbf{\specialcell{  {\it{Photometry}}: . \\ {\it{Spectroscopy}}: }} & \textbf{SM-star} & \textbf{\specialcell{}} & \textbf{\scriptsize{--}} & \textbf{\specialcell{}}\\
\tabledashline 
MW/SN 1006 & \citet{gonzalez2012} & \specialcell{  {\it{Photometry}}: 2MASS $R$, $B$, $J$, $H$, $K$. \\ {\it{Spectroscopy}}: VLT-UT2 Kueyen 8.2 m UVES} & None & \specialcell{Radial velocity, rotational velocity,\\ distance, and abundances  } & \scriptsize{--} & \specialcell{SD (WD+MS$<$\msun) or DD (WD+WD)}\\
%\tabledashline  
\rule{0pt}{5ex}. & \citet{kerzendorf2012} & \specialcell{  {\it{Photometry}}: ANU 2.3 m Imager $U$, $B$, $V$, $I$. \\ {\it{Spectroscopy}}: VLT-UT2 FLAMES } & None & \specialcell{ Radial velocity, rotational velocity,\\ and stellar parameters } & \scriptsize{--} & \specialcell{Non-classical SD or DD (WD+WD) }\\
%\tabledashline  
\rule{0pt}{6ex}. & \citet{kerzendorf2018a} & \specialcell{  {\it{Photometry}}: CTIO Blanco 4 m DECam\\ $u$, $g$, $r$, $z$ } & None & \specialcell{WD models in CMD  } & \scriptsize{--} & \specialcell{Non-classical progenitors\\ (WD+anomalously red/dim WD)\\ or DD (WD+WD). Young\\ WDs models $\leq$10$^8$ yr excluded}\\
%\tabledashline 
%\rule{0pt}{6ex} MW/Kepler & \citet{chiotellis2012} & \textbf{\specialcell{ {\it{SNR imaging}}:   }} & \textbf{.} & \textbf{\specialcell{Kinematics, morphology,\\ and chemical abundances}} & \textbf{\scriptsize{.}} & \textbf{SD (WD+AGB)}\\
\tabledashline  
\rule{0pt}{6ex} MW/Kepler & \citet{kerzendorf2014} & \specialcell{  {\it{Photometry}}: \hst ACS/WFC F550M,\\ NOMAD catalog $B$, $V$, $R$, $J$, $H$, $K$.\\  {\it{Spectroscopy}}: ANU 2.3 m WiFeS} & Many & \specialcell{ Radial velocity and stellar luminosity } & \scriptsize{?} & \specialcell{RG excluded}\\
%\tabledashline   
\rule{0pt}{9ex}. & \citet{ruiz-lapuente2018b} & \specialcell{  {\it{Photometry and astrometry}}: \hst ACS/WFC\\ F502N, F660N, F658N, F550M, WFC3/UVIS\\ F336W, F438W, F547M, F656N, F658N,\\ F814W.  {\it{Spectroscopy}}: VLT-UT2 FLAMES\\ UVES and Giraffe} & None & \specialcell{ Radial velocity, rotational velocity,\\ and proper motion} & \scriptsize{--} & \specialcell{DD (WD+WD) or CD (WD+AGB) }\\
\tabledashline  
\rule{0pt}{7ex}\specialcell{LMC/SNR\\ 0509-67.5} & \citet{schaefer2012} & \specialcell{  {\it{SNR imaging}}: \hst WFPC2 F656N,\\ \chandra  ACIS.  {\it{Photometry}}:  \hst\\ WFC3/UVIS F475W, F555W, F814W } & None & \specialcell{ Stellar luminosity} & \scriptsize{--} & \specialcell{DD (WD+WD)}\\
%\tabledashline   
%\rule{0pt}{6ex} . & \citet{distefano2012} & \textbf{\specialcell{ {\it{}}:   }} & \textbf{.} & \textbf{\specialcell{}} & \textbf{\scriptsize{.}} & \textbf{SD not excluded}\\
%\tabledashline   
\rule{0pt}{6ex} . & \citet{hovey2016} & \specialcell{ {\it{SNR imaging}}:  \hst ACS/WFC F656N.\\ {\it{Photometry}}:  \hst WFC3/UVIS F110W, \\F160W, F475W, F555W,F814W } & Many & \specialcell{Position in the SNR and \\stellar color and luminosity } & \scriptsize{?} & SD not excluded\\
%\tabledashline  
\rule{0pt}{9ex}. & \specialcell{ \citet{litke2017}} & \specialcell{  {\it{SNR imaging}}: \hst ACS/WFC F658N,\\ WFPC2 F656N.  {\it{Photometry}}: \hst\\ WFC3/UVIS F475W, F555W, F814W,\\ WFC3/IR F110W, F160W,\\ {\it{Spitzer}} IRAC 3.6, 4.5, 5.8, 8.0 $\mu$m} & None & \specialcell{Models of surviving companions\\ in CMDs } & \scriptsize{--} & \specialcell{DD (WD+WD)}\\
\tabledashline  
\specialcell{LMC/SNR\\ 0519-69.0} & \citet{edwards2012} & \specialcell{ {\it{SNR imaging}}: \hst ACS/WFC F658N,\\ \chandra ACIS.  {\it{Photometry}}: \hst\\ ACS/WFC F550M, F658N} & Many & \specialcell{Stellar color and luminosity  } & \scriptsize{?} & \specialcell{SD (WD+MS) or DD (WD+WD)}\\
%\tabledashline   
\rule{0pt}{7ex}. & \specialcell{ This paper } & \specialcell{  {\it{SNR imaging}}: \hst ACS/WFC F658N.\\  {\it{Photometry}}: \hst ACS/WFC F550M,\\ WFC3/UVIS F475W, F814W.\\  {\it{Spectroscopy}}: VLT-UT4 MUSE  } & Star 5 & \specialcell{ Radial velocity and models of\\ surviving companions in CMDs} & \scriptsize{+} & \specialcell{SD (WD+star) or DD (WD+WD) }\\
\tabledashline  
LMC/N103B & \citet{pagnotta2015} & \specialcell{  {\it{SNR imaging}}: Gemini 8.1 m GMOS H$\alpha$,\\ \chandra ACIS, Australia Telescope.\\  {\it{Photometry}}: Gemini 8.1 m GMOS $g'$, $r'$, $i'$} & Many & \specialcell{Stellar color and luminosity } & \scriptsize{?} & \specialcell{SD (WD+star) or DD (WD+WD) }\\   
\rule{0pt}{9ex}. & \specialcell{ \citet{li2017}\\ } & \specialcell{  {\it{SNR imaging}}: \hst WFC3/UVIS F656N,\\ CTIO Blanco 4 m MOSAIC II H$\alpha$.\\  {\it{Photometry}}: \hst WFC3/UVIS F475W,\\ F555W, F814W. Spectroscopy: CTIO 4 m\\ echelle, CTIO SMARTS 1.5 m CHIRON} & Star 1 & \specialcell{ Stellar colors and luminosity, \\models of surviving companions in\\ CMDs, and position in the SNR} & \scriptsize{+} & \specialcell{SD (WD+star)}\\
\tabledashline  
LMC/DEM L71 & \citet{pagnotta2015} & \specialcell{ {\it{SNR imaging}}: Gemini 8.1 m GMOS H$\alpha$,\\ \chandra ACIS.  {\it{Photometry}}: Gemini\\ 8.1 m GMOS $g'$, $r'$, $i'$} & Many & \specialcell{Stellar color and luminosity } & \scriptsize{?} & \specialcell{SD (WD+star) or DD (WD+WD) }\\
%\tabledashline   
\rule{0pt}{7ex}. & \specialcell{This paper} & \specialcell{  {\it{SNR imaging}}: \hst WFC3/UVIS F656N.\\  {\it{Photometry}}: \hst WFC3/UVIS F475W,\\ F555W, F814W.  {\it{Spectroscopy}}: VLT-UT4 MUSE } & Star 80 & \specialcell{Radial velocity and models of\\ surviving companions in CMDs } & \scriptsize{+} & \specialcell{SD (WD+star) or DD (WD+WD) }\\
\tabledashline  
\rule{0pt}{6ex} \specialcell{LMC/SNR\\ 0548-70.4} & \specialcell{This paper}  & \specialcell{  {\it{SNR imaging}}: \hst WFC3/UVIS F656N.\\  {\it{Photometry}}: \hst WFC3/UVIS F475W,\\ F555W, F814W } & Star 11 & \specialcell{ Models of surviving companions\\ in CMDs } & \scriptsize{?} & \specialcell{SD (WD+star) or DD (WD+WD) }

\enddata
\tablenotetext{*}{ The symbol "+" and "--" represents a positive and negative result, respectively.}
\tablenotetext{$\dagger$}{See \citet{williams2016} for discussions.}
\label{table:summary}
\end{deluxetable*}
\clearpage
\end{landscape}

\section{Appendix B}
\label{appendix:0548}

%--------------------------------------------
%             Table 8
\LongTables
\begin{deluxetable*}{ccccccccc}
\tablecolumns{9}
\tabletypesize{\scriptsize}
%\rotate
\tablewidth{0pc}
\tablecaption{Stars Brighter Than $V = 23.0$ mag near Central Region in the SNR 0548--70.4}
\tablehead{Star & R.A. (J2000) & Decl. (J2000) & B & V & I & B-V & V-I & $r$}
\startdata
1 & 05:47:48.339 & -70:24:51.19 & 22.62 $\pm$ 0.01 & 22.32 $\pm$ 0.01 & 21.59 $\pm$ 0.01 & 0.30 $\pm$ 0.01 & 0.73 $\pm$ 0.01 & 1.39\arcsec\\
2 & 05:47:47.950 & -70:24:53.02 & 22.81 $\pm$ 0.01 & 22.53 $\pm$ 0.01 & 21.81 $\pm$ 0.01 & 0.28 $\pm$ 0.01 & 0.72 $\pm$ 0.01 & 2.65\arcsec\\
3 & 05:47:48.004 & -70:24:50.30 & 21.77 $\pm$ 0.01 & 21.55 $\pm$ 0.01 & 20.96 $\pm$ 0.01 & 0.22 $\pm$ 0.01 & 0.59 $\pm$ 0.01 & 3.14\arcsec\\
4 & 05:47:48.290 & -70:24:49.32 & 22.37 $\pm$ 0.01 & 22.15 $\pm$ 0.01 & 21.55 $\pm$ 0.01 & 0.22 $\pm$ 0.01 & 0.60 $\pm$ 0.01 & 3.23\arcsec\\
5 & 05:47:47.951 & -70:24:50.01 & 23.12 $\pm$ 0.01 & 22.87 $\pm$ 0.01 & 22.09 $\pm$ 0.01 & 0.25 $\pm$ 0.02 & 0.78 $\pm$ 0.02 & 3.54\arcsec\\
6 & 05:47:48.103 & -70:24:55.48 & 21.64 $\pm$ 0.01 & 21.49 $\pm$ 0.01 & 21.03 $\pm$ 0.01 & 0.15 $\pm$ 0.01 & 0.46 $\pm$ 0.01 & 3.55\arcsec\\
7 & 05:47:48.863 & -70:24:48.68 & 21.13 $\pm$ 0.01 & 21.03 $\pm$ 0.00 & 20.73 $\pm$ 0.01 & 0.10 $\pm$ 0.01 & 0.30 $\pm$ 0.01 & 4.25\arcsec\\
8 & 05:47:49.342 & -70:24:52.83 & 23.13 $\pm$ 0.01 & 22.85 $\pm$ 0.01 & 22.05 $\pm$ 0.01 & 0.28 $\pm$ 0.02 & 0.80 $\pm$ 0.02 & 4.43\arcsec\\
9 & 05:47:47.979 & -70:24:56.27 & 18.95 $\pm$ 0.00 & 18.56 $\pm$ 0.00 & 17.48 $\pm$ 0.00 & 0.39 $\pm$ 0.00 & 1.08 $\pm$ 0.00 & 4.55\arcsec\\
10 & 05:47:48.953 & -70:24:48.59 & 22.87 $\pm$ 0.01 & 22.62 $\pm$ 0.01 & 21.89 $\pm$ 0.01 & 0.25 $\pm$ 0.01 & 0.73 $\pm$ 0.01 & 4.56\arcsec
\enddata
\tablenotetext{*}{This table only contains a 10-line example. The full data that contains 973 data lines only appears in the machine readable format in the HTML.}
\label{table:photometry0548}
\end{deluxetable*}

\end{document}